\documentclass[12pt,preprint]{aastex}
\usepackage{amssymb}
\usepackage{amsmath}
\usepackage{graphicx}

\newcommand\kms{km~s$^{-1}$}

\newcommand\co{$^{12}$CO}
\newcommand\mjb{mJy~beam$^{-1}$}
\newcommand\jb{Jy~beam$^{-1}$}

\newcommand\pp{$^{\prime\prime}$}
\newcommand\um{$\mu$m}

\newcommand\q{$\sim$}

\newcommand\h{H$_{2}$}
\newcommand\msun{M$_{\odot}$}  

\newcommand\lsun{L$_{\odot}$}

\newcommand{\methanol}{CH$_3$OH}
\newcommand{\ammonia}{NH$_3$}
\newcommand{\eup}{E$_{upper}$}

\newcommand{\hco}{HCO$^{+}$}
\newcommand{\hisoco}{H$^{13}$CO$^{+}$}
\newcommand\water{H$_{2}$O}

\newcommand{\meth}{CH$_3$OH}

\newcommand{\vlsr}{$v_{LSR}$}
\newcommand{\methcyn}{CH$_{3}$CN}
\newcommand{\degree}{$^{\circ}$}

\newcommand{\noprint}[1]{}

\begin{document}
\shortauthors{Cyganowski et al.}

\title{Bipolar Molecular Outflows and Hot Cores in GLIMPSE Extended Green Objects (EGOs)}
\author{C.J. Cyganowski\altaffilmark{1,3}, C.L. Brogan\altaffilmark{2},
  T.R. Hunter\altaffilmark{2}, E. Churchwell\altaffilmark{1}, Q. Zhang\altaffilmark{3}}

\email{ccyganowski@cfa.harvard.edu}

\altaffiltext{1}{University of Wisconsin, Madison, WI 53706}
\altaffiltext{2}{NRAO, 520 Edgemont Rd, Charlottesville, VA 22903}
\altaffiltext{3}{NSF Astronomy and Astrophysics Postdoctoral Fellow, Harvard-Smithsonian Center for Astrophysics, Cambridge, MA 02138}

\begin{abstract}

We present high angular resolution Submillimeter Array (SMA) and
Combined Array for Research in Millimeter-wave Astronomy (CARMA)
observations of two GLIMPSE Extended Green Objects (EGOs)--massive
young stellar object (MYSO) outflow candidates identified based on
their extended 4.5 \um\/ emission in \emph{Spitzer} images.  The mm
observations reveal bipolar molecular outflows, traced by
high-velocity \co(2-1) and \hco(1-0) emission, coincident with the 4.5
\um\/ lobes in both sources.  SiO(2-1) emission 
confirms that the extended 4.5 \um\/ emission traces active
outflows.  A single dominant outflow is identified in each EGO, with
tentative evidence for multiple flows in one source (G11.92$-$0.61).
The outflow driving sources are compact millimeter continuum cores,
which exhibit hot-core spectral line emission and are associated with
6.7 GHz Class II \meth\/ masers.  G11.92$-$0.61 is associated with at
least three compact cores: the outflow driving source, and two cores
that are largely devoid of line emission.  In contrast, G19.01$-$0.03
appears as a single MYSO.  The difference in multiplicity, the
comparative weakness of its hot core emission, and the dominance of
its extended envelope of molecular gas all suggest that G19.01$-$0.03
may be in an earlier evolutionary stage than G11.92$-$0.61.  Modeling
of the G19.01$-$0.03 spectral energy distribution suggests that a
central (proto)star (M \q10 \msun) has formed in the compact mm core
(M$_{gas}$ \q 12-16\msun), and that accretion is ongoing at a rate of
\q10$^{-3}$ \msun\/ year$^{-1}$.  Our observations confirm that these
EGOs are young MYSOs driving massive bipolar molecular outflows, and
demonstrate that considerable chemical and evolutionary diversity are
present within the EGO sample.

\end{abstract}


\section{Introduction}\label{introduction}

Massive star formation remains a poorly understood phenomenon, largely due to
the difficulty of identifying and studying massive
young stellar objects (MYSOs)\footnote{We define MYSOs as young stellar
  objects (YSOs) that will become main sequence stars of M$>$8 \msun\/ (O or
  early-B type ZAMS stars).} in the crucial early active accretion and outflow
phase.  
During the earliest stages of their evolution, young MYSOs remain deeply embedded in their natal clouds.
Most massive star-forming regions are also distant ($>$ 1kpc) and crowded,
with massive stars forming in close proximity
to other MYSOs and to large numbers of lower-mass YSOs.  
Studying the early stages of massive star formation thus requires high angular
resolution observations (to resolve individual objects in crowded regions) at
long wavelengths unaffected by extinction.




Large-scale \emph{Spitzer} surveys of the Galactic Plane have yielded a
promising new sample of young MYSOs with \emph{active outflows}, which may be
inferred to be actively accreting.  Identified based on their extended 4.5
\um\/ emission in \emph{Spitzer} images, these sources are known as ``Extended
Green Objects (EGOs)'' \citep{egocat,maserpap} or ``green fuzzies''
\citep{Chambers09} from the common coding of the 4.5 \um\/ band as green in
3-color IRAC images.  In active protostellar outflows, the \emph{Spitzer} 4.5
\um\/ broadband flux can be dominated by emission from shock-excited molecular
lines \citep[predominantly \h:][]{SmithRosen05,Smith06,Davis07,Ybarra09,Ybarra10,DeBuizer10}.  The
resolution of \emph{Spitzer} at 4.5 \um\/ ($\sim$2\pp) is sufficient to
resolve the extended emission from outflows in massive star forming regions
nearer than $\sim$7 kpc.
Over 300 EGOs have been cataloged in the Galactic Legacy
Infrared Mid-Plane Survey Extraordinaire (GLIMPSE-I) survey area by \citet{egocat}.  
The mid-infrared (MIR) colors of EGOs are consistent with those of young protostars
still embedded in infalling envelopes \citep{egocat}.
A majority of EGOs are also associated with infrared dark clouds (IRDCs),
identified by recent studies as sites of the earliest stages of massive star
and cluster formation \citep[e.g.][]{Rathborne07,Chambers09}.


Remarkably high detection rates for two diagnostic types of \meth\/
masers in high-resolution Very Large Array (VLA) surveys provide
strong evidence that GLIMPSE EGOs are indeed \emph{massive} YSOs with
active outflows \citep{maserpap}.  There are two Classes of \meth\/
masers, both associated with star formation, but excited under different
conditions by different mechanisms.  Class II 6.7 GHz \meth\/ masers
are radiatively pumped by IR emission from warm dust \citep[e.g.][and
references therein]{Cragg05} and are associated exclusively with
massive YSOs \citep[e.g.][]{Minier03,Bourke05,Xu08,Pandian08}.  Class
I 44 GHz \meth\/ masers are collisionally excited in molecular
outflows, and particularly at interfaces between outflows and the
surrounding ambient cloud \citep[e.g.][]{PlambeckMenten90, Kurtz04}.
Of a sample of 28 EGOs, $>$64\% have 6.7 GHz Class II \meth\/ masers
(nearly double the detection rate of surveys using other MYSO
selection criteria), and of these 6.7 GHz maser sources, $\sim$89\%
also have 44 GHz masers \citep{maserpap}.

A complementary James Clerk Maxwell Telescope (JCMT; resolution \q20\pp) molecular line survey towards EGOs with 6.7
GHz \meth\/ maser detections found SiO(5-4) emission and \hco(3-2) line profiles
consistent with the presence of active molecular outflows \citep{maserpap}.  SiO is particularly well-suited to
tracing \emph{active} outflows, as it persists in the gas phase for
only $\sim$10$^{4}$ years after being released by shocks \citep[e.g.][]{pdf97}. 
A single-dish (resolution \q80\pp) 3 mm spectral line survey of all EGOs visible
from the northern hemisphere by \citet{Chen10} found associated gas/dust clumps
of mass 69-29000 \msun, consistent with the
identification of EGOs as MYSOs.   
The nature of the driving sources of the 4.5 \um\/ outflows is only loosely
constrained by the survey results.  Bright ultracompact (UC) HII regions are, in most cases,
ruled out as powering sources by the lack of VLA 44 GHz continuum detections \citep{maserpap}.
A high detection rate (83\%) for thermal \meth\/ emission in the
\citet{maserpap} JCMT survey indicates the presence of warm dense gas, and possible hot
core line emission.


Further understanding of the nature of EGOs, and their implications for the
mode(s) of high-mass star formation, requires identifying 
the driving source(s) and characterizing their physical properties, as well as
those of the outflows associated with EGOs.
Interferometric millimeter-wavelength line and continuum observations provide access to
\emph{direct} tracers of molecular outflows and dense, compact gas and dust cores,
including a wealth of chemical diagnostics.
In this paper, we present Submillimeter Array (SMA)\footnote{The Submillimeter
Array is a joint project between the Smithsonian Astrophysical Observatory and
the Academia Sinica Institute of Astronomy and Astrophysics and is funded by
the Smithsonian Institution and the Academia Sinica.}  and Combined Array for
Research in Millimeter-wave Astronomy (CARMA)\footnote{Support for CARMA
construction was derived from the Gordon and Betty Moore Foundation, the
Kenneth T. and Eileen L. Norris Foundation, the James S. McDonnell Foundation,
the Associates of the California Institute of Technology, the University of
Chicago, the states of California, Illinois, and Maryland, and the National
Science Foundation. Ongoing CARMA development and operations are supported by
the National Science Foundation under a cooperative agreement, and by the
CARMA partner universities.} observations at 1 and 3 mm of two EGOs from the
\citet{maserpap} sample: G11.92$-$0.61 and G19.01$-$0.03.  The targets were
chosen to have bipolar (and in some cases quadrupolar) 4.5 \um\/ morphology, associated 24 \um\/ emission,
associated (sub)mm continuum emission in single-dish surveys, and 6.7 Class II and
44 GHz Class I \methanol\/ maser detections in the \citet{maserpap} survey.
The promise of extended 4.5 \um\/ emission as a MYSO diagnostic lies largely
in its ability to identify very young sources with ongoing accretion and
outflow that are missed by other sample selection
methods.  These sources had not been targeted for study prior to their
identification as EGOs and
inclusion in the \citet{maserpap} sample, and very little is known about them
beyond the results of that survey (see also \S\ref{g11_previous} and \S\ref{g19_previous}).
In \S\ref{obs} we describe our observations, and in \S\ref{results} we present
our results.  In \S\ref{discussion} we discuss the physical properties of the
compact cores and outflows associated with our target EGOs, and in
\S\ref{conclusions} we summarize our conclusions.

\section{Observations}\label{obs}

\subsection{Submillimeter Array (SMA)}\label{smaobs}
SMA observations of our target EGOs were obtained on 23 June 2008 with
eight antennas in the compact-north configuration.  The observational
parameters, including calibrators, are summarized in
Table~\ref{obstable}.  Two pointings were observed in a single track:
G11.92$-$0.61 at $\alpha=$18$^{\rm h}$13$^{\rm m}$58$^{\rm s}$.1,
$\delta=-$18$^{\circ}$54\arcmin16\farcs7, and G19.01$-$0.03 at $\alpha$=18$^{\rm
h}$25$^{\rm m}$44$^{\rm s}$.8, $\delta=-$12$^{\circ}$22\arcmin45\farcs8
(J2000).  The average 225 GHz opacity during the observations was
\q0.26, with typical system temperatures at source transit of 220 K.
In the compact-north configuration, the array is insensitive to smooth
structures larger than \q20\pp.
The projected baseline lengths ranged from 7 to 88 k$\lambda$.  The
double-sideband SIS receivers were tuned to a local oscillator
frequency of 225.11 GHz, providing coverage of 219.1-221.1 GHz in the
lower sideband (LSB) and 229.1-231.1 GHz in the upper sideband (USB).
The spectral lines detected are reported in \S\ref{g11compact} and
\S\ref{g19_compact}.
   
Initial calibration of the data was performed in MIRIAD.  Each sideband was
reduced independently, and the calibrated data were exported to AIPS.  The
AIPS task UVLSF was used to separate the line and continuum emission, using
only line-free channels to estimate the continuum.  The continuum data were
then self-calibrated, and the solutions transferred to the line data.  After
self-calibration, the continuum data from the lower and upper sidebands were
combined.  Imaging was performed in CASA using Briggs weighting and a robust
parameter of 0.5.  The synthesized beam size is 3\farcs2$\times$1\farcs8
(P.A.$=$59$^{\circ}$) for G11.92 and 3\farcs2$\times$1\farcs7
(P.A.$=$63$^{\circ}$) for G19.01.  The 1$\sigma$ rms noise level in the
continuum images is 3.5 \mjb.  
The correlator was configured to provide a uniform spectral resolution of
0.8125 MHz.  The line data were resampled to a velocity resolution of 1.1
\kms, then Hanning smoothed.  The typical noise level in a single channel of the
Hanning-smoothed spectral line images is 100 \mjb.  The \co\/ data were
further smoothed to a resolution of 3.3 \kms; the noise in a single channel is
60 \mjb.  All measurements were made from images corrected for the primary
beam response.

Flux calibration was based on observations of Uranus
and a model of its brightness distribution using the MIRIAD task smaflux.
Comparison of the derived fluxes of the observed quasars (including 3C279, included as
an alternate bandpass calibrator) with SMA flux monitoring suggests that the
absolute flux calibration is good to $\lesssim$15\%.  The absolute position
uncertainty is estimated to be 0\farcs3.

\subsection{Combined Array for Research in Millimeter-wave Astronomy
  (CARMA)}\label{carmaobs}

Our 3 mm CARMA observations were obtained on 29 July 2008 (G11.92$-$0.61) and
30 July 2008 (G19.01$-$0.03) in the D-configuration with 15 antennas (six 10.4
m and nine 6.1 m antennas).  The observational parameters, including calibrators, are summarized in
Table~\ref{obstable}.  The projected baselines ranged from 1.5 to 31
k$\lambda$ for the 29 July observations and from 1.5 to 36.5 k$\lambda$ for
the 30 July observations.  The correlator was configured to cover SO
(2$_2$-1$_1$) at 86.094 GHz (\eup=19.3 K) in LSB, SiO (2-1) at 86.846 GHz
(\eup=6.3 K) in LSB and
\hco(1-0) at 89.189 GHz (\eup=4.3 K) in USB with 31 MHz windows.  Each 31 MHz window
consisted of 63 channels, providing a spectral resolution of 0.488 MHz (\q1.7
\kms) and velocity coverage of \q100 \kms.  In addition, the correlator setup
included two 500 MHz (pseudo)continuum bands, each comprised of 15 channels:
one in LSB centered at \q85.7 GHz and one in USB centered at \q90.3 GHz.  In the D-configuration, the array is
insensitive to smooth structures larger than \q50\pp.
During the observations, the 230 GHz
opacity ranged from \q0.47 to 0.5 on 29 July and from \q0.46-0.54 on 30 July.
Typical (SSB) system temperatures at source transit were \q230-280 K on 29
July and \q190-240 K on 30 July.  The phase center was $\alpha=$18$^{\rm h}$13$^{\rm
m}$58$^{\rm s}$.1, $\delta=-$18$^{\circ}$54\arcmin16\farcs7 (J2000) on 29 July
(G11.92$-$0.61), and $\alpha=$18$^{\rm h}$25$^{\rm m}$44$^{\rm s}$.8,
$\delta=-$12$^{\circ}$22\arcmin45\farcs8 (J2000) on 30 July (G19.01$-$0.03).

The data were calibrated in MIRIAD and imaged in CASA, using Briggs
weighting and a robust parameter of 0.5.  The synthesized beamsize is
6\farcs8$\times$5\farcs1 (P.A.$=$11$^{\circ}$) for G11.92 and
5\farcs7$\times$5\farcs1 (P.A.$=$-27$^{\circ}$) for G19.01.  Each band
was reduced independently.  The 3 mm continuum data for G11.92$-$0.61
are not presented here because of aliasing from \emph{IRAS}
18110-1854, which is a much brighter continuum source than the target
EGO.  The \emph{IRAS} source is \q1\arcmin\/ northeast of the EGO,
within the primary beam FWHP of the 6.1 m antennas (\q130\pp) but only
at the 20\% response level of the primary beam of the 10.4 m antennas
(FWHP \q80\pp).  For G19.01$-$0.03, the continuum data from the upper
and lower sidebands were combined to make a final continuum image with
a 1$\sigma$ rms noise level of 0.5 \mjb.  All line data were Hanning
smoothed to improve the signal to noise.  The noise level in a single
channel of the Hanning-smoothed spectral line images is \q 12 \mjb\/
for G11.92$-$0.61 and \q 9 \mjb\/ for G19.01$-$0.03.  No continuum
subtraction was performed, as the continuum contribution to each
channel of the line data is negligible.  For G19.01$-$0.03, the 3.4 mm
continuum peak intensity (9.4 \mjb) is less than the 1$\sigma$ rms
level in the line datacube (10 \mjb).  For G11.92$-$0.61,
extrapolating the 1.3 mm continuum peak intensity to 3.4 mm (assuming
a spectral index of three) likewise predicts a continuum contribution
to the line data at the $\lesssim$1$\sigma$ level.  All measurements
were made from images corrected for the convolved primary beam
response of the heterogeneous CARMA array; in CASA this calculation is
done in the visibility domain.



Flux calibration was based on
observations of Uranus and a model of its brightness distribution using the
MIRIAD task smaflux.  Comparison of the derived fluxes of the observed quasars
(including 3C273, observed as an alternate bandpass calibrator) with CARMA
flux monitoring suggests that the absolute flux calibration is good to
$\lesssim$15\%.  

In addition to the 3 mm observations described above, we obtained 1 mm
observations of G11.92$-$0.61 with CARMA in the C-configuration on 25 April
2008 with eleven antennas (three 10.4 m antennas and eight 6.1 m antennas).
The
projected baselines ranged from 9 to 184.5 k$\lambda$, and the phase center
was $\alpha=$18$^{\rm h}$13$^{\rm m}$58$^{\rm s}$.1 $\delta=-$18$^{\circ}$54\arcmin16\farcs7
(J2000).  In the C-configuration, the array is insensitive to smooth structures larger than \q15\pp.  
The (SSB) system temperature ranged from \q400-600 K during the
observations.  
The correlator was configured to cover SiO(5-4) at 217.105 GHz
and DCN(3-2) at 217.239 GHz in LSB and SO(5$_{6}$-4$_{5}$) at 219.949 GHz and
\meth(8$_{0,8}$-7$_{1,6}$) at 220.078 GHz in USB with 62 MHz windows and two
500 MHz (pseudo)continuum windows centered at \q216.45 GHz (LSB) and 220.75
GHz (USB).  Due to the high system temperatures and limited integration time,
only the 500 MHz bands have sufficient
signal-to-noise.  The USB 500 MHz window encompassed the \methcyn(J=12-11)
ladder, thus only the LSB (\q216.5 GHz) (psuedo)continuum data are presented here.
The absolute flux scale was set assuming a flux of 2.69 Jy for J1733-130 based
on quasar flux monitoring.  The uncertainty in the absolute amplitude
calibration is estimated to be \q20\%.  The data were calibrated in MIRIAD and
imaged in CASA, using Briggs weighting and a robust parameter of 0.5.  The
resulting 1 mm continuum image has a synthesized beamsize of
1\farcs44$\times$0\farcs87 (P.A.$=$25$^{\circ}$) and a 1$\sigma$ rms noise
level of 4.3 \mjb.  As with the 3 mm data, all measurements were
made from images corrected for the response of the heterogeneous primary
beams.

\section{Results}\label{results}

\subsection{G11.92$-$0.61}\label{g11results}

\subsubsection{Previous Observations: G11.92$-$0.61}\label{g11_previous}

The EGO G11.92$-$0.61 is \q1\arcmin\/ SE of the more evolved massive star
forming region \emph{IRAS} 18110-1854.  Single dish (sub)mm continuum maps
targeting the \emph{IRAS} source show a millimeter/submillimeter clump
coincident with the EGO \citep{Walsh03,Faundez04,Thompson06}.  
Strong ($>$240 Jy) \water\/ maser emission associated with the EGO was
likewise serendipitously detected in VLA observations targeting the
\emph{IRAS} source \citep{HC96}.  The \water\/ maser source was
subsequently included in the large single-dish \co\/ survey of \citet{ShC96}, who
detected broad \co\/ line wings.  

The MIR emission of G11.92$-$0.61 is characterized by a bipolar 4.5 \um\/
morphology, with a NE and a SW lobe.  The EGO is located in an IRDC \citep[][see also Fig.~\ref{contfig}a]{egocat}.  The SW
lobe is coincident with strong, blueshifted 44 GHz Class I \meth\/ masers,
while the NE lobe is surrounded by an arc of systemic to slightly redshifted
44 GHz masers \citep{maserpap}.  Elongated 24 \um\/ emission is coincident
with the NE 4.5 \um\/ lobe, as are two 6.7 GHz Class II \meth\/ masers (Fig.~\ref{contfig}).  This
EGO is unique among the \citet{maserpap} sample in having multiple spatially
and kinematically distinct loci of 6.7 GHz Class II \meth\/ maser emission.
The extended 24 \um\/ morphology and multiple 6.7 GHz maser spots suggest the
possible presence of multiple MYSOs.  SiO(5-4) and thermal
\meth(5$_{2,3}$-4$_{1,3}$) emission were detected towards G11.92$-$0.61 in
single-pointing JCMT observations (resolution \q 20\pp) targeted at the NE
lobe/24 \um\/ source \citep{maserpap}.  No 44 GHz continuum emission was
detected towards the EGO to a 5$\sigma$ sensitivity limit of 7 \mjb\/
\citep[resolution 0\farcs99 $\times$ 0\farcs44,][]{maserpap}.  We adopt the
near kinematic distance from \citet{maserpap} for G11.92$-$0.61: 3.8 kpc.

\subsubsection{Continuum Emission: G11.92$-$0.61}\label{g11cont}

\begin{figure}
\plotone{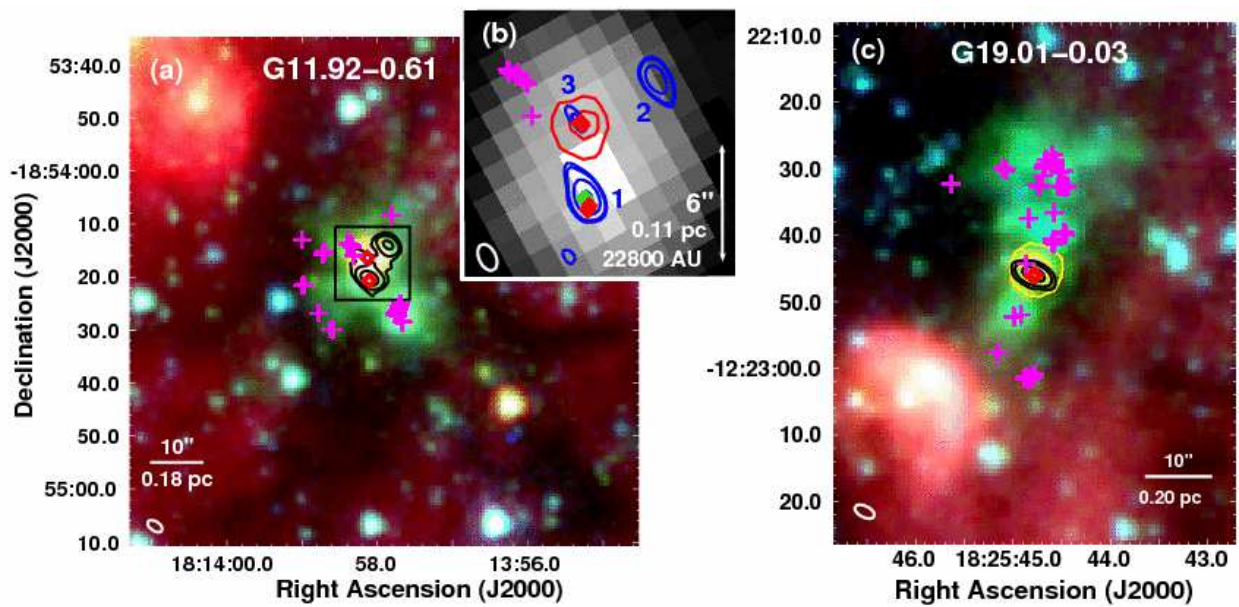}
\caption{Three color \emph{Spitzer} images (3.6, 4.5, 8.0
$\mu$m: blue, green, red) of the EGOs (a) G11.92$-$0.61 and (c) G19.01$-$0.03
overlaid with contours of 1.3 mm continuum emission from the SMA (black
contours, levels 5,10,30 $\times$ $\sigma=$ 3.5 \mjb).  The SMA beam is shown at lower
left in each panel.  In all panels, red diamonds mark
positions of 6.7 GHz \meth\/ masers and magenta crosses mark 
positions of 44 GHz \meth\/ masers from
\citet{maserpap}.
The 8 \um\/ nebula to the NE of G11.92$-$0.61 (extreme upper left in panel (a)) is associated with \emph{IRAS} 18110-1854.
The black rectangle in (a) 
indicates the field of view of (b).  (b) The inset shows the 24 \um\/
emission towards G11.92$-$0.61 in greyscale, overlaid with contours of
1.4 mm continuum emission (blue) from CARMA (levels 4,5,10,20
$\times$ $\sigma=$ 4.3 \mjb) and 8 \um\/ emission (red) from GLIMPSE (levels 250,500 MJy sr$^{-1}$).
G11.92$-$0.61-MM1, MM2, and MM3 are labeled ``1'', ''2'', and ``3'', respectively.
The resolution of the 24 \um\/ image is 6\pp\/ (scalebar) and 
the CARMA beam is shown at lower left.  The position of the \water\/ maser from
\citet{HC96} is marked with a filled green diamond.  The 24 \um\/ emission
towards G11.92$-$0.61 is saturated in the MIPS image (white pixels).
In (c), the MIPS 24 \um\/ emission towards G19.01$-$0.03 is shown as yellow
contours (levels 900, 1500 MJy sr$^{-1}$).
}
\label{contfig}
\end{figure}

Our 1.3 mm SMA and 1.4 mm CARMA data resolve three distinct compact continuum
sources (Fig.~\ref{contfig}a,b).  These sources are designated MM1, MM2, and
MM3 in order of descending peak intensity.  Table~\ref{conttable} lists the
observed properties of each source, including the integrated flux density and
deconvolved source size determined from a single two-dimensional Gaussian fit.  The
(sub)mm clump coincident with the EGO is visible in the 1.2 mm SEST/SIMBA map
(resolution 24\pp) of \citet{Faundez04}.  No parameters are tabulated,
however, so we cannot compute the fraction of the single-dish flux recovered
by the interferometers.\footnote{Two major blind (sub)mm surveys of the
  Galactic Plane have recently been completed: the Bolocam Galactic Plane
  Survey (BGPS) at 1.1 mm and the ATLASGAL survey at 870 \um.  Unfortunately,   
G11.92$-$0.61 falls outside the coverage of both the BGPS \citep[$|$b$|<$0.5,
][]{Rosolowsky09} and the 2007 ATLASGAL campaign presented in \citet{atlasgal}.}

All three mm continuum sources are coincident with the NE 4.5 \um\/ lobe.  As
shown in Figure~\ref{contfig}b, the MIPS 24 \um\/ emission associated with the EGO is
elongated along a N-S axis, and encompasses both MM1 and MM3.  The MIPS 24
\um\/ image is saturated, introducing considerable (\q2-4\pp) uncertainty into
the determination of the 24 \um\/ centroid position \citep[see
also][]{maserpap}.  Figure~\ref{contfig}b shows, however, that the (saturated) 24\um\/
peak lies \q1-2\pp\/ N of MM1, and roughly between MM1 and MM3.  The MIPS
24 \um\/ counterpart thus likely consists of blended emission from these two
sources.  MM1 and MM3 are also coincident with 6.7 GHz Class II \methanol\/
masers reported by \citet{maserpap} (Fig.~\ref{contfig}a,b).  The CARMA 1.4 mm centroid
position of MM1 is offset \q0\farcs6 ($\gtrsim$ 2100 AU) to the north of
the intensity-weighted position of the southern 6.7 GHz \methanol\/ maser
group (G11.918-0.613), and the CARMA 1.4 mm centroid position for MM3 is
offset \q0\farcs4 ($\gtrsim$ 1500 AU) to the northeast of the
intensity-weighted position of the northern 6.7 GHz \methanol\/ maser group
(G11.919$-$0.613).  
The \water\/ maser reported by \citet{HC96} is also coincident with MM1 to within the astrometric uncertainties (Fig.~\ref{contfig}b).
Notably, MM2 is offset to the northwest by $\gtrsim$4\pp\/
($\gtrsim$ 14800 AU) from the 24 \um\/ peak, and is not associated with a 6.7
GHz \methanol\/ maser.


\subsubsection{Compact Molecular Line Emission: G11.92$-$0.61}\label{g11compact}

\begin{figure}
\epsscale{0.9}
\plotone{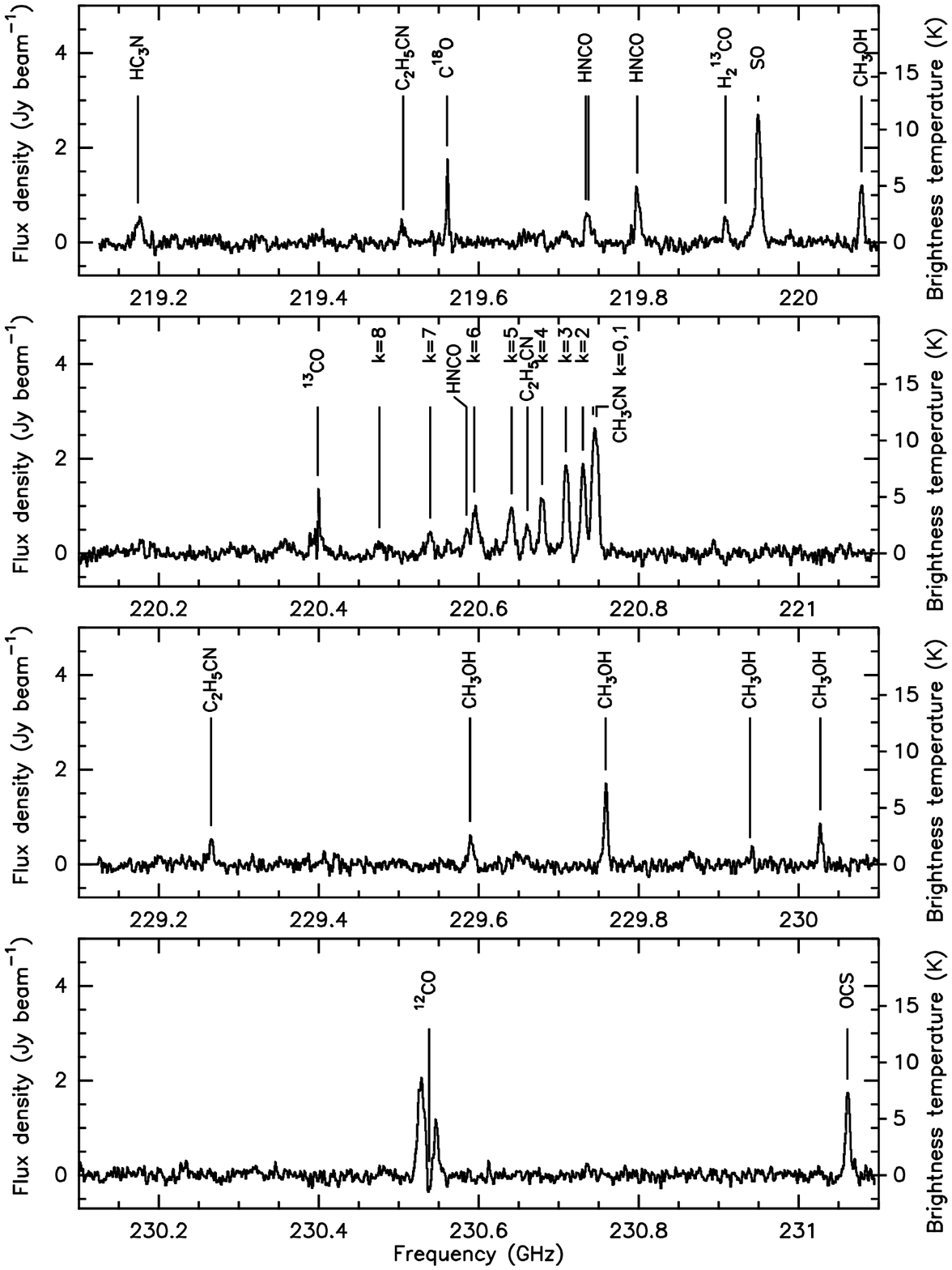}
\caption{SMA LSB and USB spectra towards the 1.3 mm continuum peak of 
  G11.92$-$0.61MM1.  The spectra have been Hanning smoothed. Lines detected at
$\ge$3$\sigma$ are labeled and listed in Table~\ref{g11trans}.}
\label{g1192_smaspec}
\end{figure}

The continuum source MM1 is associated with the richest molecular line
emission in the G11.92$-$0.61 complex.  Emission in 27 lines of 11 species is
detected towards G11.92$-$0.61-MM1 in our SMA observations.
Table~\ref{g11trans} lists the specific transitions, frequencies, and
upper-state energies of lines detected at $\ge$3$\sigma$.
Figure~\ref{g1192_smaspec} shows the spectrum at the MM1 continuum peak across
the 4 GHz bandwidth observed with the SMA, with the transitions listed in
Table~\ref{g11trans} labeled.  
Table~\ref{g11trans} also lists the peak line intensities, line centroid velocities,
$\Delta$v$_{FWHM}$, and integrated line intensities obtained from single
Gaussian fits to lines detected at $>$3$\sigma$ at the MM1 continuum
peak.  Some line profiles may be affected by outflowing gas; transitions with
non-Gaussian shapes are noted in Table~\ref{g11trans}.  \co\/ and
$^{13}$CO were not fit, as the line profiles are complex and strongly
self-absorbed.

The spectrum of G11.92$-$0.61-MM1 is similar to those of hot cores
observed with comparable setups with the SMA, such as AFGL 5142 MM2
\citep{Zhang07}.  The EGO spectrum is also similar to that of HH80-81 MM1, the
driving source of the HH80-81 radio jet \citep{Qiu09}, with the notable
exception that SO$_{2}$(11$_{5,7}$-12$_{4,8}$) emission (229.347
GHz, \eup=122 K) is not detected towards G11.91$-$0.61-MM1.  Emission from
complex oxygen-rich organic
molecules characteristic of strong hot cores (such as HCOOCH$_{3}$) also is not
detected towards G11.92$-$0.61. 

\begin{figure}
\plotone{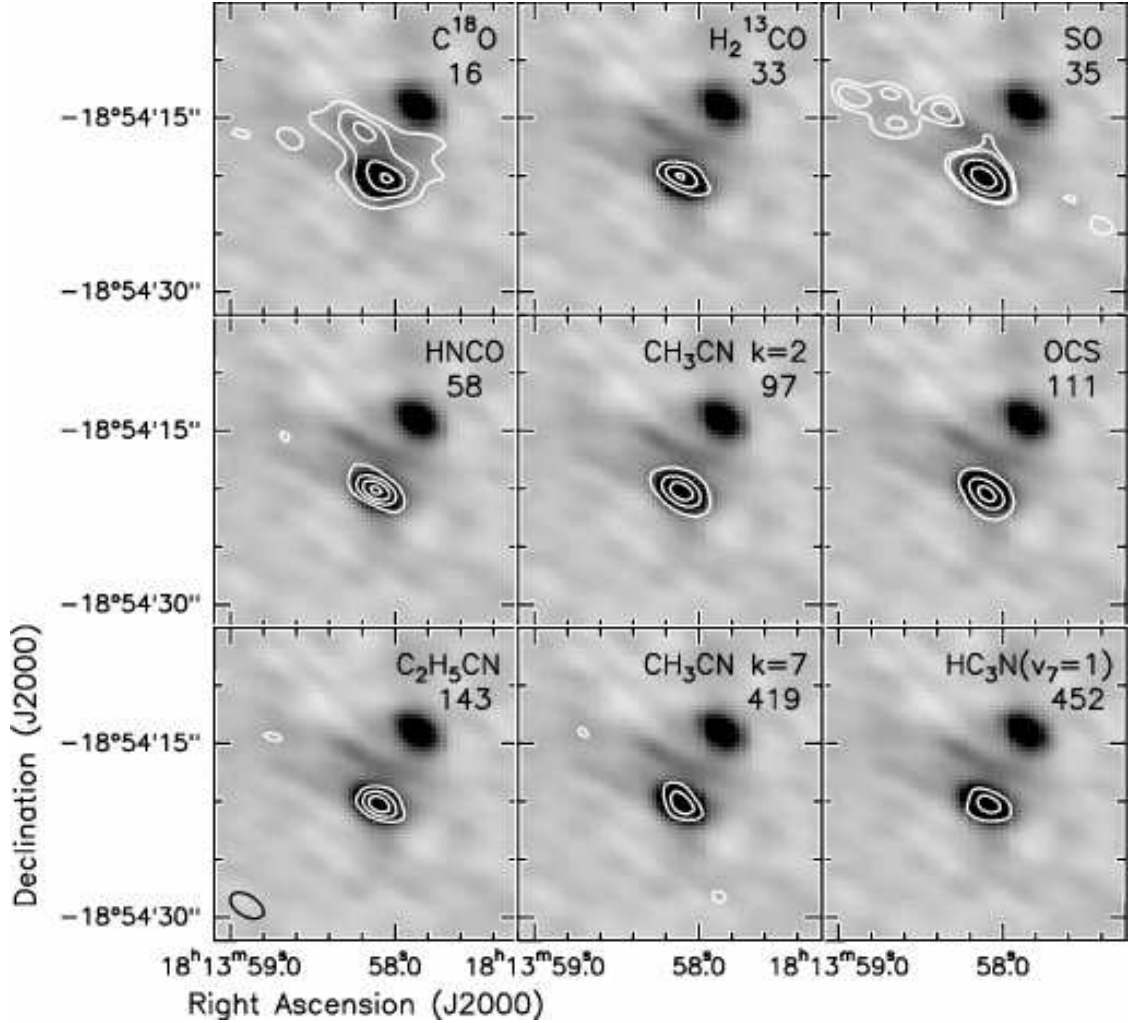}
\caption{G11.92$-$0.61: The greyscale shows the SMA 1.3 mm continuum
  emission.  Contours are drawn from integrated intensity (moment
  zero) maps of the indicated molecular line.  The species and upper
  state energy in K of the transition are listed at upper right in
  each panel.  Contour levels are: C$^{18}$O: 1.92, 3.84, 5.76, 7.68
  Jy beam$^{-1}$ \kms; H$_{2}^{13}$CO: 1.65, 3.3, 4.95 Jy beam$^{-1}$
  \kms; SO: 3.3, 4.95, 13.2, 23.1 Jy beam$^{-1}$ \kms; HNCO: 2.76,
  5.52, 8.28, 11.04 Jy beam$^{-1}$ \kms; CH$_{3}$CN (k=2): 2.82, 8.46,
  14.1 Jy beam$^{-1}$ \kms; OCS: 2.4, 7.2, 12 Jy beam$^{-1}$ \kms;
  C$_{2}$H$_{5}$CN: 1.62, 3.24, 4.86 Jy beam$^{-1}$ \kms; CH$_{3}$CN
  (k=7): 1.8, 3.6 Jy beam$^{-1}$ \kms; HC$_{3}$N(v$_{7}$=1): 2.52,
  5.04 Jy beam$^{-1}$ \kms. These contour levels are (3,6,9,12)
  $\times \sigma$ for C$^{18}$O and HNCO; (3,6,9) $\times \sigma$ for
  H$_{2}^{13}$CO and C$_{2}$H$_{5}$CN; (3,9,15) $\times \sigma$ for
  CH$_{3}$CN (k=2) and OCS; (3,6) $\times \sigma$ for CH$_{3}$CN (k=7)
  and HC$_{3}$N; and (3,4.5,12,21) $\times \sigma$ for SO.  The SMA
  beam is shown at lower left.}
\label{g1192_mom0}
\end{figure}

\begin{figure}
\plotone{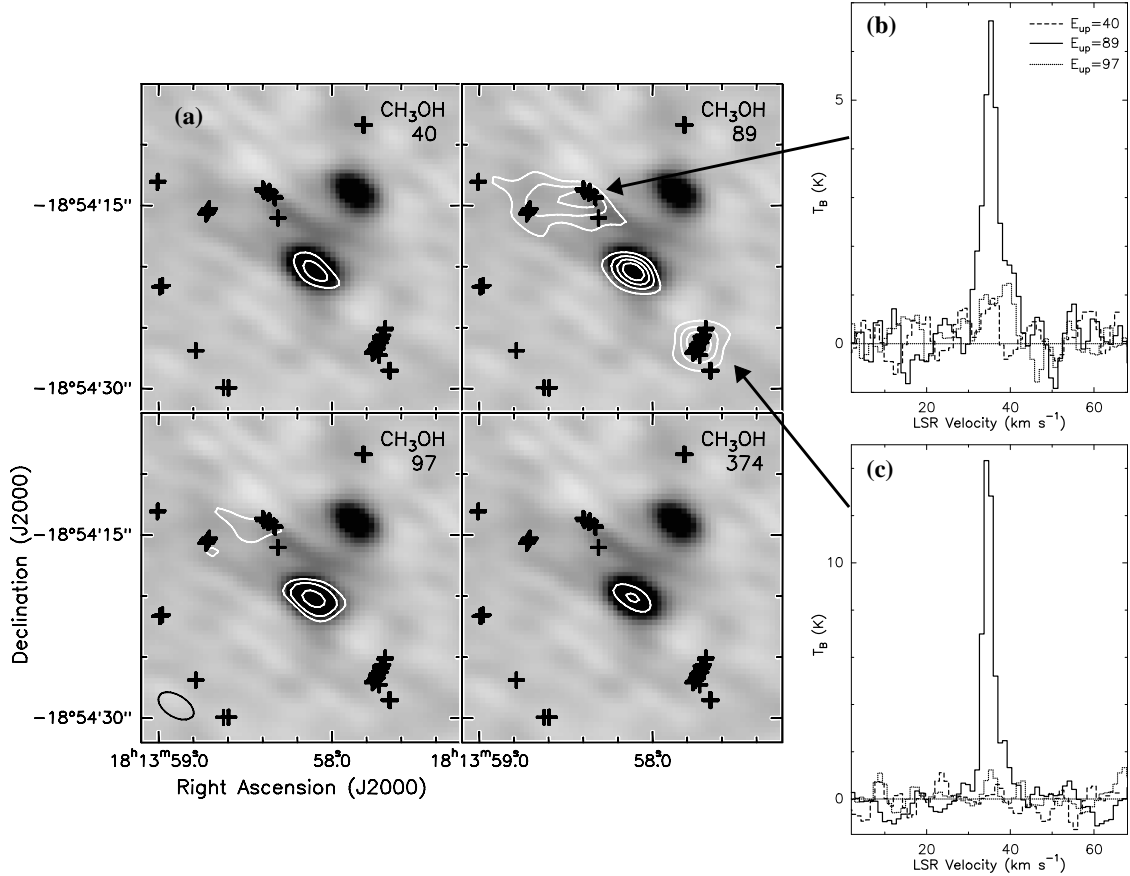}
\caption{G11.92$-$0.61: (a) The greyscale shows the SMA 1.3 mm
  continuum emission.  Contours are drawn from integrated intensity
  (moment zero) maps of the indicated \methanol\/ line.  The upper
  state energy in K of the transition is given at upper right in each
  panel.  Contour levels are: \eup=40 K: 2.2, 4.4 Jy beam$^{-1}$ \kms;
  \eup=89K: 2.52, 5.04, 7.56, 10.08 Jy beam$^{-1}$ \kms; \eup=97 K:
  2,4,8 Jy beam$^{-1}$ \kms; \eup=374 K: 2.8, 5.6 Jy beam$^{-1}$
  \kms. These contour levels are (4,8) $\times \sigma$ for the \eup=40
  K and \eup=374 K transitions, (4,8,12,16) $\times \sigma$ for the
  \eup=89 K transition, and (4,8,16) $\times \sigma$ for the \eup=97 K
  transition.  Black crosses mark the positions of 44 GHz Class I
  \methanol\/ masers from \citet{maserpap}.  The SMA beam is shown at
  lower left.  (b) and (c): Spectra at the peak positions of the
  indicated \methanol\/ (8$_{-1,8}$-7$_{0,7}$) emission.  Solid line:
  8$_{-1,8}$-7$_{0,7}$ transition (229.759 GHz, \eup=89 K); dashed
  line: 3$_{-2,2}$-4$_{-1,4}$ (230.027 GHz, \eup=40 K); dotted line:
  8$_{0,8}$-7$_{1,6}$ (220.078 GHz, \eup=97 K).}
\label{g1192_mom0_meth}
\end{figure}

Figures~\ref{g1192_mom0} and \ref{g1192_mom0_meth} present integrated
intensity (moment zero) maps for selected transitions from Table~\ref{g11trans}.  As shown
in Figures~\ref{g1192_mom0} and \ref{g1192_mom0_meth}, emission from most
species is compact and coincident with the continuum source MM1.  Emission
from high-excitation lines (\eup$\gtrsim$100 K) is detected exclusively
towards MM1.  In contrast, the continuum source MM2 is devoid of line
emission.  The only species that exhibits compact emission coincident with the
continuum source MM3 is C$^{18}$O (Fig.~\ref{g1192_mom0}).  The \meth\/
integrated intensity maps shown in Figure~\ref{g1192_mom0_meth} are discussed
further in \S\ref{g11methmasers}.

\begin{figure}
\plotone{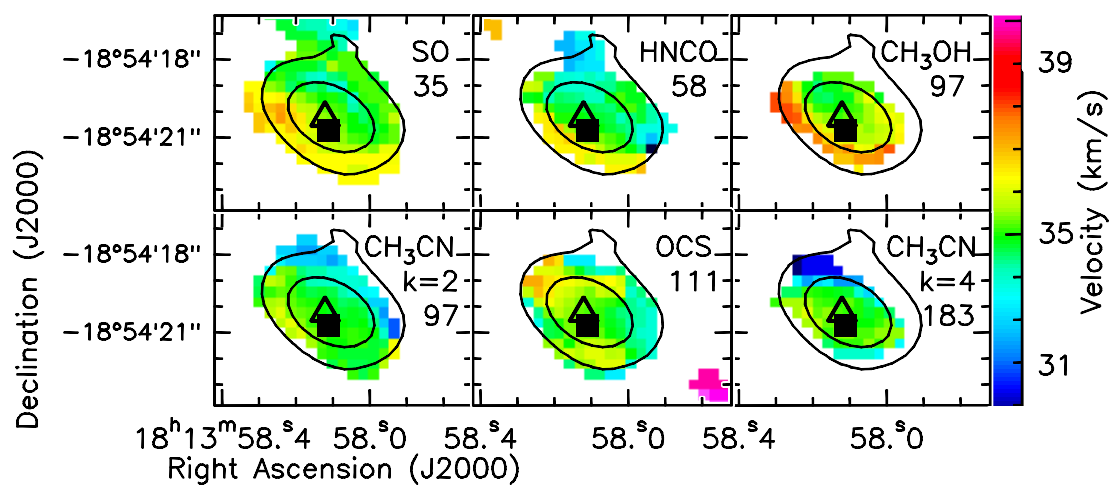}
\caption{G11.92$-$0.61: First moment maps of the indicated species (colorscale) overlaid with
contours of the SMA 1.3 mm continuum emission (levels 10, 30 $\times$ $\sigma =$ 3.5 \mjb).
The inner contour is approximately the size of the SMA beam.  \eup\/ in K is
listed under the molecule name in the upper right of each panel.  The open
triangle marks the water maser position from \citet{HC96}.  The filled square
marks the position of the 6.7 GHz Class II \methanol\/
maser from \citet{maserpap}.}
\label{g11_mom1}
\end{figure}

Most of the lines detected towards MM1 are quite broad, with
$\Delta$v$_{FWHM}$ of 8-10 \kms.  The compact molecular line emission
exhibits a velocity gradient, from SE (redshifted) to NW (blueshifted)
(Fig.~\ref{g11_mom1}).  As shown in Figure~\ref{g11_mom1}, this
gradient is consistent across species including SO, HNCO, \methanol\/,
and CH$_{3}$CN.  One possible explanation for this velocity gradient
is an unresolved disk, oriented roughly perpendicular to the outflow
axis.  Higher angular resolution data are required to investigate this
possibility.  Not all molecules detected towards the hot core show the
same velocity gradient.  One exception is OCS, which has redshifted
emission to the NE and blueshifted emission to the SW.  This is
consistent with the kinematics of the dominant outflow
(\S\ref{g11_extended}), and suggests that the inner regions of the
outflow may be contributing significantly to the observed compact OCS
emission.

Determining the \vlsr's of the mm continuum sources is complicated by the
possibility of confusion from outflowing gas or resolved-out emission from the
extended envelope.  For MM1, there is sufficient agreement among lines that
exhibit compact emission and are detected with high signal-to-noise to
estimate \vlsr(MM1)=35.2$\pm$0.4 \kms.  This is slightly blueward of the
\vlsr\/ of 36 \kms\/ estimated from the lower angular resolution \hisoco\/ observations of
\citet{maserpap}.  The systemic velocity of MM1 is also blueshifted  
relative to both the 6.7 GHz
Class II \methanol\/ masers coincident with MM1 \citep[v\q37.1-37.6
\kms;][]{maserpap}, and the peak \water\/ maser velocity \citep[v=40.7
\kms;][]{HC96}.  There is also weaker \water\/ maser emission at the 6.7 GHz
\meth\/ maser velocity.
Table~\ref{g11trans} lists a Gaussian fit
to the C$^{18}$O emission towards the MM3 continuum peak.  The emission is
narrow ($\Delta$v$_{FWHM}$=3.5 \kms), and has a line centroid velocity of 34.4
\kms.  Since no other compact line
emission is detected associated with MM3, however, it is difficult to be
certain whether this velocity represents the MM3 gas \vlsr.  If so, then
the thermal gas emission from MM3 is blueshifted by $\ge$ 4 \kms\/ relative to the
coincident 6.7 GHz Class II \methanol\/ masers, which have velocities of
\q38.6-39.5 \kms.  Since no emission centered on MM2 is detected, its \vlsr\/
cannot be determined.

\subsubsection{Extended Molecular Line Emission: G11.92$-$0.61}\label{g11_extended}

\begin{figure}
\plotone{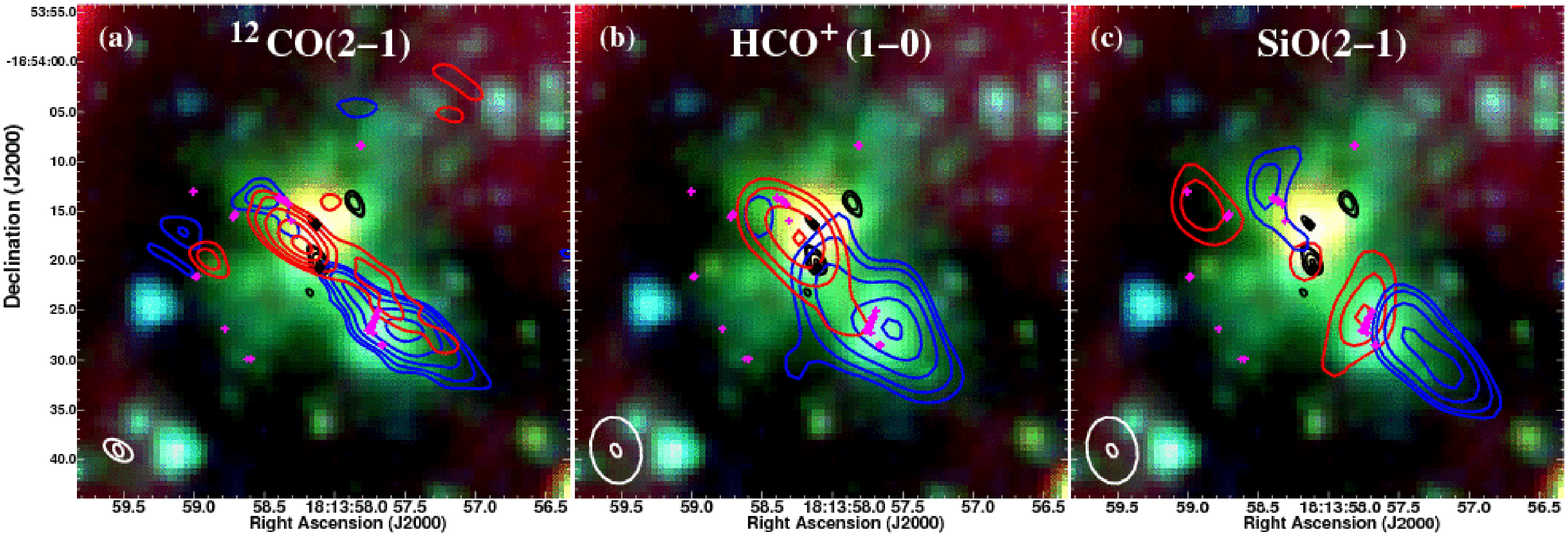}
\caption{G11.92-0.61: Three color \emph{Spitzer} images (3.6, 4.5, 8.0 $\mu$m:
blue, green, red) overlaid with contours of 1.4 mm continuum emission (black) and high velocity (a) \co(2-1), (b)
\hco(1-0), and (c) SiO(2-1) emission.  In each panel, positions of 6.7 GHz
\meth\/ masers are marked with black diamonds, and positions of 44 GHz \meth\/
masers are marked with magneta crosses \citep{maserpap}.  Continuum contour
levels are 4,5,10,20 $\times$ $\sigma =$ 4.3 \mjb.  The \vlsr\/ is \q 35
\kms\/ (\S\ref{g11compact},\ref{outflow_dis}).  (a) \co(2-1)
emission integrated over v=$-$24.4 to 21.8 \kms\/ (blue) and v=48.1 to 71.2
\kms\/ (red), e.g. \q\vlsr$\pm$13 \kms.  Levels are 5,9,17,29,45 \jb\/
\kms\/ for both red and blue contours. The SMA and CARMA beams are shown at lower left.  (b) \hco(1-0) emission integrated over
v=-14.3 to 21.8 \kms\/ (blue) and 48.1 to 71.1 \kms\/ (red), e.g. \q\vlsr$\pm$13
\kms.  Contour levels: Blue: 1.3, 1.8, 2.8, 4.3, 5.8  \jb\/
\kms; Red: 1.3, 1.8, 2.8, 3.8 \jb\/
\kms.  The CARMA beams are shown at lower left.  (c) SiO(2-1) emission
integrated over v=8.0 to 24.9 \kms\/ (blue) and v=45.1 to 53.5 \kms\/ (red),
e.g. \q\vlsr$\pm$10 \kms.  Contour levels: Blue: 0.8, 1.0, 1.4, 2.0  \jb\/
\kms; Red: 0.6, 0.8, 1.0  \jb\/
\kms.  The CARMA beams are shown at lower left.}
\label{g11redbluefig}
\end{figure}

\begin{figure}
\plotone{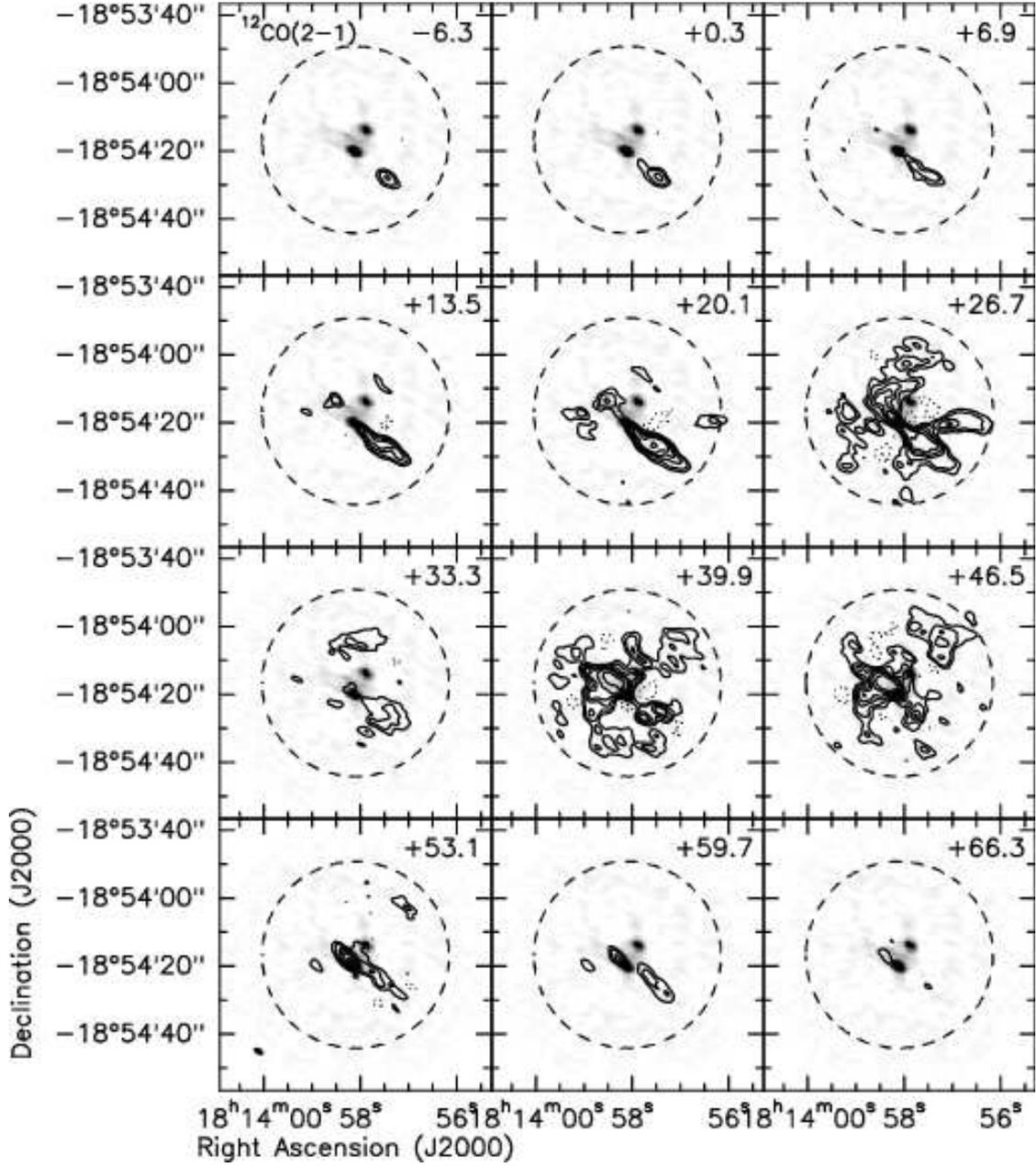}
\caption{G11.92$-$0.61: Channel maps of \co(2-1).  The greyscale is
  the 1.3 mm SMA continuum image and the contours are the \co\/
  emission (levels 0.24, 0.48, 0.96, 1.44, 2.16, 3.6 \jb (solid), $-$0.24, $-$0.48 \jb (dotted)).  Each panel
  is the average of two channels of the smoothed \co\/ cube; the
  center velocity of each panel is given at upper right.  The contour
  levels are ($-$8,$-$4,4,8,16,24,36,60) $\times$ the rms in a line-free panel.
  The field of view shown is the same as in
  Figures~\ref{g11_hcochannels} and ~\ref{g11_siochannels}.  The
  dashed circle shows the FWHP of the SMA primary beam.  The SMA
  synthesized beam is shown as a filled ellipse at lower left.}
\label{g11_12cochannels}
\end{figure}

\begin{figure}
\plotone{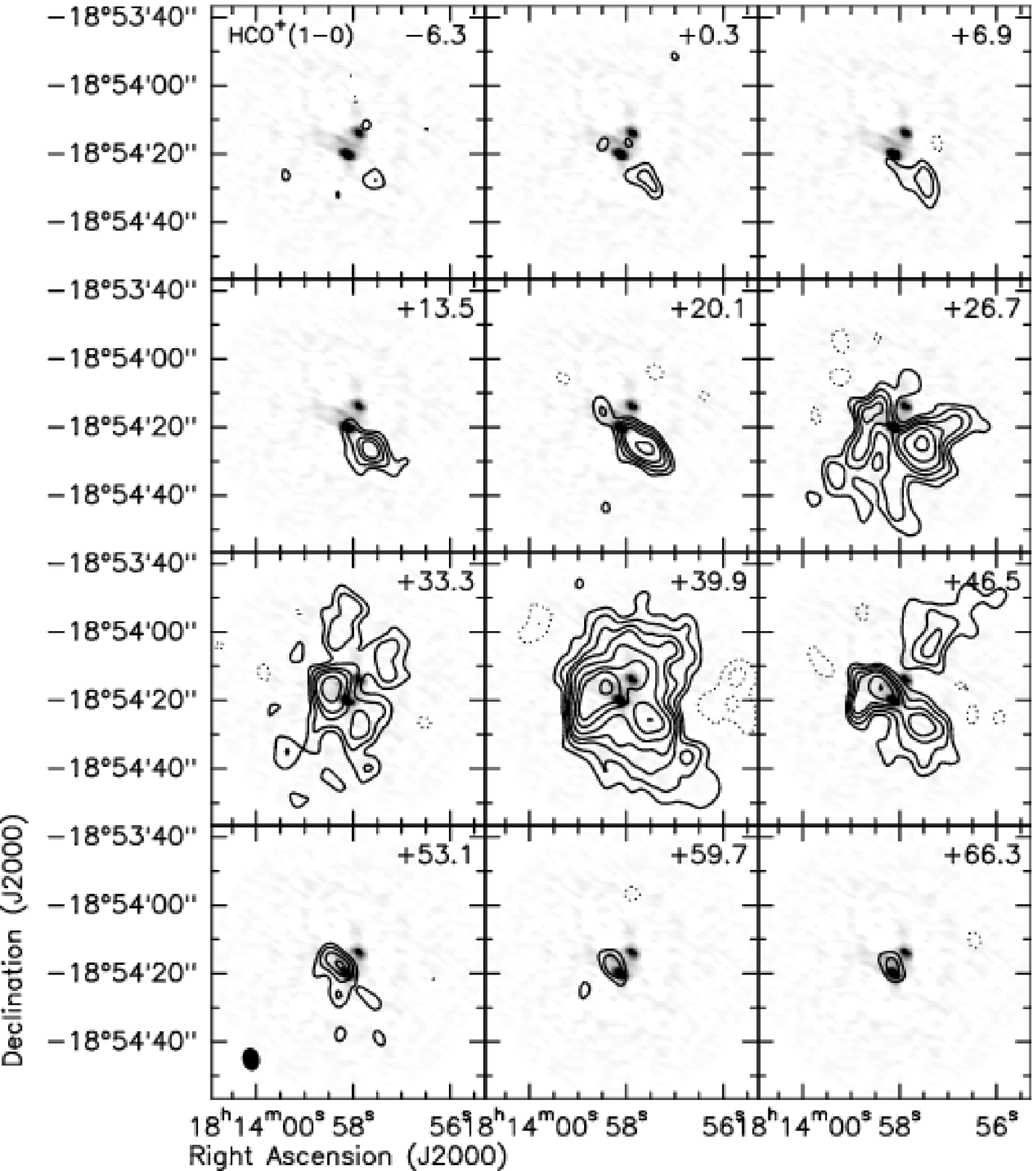}
\caption{G11.92$-$0.61: Channel maps of \hco(1-0).  The greyscale is
  the 1.3 mm SMA continuum image.  The contours are the \hco\/
  emission (levels 0.06, 0.096, 0.144, 0.192, 0.288, 0.432,
  0.576, 0.864 \jb (solid), $-$0.06, $-$0.096 (dotted)).  The contour levels are ($-$8,$-$5,5,8,12,16,24,36,48,72)
  $\times$ the rms in a line-free panel.  The center velocity of each
  panel is given at upper right.  The field of view shown is the FWHP
  primary beam of the 10 m CARMA antennas (80\pp\/ square).  The CARMA
  synthesized beam is shown as a filled ellipse at lower left.}
\label{g11_hcochannels}
\end{figure}

\begin{figure}
\plotone{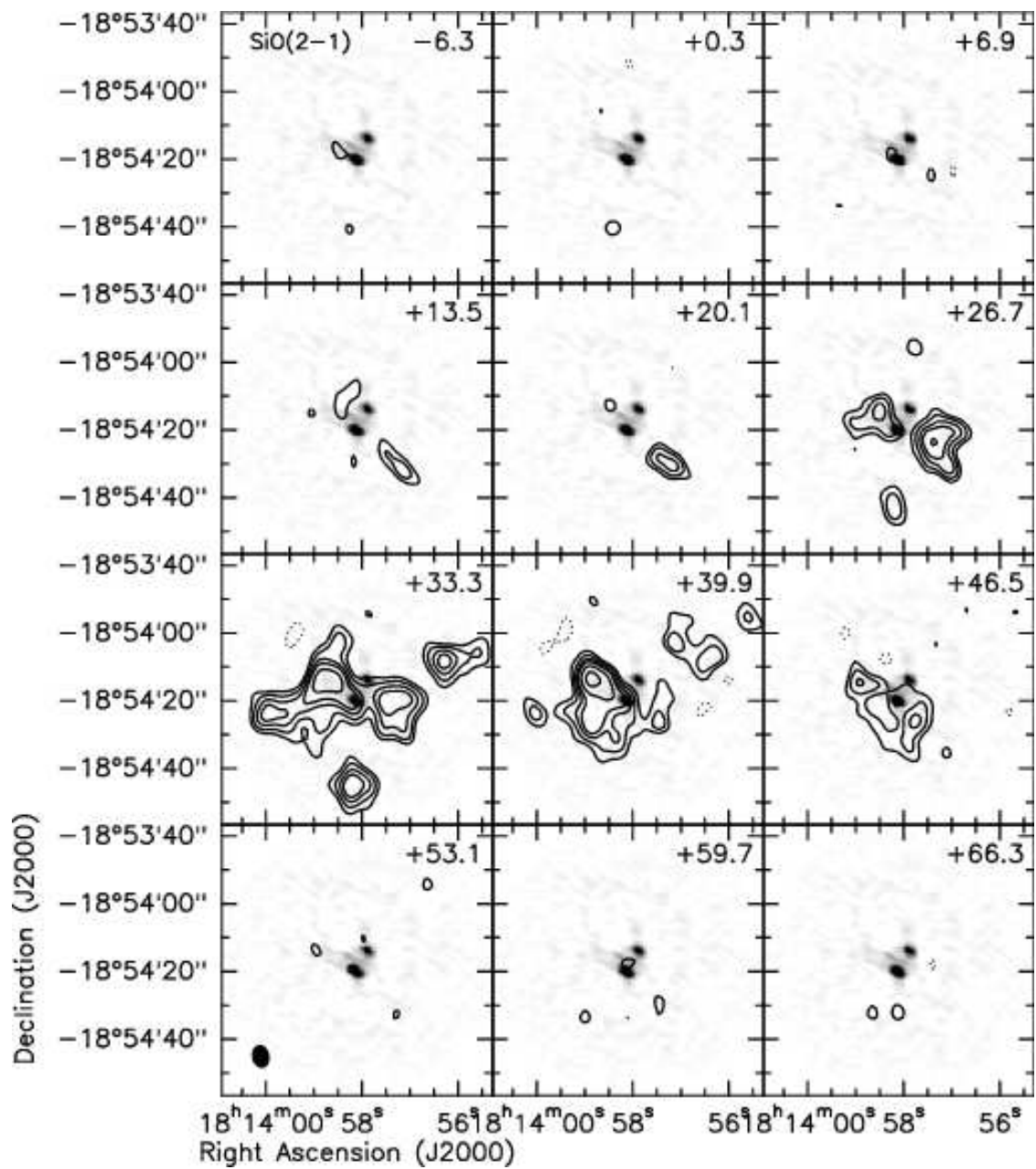}
\caption{G11.92$-$0.61: Channel maps of SiO(2-1).  The greyscale is the 1.3 mm
  SMA continuum image.  The contours are the SiO emission (levels 0.055, 0.088,
  0.132, 0.176, 0.264, 0.396 \jb (solid), $-$0.055 \jb (dotted)).  The contour levels are ($-$5,5,8,12,16,24,36)
  $\times$ the rms in a line-free panel.  The center velocity of each panel is given at
  upper right.  The field of view shown is the FWHP primary beam of the 10 m
  CARMA antennas (80\pp\/ square).  The CARMA synthesized beam is shown as a filled ellipse at
  lower left.}
\label{g11_siochannels}
\end{figure}

The observed low-excitation transitions of the abundant molecules \co\/ and \hco\/
exhibit extended emission spanning a wide velocity range ($>$ 80 \kms) and
most of the telescope field of view.  Emission in SiO (2-1) is similarly
spatially extended, but spans a narrower velocity range (\q 40 \kms).
While the kinematics of
this extended emission are complex, the high velocity
($|$v$-$\vlsr$|\gtrsim$ 13 \kms) \co\/ and \hco\/ emission are
characterized by a bipolar outflow centered on the continuum source MM1 (Fig.~\ref{g11redbluefig}).
To complement the integrated intensity maps of the high velocity gas shown in
Figure~\ref{g11redbluefig}, channel maps of the \co (2-1), \hco (1-0), and SiO
(2-1) emission are shown in
Figures~\ref{g11_12cochannels}-\ref{g11_siochannels}.  


The red and blue lobes of the molecular outflow are asymmetric, both spatially and kinematically.  The
blueshifted lobe, SW of MM1, extends to more extreme velocities
($|$v$_{max,blue}-$\vlsr$|$\q 59 \kms; $|$v$_{max,red}-$\vlsr$|$\q 36 \kms) and further
from the continuum source.
The blueshifted lobe also exhibits stronger SiO (2-1) emission.
The sense of the velocity gradient in the molecular gas agrees with that of 44
GHz Class I \meth\/ masers imaged with the VLA \citep[resolution
0\farcs99$\times$0\farcs44,][]{maserpap}.
The concentration of
blueshifted Class I masers coincides with the blueshifted molecular outflow
lobe seen in \co\/ and  \hco\/ and with the SW 4.5 \um\/
lobe (Fig.~\ref{g11redbluefig}a,b).  
The Class I masers in the arc to the NE have near-systemic or
slightly redshifted velocities, consistent with the location and more moderate
velocity of the redshifted molecular outflow lobe.  In particular, the SE section of the
maser arc is coincident with moderately redshifted \co, \hco,
and SiO emission (Figs.~\ref{g11_12cochannels}-\ref{g11_siochannels}, 39.9 and 46.5
\kms\/ panels; this relatively low-velocity gas is not included in the integrated intensity maps shown in Figure~\ref{g11redbluefig}).  


The morphology and kinematics of the SiO(2-1) emission differ from those of
the other outflow tracers, and copious SiO (2-1) emission is detected far from
the mm continuum sources (Figs.~\ref{g11redbluefig},\ref{g11_siochannels}).
The excitation of low-J rotational lines of SiO, such as the 2-1 transition,
depends primarily on the density, n$_{H_{2}}$ \citep[as opposed to the kinetic
temperature, T$_{kin}$;][]{Nisini07,js10}, and extended (parsec-scale),
quiescent ($\Delta$v\q0.8 \kms) SiO(2-1) emission has recently been observed
towards an IRDC \citep{js10}.  The comparatively broad linewidths of the
SiO(2-1) emission towards G11.92$-$0.61 indicate that the entirety of the
observed SiO emission is attributable to outflow-driven shocks, with bright
SiO(2-1) knots likely tracing the impact of these shocks on dense regions in
the surrounding cloud.


While the NE(red)-SW(blue) gradient dominates the \co\/ and \hco\/ velocity
fields, there are other features that suggest multiple outflows may be present.
In particular, blueshifted \co, \hco, and SiO emission are detected
NE of MM1 at velocities $\gtrsim$ 10 \kms 
, and redshifted emission SW of MM1 at velocities $\lesssim$ 65 \kms\/
(Figs.~\ref{g11redbluefig}-\ref{g11_siochannels}).
Near the \vlsr\/ of \q35
\kms, however, low-velocity outflow emission is confused with emission from
ambient gas.  

Emission from SO(6$_{5}$-5$_{4}$) also extends NE and SW of MM1
(Fig.~\ref{g1192_mom0}).  The morphology and kinematics are consistent with
this SO emission arising in the dominant outflow; in contrast to SiO, the SO emission
is stronger towards the redshifted (NE) outflow lobe.  The properties of the lower-excitation SO
(2$_2$-1$_1$)  emission (\eup=19.3 K) observed with CARMA (not shown) are similar
to those seen in SO(6$_{5}$-5$_{4}$) at higher spatial
and spectral resolution with the SMA.  Faint SO(6$_{5}$-5$_{4}$) emission is detected coincident with
the MM3 continuum source, at a velocity
consistent with that of the C$^{18}$O (\S\ref{g11compact}).  However, since the bipolar
outflow(s) overlap the MM3 position, it is unclear whether this SO emission is associated with MM3.



\subsubsection{Millimeter \methanol\/ masers: G11.92$-$0.61}\label{g11methmasers}

As shown in Figure~\ref{g1192_mom0_meth}, the morphology of the 229.759 GHz
\meth(8$_{-1,8}$-7$_{0,7}$)E  (\eup=89 K) line emission is strikingly different from
that of any other observed \methanol\/ transition.  There is very strong
\meth(8$_{-1,8}$-7$_{0,7}$) emission to the southwest of MM1, coincident with
the brightest 44 GHz \meth(7$_{0}$-6$_{1}$)A$^{+}$ Class I masers detected by
\citet{maserpap} (Fig.~\ref{g1192_mom0_meth}).  Strong 229.759 GHz \meth\/
emission is also observed NE of MM1, also coincident with 44 GHz Class I
masers.  The 44 GHz (7$_{0}$-6$_{1}$) and 229.759 GHz (8$_{-1,8}$-7$_{0,7}$)
lines are both Class I \methanol\/ maser transitions.  \citet{Slysh02} first
reported 229.759 GHz \methanol\/ maser emission towards DR21 (OH) and DR21
West based on observations with the IRAM 30 m telescope.  Probable maser
emission in this transition has been detected with the SMA in HH 80-81,
coincident with a 44 GHz \methanol\/ maser \citep{Qiu09}, and in \emph{IRAS} 05345+3157 \citep{Fontani09}.

The \meth(8$_{-1,8}$-7$_{0,7}$) emission observed NE and SW of G11.92$-$0.61-MM1 is
spectrally narrow (Fig.~\ref{g1192_mom0_meth}b-c).  In both cases, the velocity
of the mm emission agrees well with the velocities of the coincident 44 GHz
masers.  
The extended appearance of the northeastern 229.759 GHz emission is
consistent with emission from multiple masers being blended at the lower
spatial resolution of the SMA observations (\q3\pp\/ compared to \q0\farcs5
for the 44 GHz VLA data). 
Like the studies of \citet{Slysh02} and \citet{Qiu09}, the beamsize
of our observations is too large to definitively establish the maser nature of
the emission based on its brightness temperature.  The peak intensity of the
SW \meth(8$_{-1,8}$-7$_{0,7}$) emission is 3.5 \jb, corresponding to T$_{B}$=14.3
K.  For the NE emission, I$_{peak}$=1.6 \jb\/ (T$_{B}$=6.6 K).  \citet{Slysh02} use the
229.759/230.027 line ratio as a discrimant between
thermal and masing 229.759 GHz emission, with ratios $>$ 3 indicative of
non-thermal 229.759 GHz \meth\/ emission.  The 230.027 GHz line is a Class II
transition, so is not expected to be inverted under conditions that excite
229.759 GHz Class I maser emission \citep{Slysh02}.  The 229.759/230.027 ratios
for the SW and NE spots in G11.92$-$0.61 are 100 and 7, respectively.  For
comparison, the ratio at the MM1 1.3 mm continuum peak is 2, consistent with
thermal emission from the hot core.  
The SW and NE 229.759 GHz emission
features coincide spatially and spectrally with 44 GHz Class I \methanol\/
masers; this agreement, along with the millimeter line ratios and narrow linewidths,
strongly supports the interpretation of these emission features as mm
\methanol\/ maser emission.


\subsection{G19.01$-$0.03}\label{g19results}

\subsubsection{Previous Observations: G19.01$-$0.03}\label{g19_previous}

This source was entirely unknown prior to being cataloged as an EGO by
\citet{egocat}.  The EGO G19.01$-$0.03 has a striking MIR appearance, with
bipolar 4.5 \um\/ emission centered on a point source detected in GLIMPSE and
MIPSGAL images.  \citet{maserpap} detected kinematically
complex Class II 6.7 GHz \meth\/ maser emission coincident with the
``central'' source.  Copious 44 GHz Class I \meth\/ maser emission is
associated with the 4.5 \um\/ lobes, with blueshifted masers concentrated to
the north of the MIR point source and systemic/redshifted masers to the south
\citep{maserpap}.  The EGO is located in an IRDC \citep{egocat}, and a (sub)mm continuum
source coincident with the EGO is detected in blind single-dish surveys of the
Galactic Plane at 1.1 mm and 870 \um\/ \citep[BGPS and
ATLASGAL,][]{Rosolowsky09,atlasgal}.  SiO(5-4) and thermal
\meth(5$_{2,3}$-4$_{1,3}$) emission were detected towards G19.01$-$0.03 in
single-pointing JCMT observations (resolution \q 20\pp) targeted at the MIR
point source \citep{maserpap}.  No 44 GHz continuum emission was detected
towards the EGO to a 5$\sigma$ sensitivity limit of 5 \mjb\/ \citep[resolution
0\farcs59 $\times$ 0\farcs51,][]{maserpap}.  We adopt the
near kinematic distance from \citet{maserpap} for G19.01$-$0.03: 4.2 kpc.

\subsubsection{Continuum Emission: G19.01$-$0.03}\label{g19cont}

At the resolution of our SMA 1.3 mm and CARMA 3.4 mm observations, the
millimeter continuum emission associated with the EGO G19.01$-$0.03 appears to
arise from a single source, called MM1.  The SMA 1.3 mm continuum image is
shown in Figure~\ref{contfig}c.  The position of the peak 3.4 mm continuum emission (not
shown) coincides with the 1.3 mm continuum peak.  In our SMA 1.3 mm continuum
image, we recover 7.6$^{+1.9}_{-1.4}$\% of the single-dish flux density of 3.6$^{+0.7}_{-0.6}$ Jy
measured from the 1.1 mm Bolocam Galactic Plane Survey \citep[BGPS,
resolution \q30\pp,][]{Rosolowsky09}.\footnote{BGPS flux density from \citet{Rosolowsky09} with correction factor applied as discussed in \citet{Dunham10}.}


The fitted position of MM1 from the SMA 1.3 mm continuum image is coincident
with the intensity-weighted position of the Class II 6.7 GHz \methanol\/ maser
\citep{maserpap} and with the GLIMPSE point source SSTGLMC G019.0087-00.0293
within the absolute positional uncertainty of the mm data (0\farcs3).  The
MIPS 24 \um\/ peak is offset to the NW by \q1\farcs3 ($\gtrsim$ 5500
AU).  This is consistent, however, with MM1 being coincident with
the MIPS 24 \um\/ source within the absolute positional uncertainty of the
MIPSGAL survey \citep[median 0\farcs85, up to \q3\pp,][]{Carey09}.

\subsubsection{Compact Molecular Line Emission: G19.01$-$0.03}\label{g19_compact}

\begin{figure}
\epsscale{0.9}
\plotone{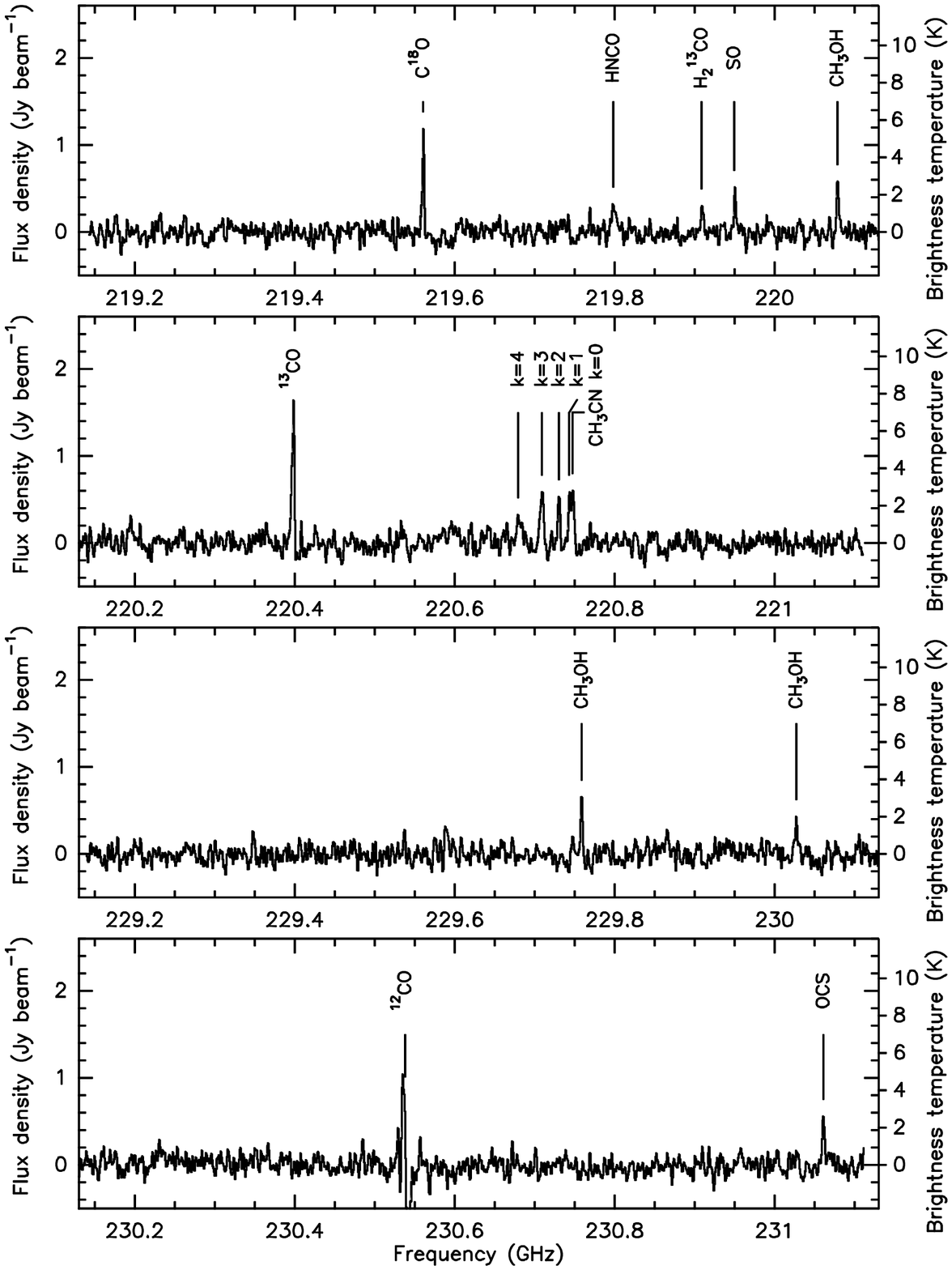}
\caption{SMA LSB and USB spectra towards the 1.3 mm continuum peak of 
  G19.01$-$0.03MM1.  The spectra have been Hanning smoothed.  Lines detected at
$\ge$3$\sigma$ are labeled and listed in Table~\ref{g19trans}.}
\label{g1901_smaspec}
\end{figure}

\begin{figure}
\plotone{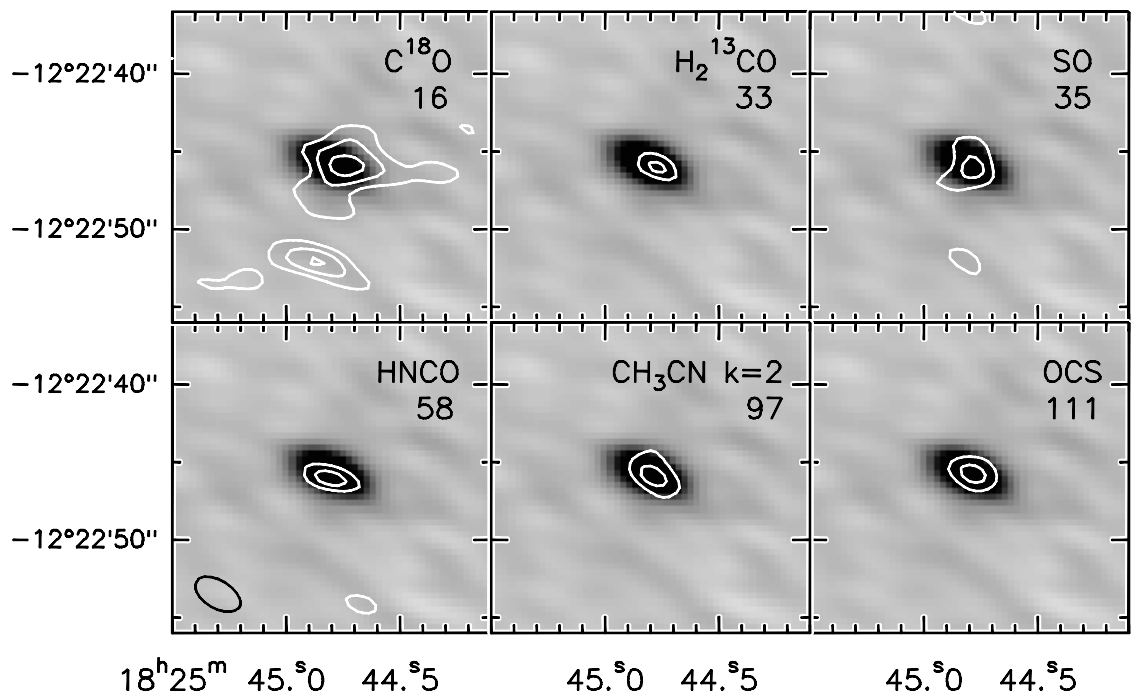}
\caption{G19.01$-$0.03: The greyscale shows the SMA 1.3 mm continuum
  emission.  Contours are drawn from integrated intensity (moment
  zero) maps of the indicated molecular line.  The species and upper
  state energy in K of the transition are listed at upper right in
  each panel.  Contour levels are: C$^{18}$O: 1.65, 3.3, 4.95 Jy
  beam$^{-1}$ \kms; H$_{2}^{13}$CO: 1.2, 2.0 Jy beam$^{-1}$ \kms; SO:
  0.9, 1.5 Jy beam$^{-1}$ \kms; HNCO: 1.32, 2.20 Jy beam$^{-1}$ \kms;
  CH$_{3}$CN (k=2): 1.11, 1.85 Jy beam$^{-1}$ \kms; OCS: 1.2, 2.4 Jy
  beam$^{-1}$ \kms.  These contour levels are (3,6,9)$\times$ $\sigma$
  for C$^{18}$O, (3,5)$\times$ $\sigma$ for H$_{2}^{13}$CO, SO, HNCO,
  and CH$_{3}$CN, and (3,6)$\times$ $\sigma$ for OCS.  The SMA beam is
  shown at lower left.}
\label{g19mom0}
\end{figure}

Emission in 15 lines of 9 species is detected towards G19.01$-$0.03-MM1 in our
SMA observations.  Table~\ref{g19trans} lists the specific transitions,
frequencies, and upper-state energies of lines detected at $\ge$3$\sigma$.
Figure~\ref{g1901_smaspec} shows the spectrum at the MM1 continuum peak across
the 4 GHz bandwidth observed with the SMA, with the transitions listed in
Table~\ref{g19trans} labeled.  As shown by a comparison of
Figures~\ref{g1192_smaspec} and \ref{g1901_smaspec}, many of the stronger lines
detected towards G11.92$-$0.61-MM1 are also detected towards
G19.01$-$0.03-MM1.  These lines are much weaker towards G19.01$-$0.03,
however, and emission from higher-energy transitions (\eup$>$ 200 K) is
notably lacking.  The highest-energy line detected towards G19.01$-$0.03-MM1
is the k=4 component of the \methcyn(12-11) ladder (\eup=183 K), and this line is
very weak (\q3.5$\sigma$).  Other than the k=3 and k=4 \methcyn\/ lines, no
line emission with \eup$>$100 K is detected towards G19.01$-$0.03-MM1.  As
shown in Figures~\ref{g19mom0} and \ref{g19methmom0}, emission from most
species is compact and coincident with the continuum source.

Table~\ref{g19trans} also lists the peak line intensities, line center velocities,
$\Delta$v$_{FWHM}$, and integrated line intensities obtained from single
Gaussian fits to lines detected at $>$3$\sigma$ at the G19.01$-$0.03 MM1
continuum peak.  Some line profiles may be affected by outflowing gas; lines
not well fit by a Gaussian are noted in Table~\ref{g19trans}.  As for
G11.92$-$0.61, the \co\/ and $^{13}$CO lines were not fit, as the line
profiles are non-Gaussian and, in the case of \co, strongly self-absorbed.  As
Table~\ref{g19trans} demonstrates, there is good agreement among the central
velocities determined from the different species and transitions.  Considering
all lines detected at $>$5$\sigma$, \vlsr(MM1)=59.9$\pm$1.1 \kms.
This is in good agreement with the central velocities of \hisoco(3-2)
(v$_{center}$=59.9$\pm$0.1 \kms) and \meth(5$_{2,3}$-4$_{1,3}$)
(v$_{center}$=59.7$\pm$0.2 \kms) observed with the JCMT \citep[resolution \q20\pp,][]
{maserpap}.

Compared to G11.92$-$0.61-MM1, the lines detected towards G19.01$-$0.03-MM1 are
relatively narrow.  Most of the transitions detected with $>$5$\sigma$ have
$\Delta$v$_{FWHM}<$ 4 \kms.  The 6.7 GHz Class II \methanol\/ masers
associated with G19.01$-$0.03-MM1 span a comparatively wide velocity range of
\q7.5 \kms, from 53.7-61.1 \kms\/ \citep{maserpap}.  The velocity range of the
6.7 GHz \meth\/ maser emission extends much further to the blue of the \vlsr\/
than it does to the red.

Figures~\ref{g19mom0} and ~\ref{g19methmom0} present integrated intensity
(moment zero) maps for selected transitions in Table~\ref{g19trans}.
The only species in Figure~\ref{g19mom0} that exhibits significant emission
not coincident with the continuum source is C$^{18}$O (2-1).  
(The properties of the SO (2$_2$-1$_1$) (\eup=19.3
K) emission observed with CARMA (not shown) are similar to those of the
SO(6$_5$-5$_4$) emission observed at higher spatial and spectral resolution
with the SMA.)  
The C$^{18}$O emission coincident with the continuum source has near-systemic
velocities, while the two knots of C$^{18}$O emission to the south of MM1 are
redshifted and likely associated with knots in the outflow.  

\subsubsection{Extended Molecular Line Emission: G19.01$-$0.03}\label{g19_extended}

\begin{figure}
\plotone{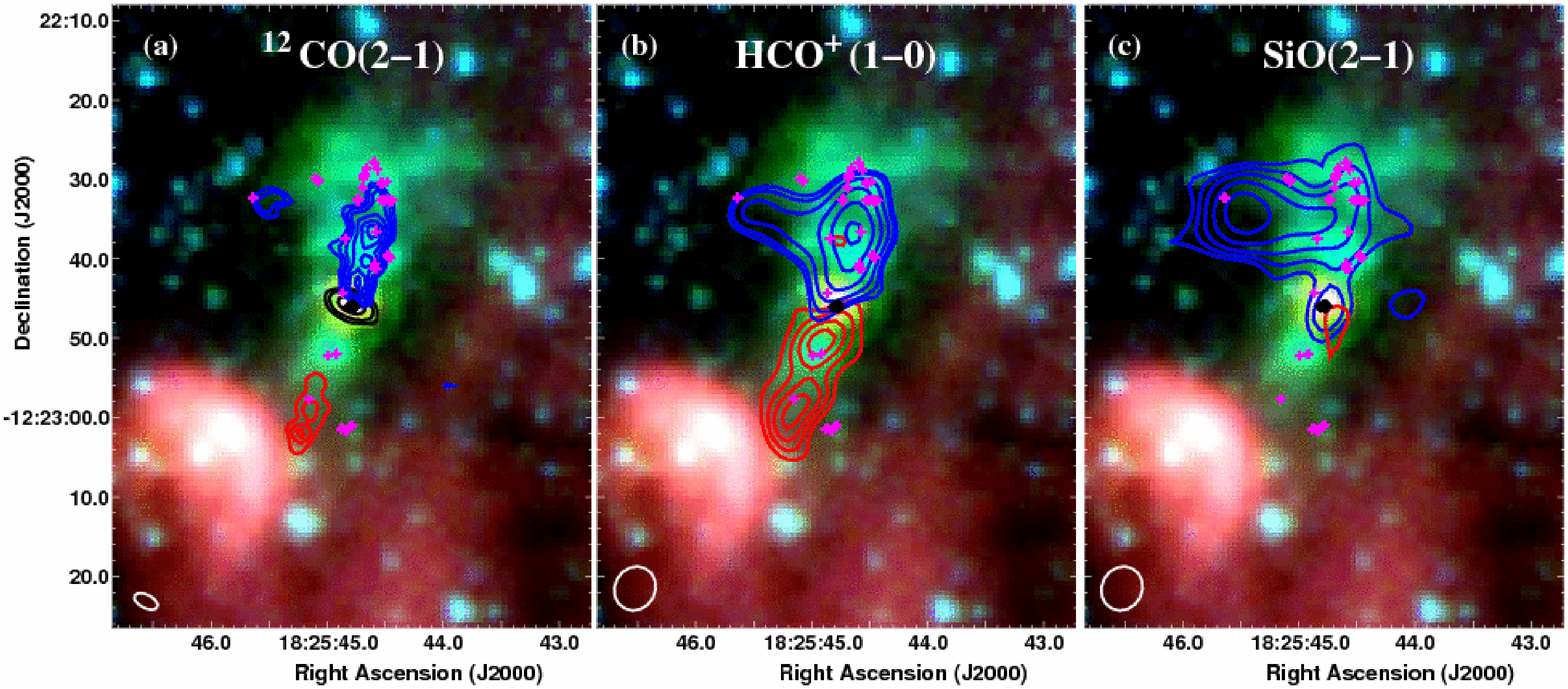}
\caption{G19.01-0.03: Three color \emph{Spitzer} images (3.6, 4.5, 8.0
$\mu$m: blue, green, red) overlaid with contours of high velocity (a)
\co(2-1), (b) \hco(1-0), and (c) SiO(2-1) emission.  In each panel,
positions of 6.7 GHz \meth\/ masers are marked with black diamonds,
and positions of 44 GHz \meth\/ masers are marked with magneta crosses
\citep{maserpap}.  The \vlsr\/ is \q 60 \kms\/ (\S\ref{g19_compact}).
(a) \co(2-1) emission integrated over v=$-$46.4 to 19.5 \kms\/ (blue)
and v=75.6 to 88.8 \kms\/ (red).  Contour levels: Blue: 7.2, 9.6, 12,
15.6, 19.2, 22.8 \jb\/ \kms; Red: 4.8, 7.2, 9.6 \jb\/ \kms.  Black
contours: SMA 1.3 mm continuum emission (levels 5,10,30 $\times$ $\sigma =$ 3.5
\mjb).  The SMA beam is shown at lower left.  (b) \hco(1-0) emission
integrated over v=7.8 to 52.2 \kms\/ (blue) and 68.6 to 83.4 \kms\/
(red), e.g. \q\vlsr$\pm$8 \kms.  Contour levels: Blue: 1.0, 1.2, 1.6,
2.2, 3.0, 4.0 \jb\/ \kms; Red: 0.6, 0.8, 1.0, 1.2 \jb\/ \kms.  The
CARMA beam is shown at lower left.  (c) SiO(2-1) emission integrated
over v=26.0 to 52.9 \kms\/ (blue) and v=66.4 to 83.3 \kms\/ (red),
e.g. \q\vlsr$\pm$7 \kms.  Contour levels: Blue: 0.5, 0.8, 1.2, 1.8,
2.6 \jb\/ \kms; Red: 0.5 \jb\/ \kms.  The CARMA beam is shown at lower
left.}
\label{g19redbluefig}
\end{figure}

\begin{figure}
\plotone{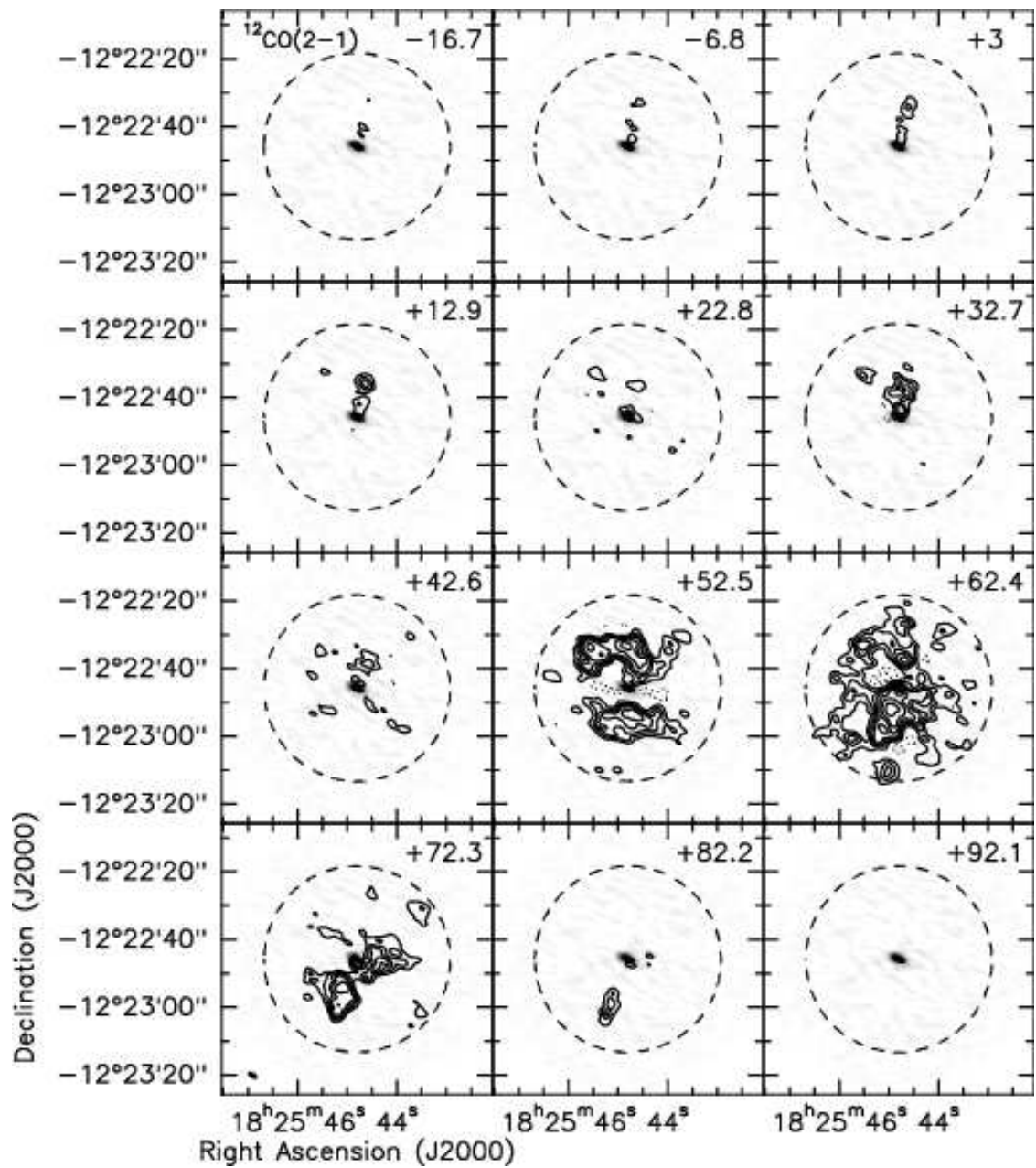}
\caption{G19.01$-$0.03: Channel maps of \co(2-1).  The greyscale is
  the 1.3 mm SMA continuum image and the contours are the \co\/
  emission (levels 0.2, 0.4, 0.6, 0.8, 1.2, 1.8, 3 \jb (solid), $-$0.2, $-$0.4 \jb (dotted)).  Each panel
  is the average of three channels of the smoothed \co\/ cube; the
  center velocity of each panel is given at upper right.  The contour
  levels are ($-$8,$-$4,4,8,16,24,36,60) $\times$ the rms in a line-free panel.
  The field of view shown is the same as in Figures~\ref{g19_hcochan}
  and ~\ref{g19_siochan}.  The dashed circle shows the FWHP of the SMA
  primary beam.  The SMA synthesized beam is shown as a filled ellipse
  at lower left.}
\label{g19_12cochan}
\end{figure}

\begin{figure}
\plotone{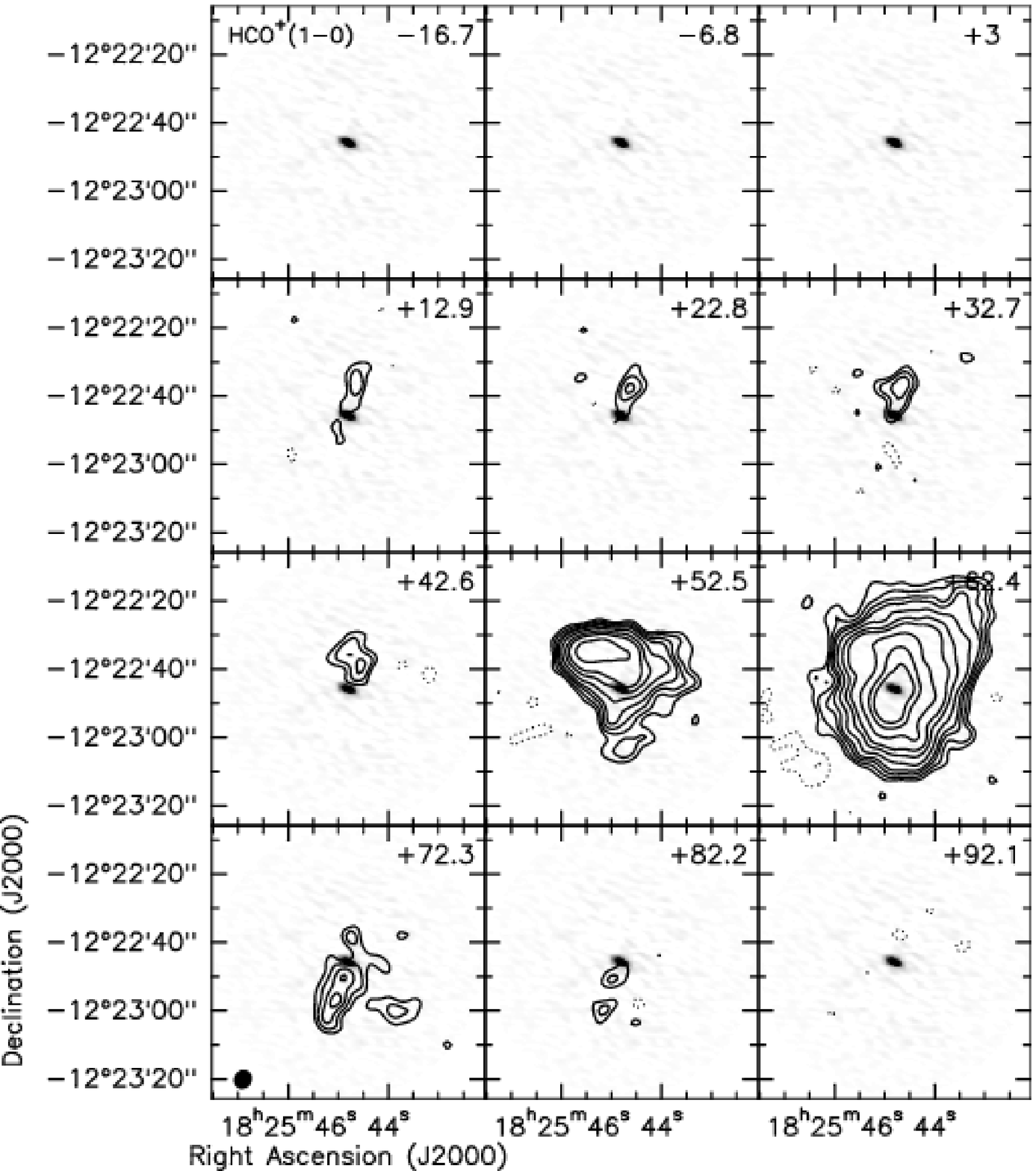}
\caption{G19.01$-$0.03: Channel maps of \hco(1-0).  The greyscale is
the 1.3 mm SMA continuum image.  The contours are the \hco\/ emission
(levels 0.03, 0.048, 0.072, 0.096, 0.144, 0.216, 0.288, 0.432, 0.648,
0.864 \jb (solid), $-$0.03, $-$0.048 \jb (dotted)).  The contour levels are ($-$8,$-$5,5,8,12,16,24,36,48,72,108,144)
$\times$ the rms in a line-free panel.  The center velocity of each
panel is given at upper right.  The velocity range shown matches that
of Figure~\ref{g19_12cochan}.  Note that the velocity coverage of the
CARMA \hco\/ observations does not extend blueward of \q 8 \kms.  The
field of view shown is the FWHP primary beam of the 10 m CARMA
antennas (80\pp\/ square).  The CARMA synthesized beam is shown as a
filled ellipse at lower left.}
\label{g19_hcochan}
\end{figure}

\begin{figure}
\plotone{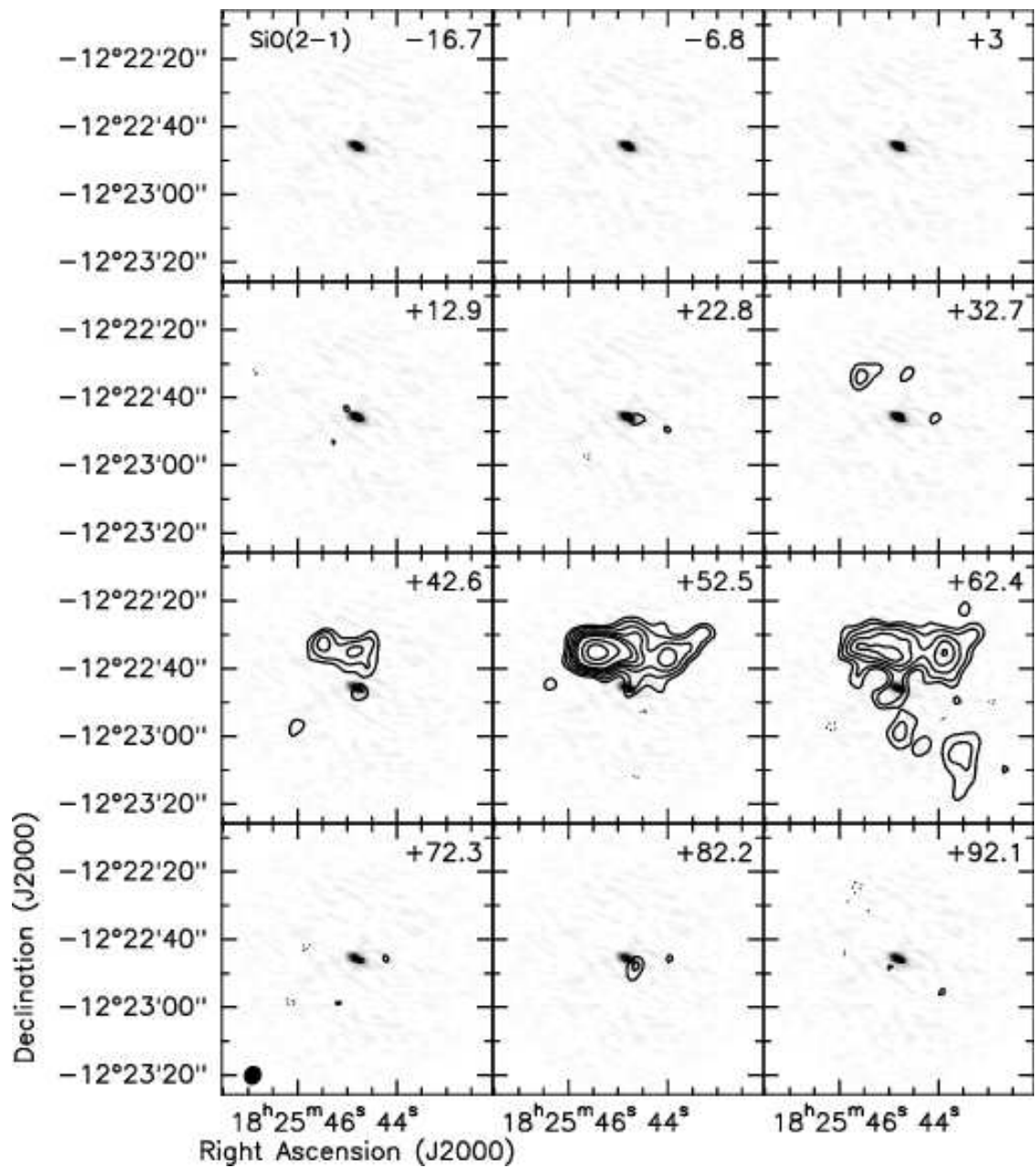}
\caption{G19.01$-$0.03: Channel maps of SiO(2-1).  The greyscale is
the 1.3 mm SMA continuum image.  The contours are the SiO emission
(levels 0.03, 0.048, 0.072, 0.096, 0.12, 0.144,0.216,0.288 \jb (solid), $-$0.03 \jb (dotted)).  The
contour levels are ($-$5,5,8,12,16,20,24,36,48) $\times$ the rms in a
line-free panel.  The center velocity of each panel is given at upper
right.  The velocity range shown matches that of
Figure~\ref{g19_12cochan}.  Note that the velocity coverage of the
CARMA SiO observations does not extend blueward of \q 8 \kms. The
field of view shown is the FWHP primary beam of the 10 m CARMA
antennas (80\pp\/ square).  The CARMA synthesized beam is shown as a
filled ellipse at lower left.}
\label{g19_siochan}
\end{figure}

Extended molecular line emission, spanning most of the telescope field
of view, is exhibited by \co(2-1) and \hco(1-0).  Extended SiO(2-1)
emission is also observed.  Figure~\ref{g19redbluefig} presents
integrated intensity images of high-velocity gas, while
Figures~\ref{g19_12cochan}, \ref{g19_hcochan}, and \ref{g19_siochan}
show channel maps of the \co(2-1), \hco(1-0), and SiO(2-1) emission,
respectively.  

As shown in Figures~\ref{g19redbluefig}-\ref{g19_hcochan}, \co\/ and \hco\/
trace a bipolar molecular outflow centered on the continuum source MM1.  The
outflow axis is roughly N-S, with the blueshifted lobe to the north and the
redshifted lobe to the south.  
This is consistent with the velocity gradient
of the 44 GHz Class I \meth\/ masers \citep{maserpap}.  Blueshifted 44 GHz
\meth\/ masers are concentrated towards the northern 4.5 \um\/ lobe, while
systemic and redshifted 44 GHz \meth\/ masers are concentrated to the south of
the central source.  The 44 GHz Class I \meth\/ masers trace the edges of the
high-velocity \co\/ and \hco\/ lobes remarkably well (Fig.~\ref{g19redbluefig}a,b).
In particular, an arc of 44 GHz masers
appears to trace the edges and terminus of the blueshifted \co\/ jet.

The full velocity range of the \co\/ outflow is
\q 135 \kms.  The kinematics of the outflow are notably asymmetric.  The
highest-velocity blueshifted gas has $|$v$_{max,blue}-$\vlsr$|$\q 106 \kms,
while for the redshifted lobe, $|$v$_{max,red}-$\vlsr$|$ is only \q 29 \kms.
The velocity distribution of the 44 GHz \meth\/ masers
is also asymmetric with respect to the \vlsr: $|$v$_{max,blue,maser}-$\vlsr$|$\q
6.5 \kms\/ while $|$v$_{max,red,maser}-$\vlsr$|$\q
2.2 \kms\/ \citep{maserpap}.
The blueshifted lobe of the molecular outflow traced by \co\/ and \hco\/ is
coincident with the northern lobe of extended 4.5 \um\/ emission.  In
contrast, the most highly redshifted molecular gas is found south of the
brightest 4.5 \um\/ emission in the 
southern lobe.  The high velocity outflow gas is clumpy.  Both the
red and blue lobes are characterized by strings of bright knots.

The SiO(2-1) emission differs in kinematics and morphology from the
\co\/ and \hco\/ emission (Fig.~\ref{g19redbluefig}).  
Very little redshifted SiO emission is detected.
Blueshifted SiO emission is concentrated north of the continuum source,
consistent with the orientation of the \co/\hco\/ outflow.  The morphology of
the blueshifted SiO emission, however, is linearly extended along an E-W axis.  Near the
systemic velocity (\q60 \kms), the SiO emission extends to the south towards the
continuum source (Fig.~\ref{g19_siochan}).  
Like the high-velocity
\co\/ and \hco\/ emission, the SiO emission is characterized by clumps and
knots.  The strongest blueshifted SiO emission arises from a clump offset to
the east of the extended 4.5 \um\/ emission.

Near the \vlsr (\q 60 \kms), the \co\/ image cube shows artifacts from
large-scale emission resolved out by the interferometer (Fig.~\ref{g19_12cochan}).
This suggests that near the systemic velocity, the \co(2-1) emission is
dominated by emission from a large-scale extended envelope.  This
interpretation is consistent with the large spatial extent of the \hco(1-0)
emission near the \vlsr\/ (Fig.~\ref{g19_hcochan}).

\subsubsection{Millimeter \methanol\/ masers: G19.01$-$0.03}

\begin{figure}
\plotone{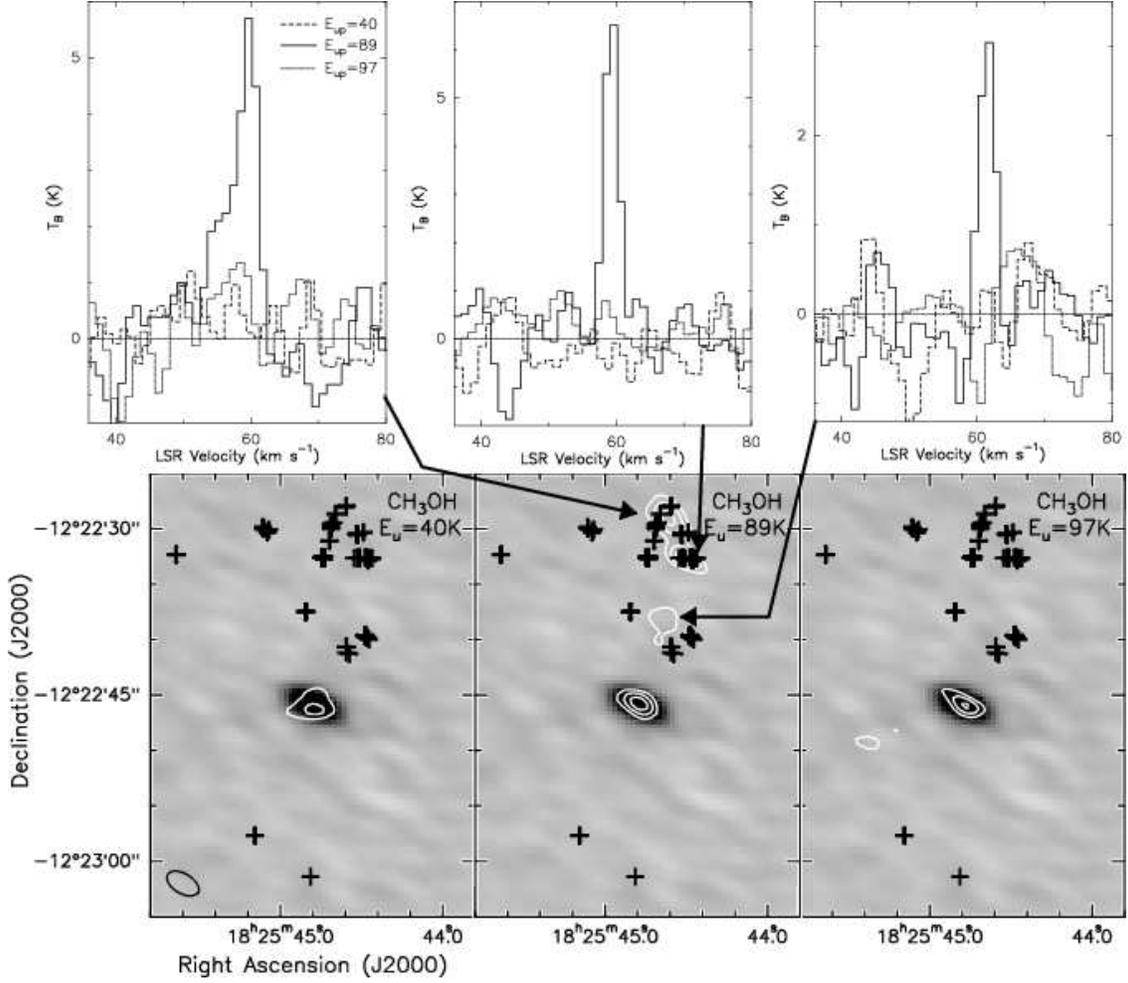}
\caption{G19.01$-$0.03: The greyscale shows the SMA 1.3 mm continuum
  emission.  Contours are drawn from integrated intensity (moment
  zero) maps of the indicated \methanol\/ line.  The upper state
  energy in K of the transition is given at upper right in each panel.
  Contour levels are: \eup=40 K: 1.11, 1.85 Jy beam$^{-1}$ \kms;
  \eup=89 K: 1.02, 1.7, 2.38 Jy beam$^{-1}$ \kms; \eup=97 K: 1.32,
  2.20, 3.08 Jy beam$^{-1}$ \kms.  These contour levels are
  (3,5)$\times$ $\sigma$ for the \eup=40 K transition and
  (3,5,7)$\times$ $\sigma$ for the \eup=89 K and \eup=97 K
  transitions.  Black crosses mark the positions of 44 GHz Class I
  \methanol\/ masers from \citet{maserpap}.  The SMA beam is shown at
  lower left.  Spectra are shown at the peak position of the indicated
  \methanol\/ (8$_{-1,8}$-7$_{0,7}$) emission spot.  Solid line:
  8$_{-1,8}$-7$_{0,7}$ transition (229.759 GHz, \eup=89 K); dashed
  line: 3$_{-2,2}$-4$_{-1,4}$ (230.027 GHz, \eup=40 K); dotted line:
  8$_{0,8}$-7$_{1,6}$ (220.078 GHz, \eup=97 K).}
\label{g19methmom0}
\end{figure}

As seen in G11.92$-$0.61, the 229.759 GHz \meth(8$_{-1,8}$-7$_{0,7}$)E
(\eup=89 K) emission towards G19.01$-$0.03 has a very different morphology
than any other observed \meth\/ transition.  Figure~\ref{g19methmom0} presents
integrated intensity maps of the 230.027 GHz (\eup=40 K), 229.759 GHz, and
220.078 GHz (\eup=97 K) \meth\/ emission towards G19.01$-$0.03, with the
positions of 44 GHz Class I masers from \citet{maserpap} marked.  There are
three compact loci of 229.759 GHz \meth\/ emission to the north of MM1, two of
which are coincident with numerous 44 GHz masers.  Figure~\ref{g19methmom0}
also shows the profiles of the 230.027 GHz, 229.759 GHz, and 220.078 GHz
\meth\/ emission towards the peak of each of these spots.  The 229.759 GHz
\meth\/ emission towards these loci is spectrally narrow.  Towards the two
northern spots, the velocity of the 229.759 GHz \meth\/ emission agrees well
with the velocity range of the coincident 44 GHz \meth\/ masers.  In the
integrated intensity map shown in Figure~\ref{g19methmom0}, the southernmost
of the three 229.759 GHz \meth\/ emission spots appears to lie between
three clusters of 44 GHz \meth\/ masers.  The SMA data has much lower spatial
and spectral resolution than the 44 GHz VLA data (SMA: 3\pp, 1.6 \kms; VLA:
0\farcs75, 0.24 \kms).  Examination of the data cubes shows that the 229.759
GHz \meth\/ emission near this location is consistent with being a blend of
masers seen at 44 GHz, given the lower spatial and spectral resolution of the
mm data.  As shown in Figure~\ref{g19methmom0}, the 229.759/230.027 line
ratios towards the three spots are all $>$3, consistent with non-thermal
229.759 GHz \meth\/ emission \citep[][see also
\S\ref{g11methmasers}]{Slysh02}.  Towards the MM1 continuum peak, this line
ratio is 2, consistent with thermal emission.

\section{Discussion}\label{discussion}

\subsection{Continuum Sources}

\subsubsection{Spectral Energy Distributions (SEDs)}\label{seds}

\begin{figure}
\epsscale{1.0}
\plotone{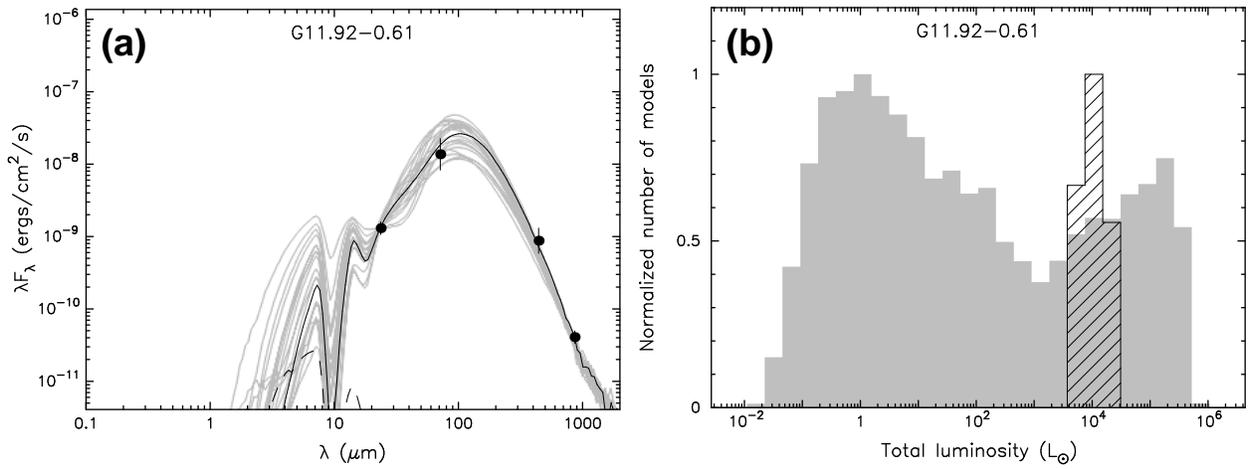}
\caption{G11.92$-$0.61: (a) Data (black circles) overlaid with the best fit model (solid
  black line) and other good fits (grey lines) from the \citet{Robitaille07}
  model grid.  The twenty best fits are shown.  (b) Histogram of the
  bolometric luminosity of the 20 best-fit models (hatched histogram) and all
  models in the grid (grey histogram).}
\label{g1192_robitaille_fits}
\end{figure}

Unfortunately, the three members of the G11.92$-$0.61 (proto)cluster are not
resolved in existing data at wavelengths shorter than 1.3 mm.  To better
constrain the SED of the protocluster as a whole, we measured the 70 \um\/
flux density from MIPSGAL images \citep{Carey09}.  We find a flux density of
369 Jy, with an estimated uncertainty of \q50\% due to artifacts in the
publicly available BCD images \citep{Carey09}.  \citet{Walsh03} report
integrated flux densities of 140 Jy (450 \um) and 12 Jy (850 \um) for the
(sub)mm clump (G11.92$-$0.64B in their nomenclature).  Using these fluxes and
the 24 \um\/ flux from \citet{maserpap}, 
we fit the 24-850 \um\/ data using the model fitter of \citet{Robitaille07}
\footnote{http://caravan.astro.wisc.edu/protostars/}.  We do not include IRAC
data in the SED because emission mechanisms not included in the models
(e.g. line emission from shocked molecular gas and PAHs) may contribute
significantly to the IRAC bands in this source.  The fits, shown in
Figure~\ref{g1192_robitaille_fits}a, are consistent with a bolometric
luminosity of \q10$^{4}$ \lsun\/ for the cluster as a whole
(Fig.~\ref{g1192_robitaille_fits}b).  The models in the \citet{Robitaille07}
grid assume a single central object in determing source properties (such as
stellar mass) from the SED.  Since G11.92$-$0.61 is a (proto)cluster and
unresolved at $\lambda<$ 1.3 mm, the model fitter cannot be used to constrain
the properties of the cluster members.  

\begin{figure}
\plotone{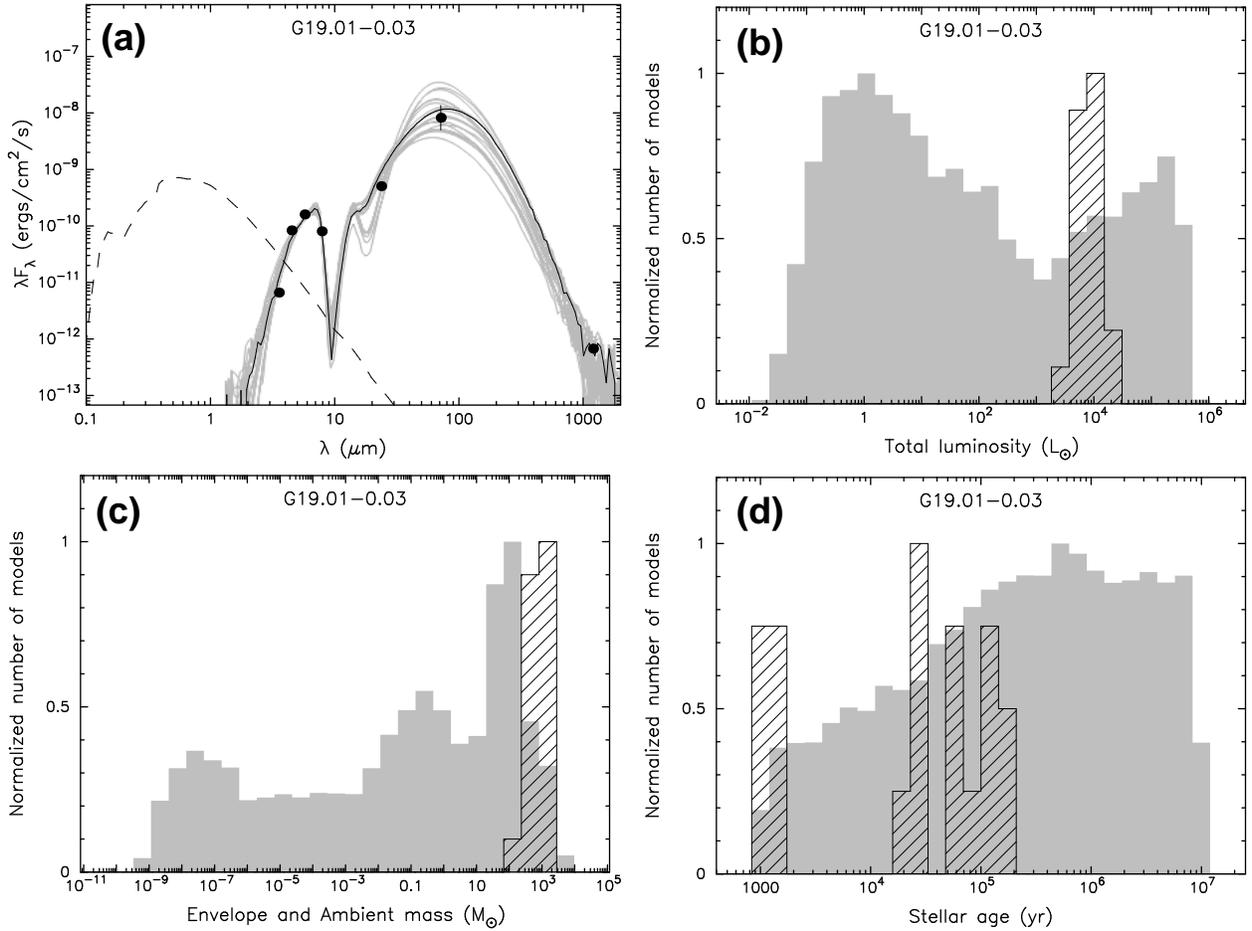}
\caption{G19.01$-$0.03: (a) Data (black circles) overlaid with the best fit
model (solid black line) and other good fits (grey lines) from the
\citet{Robitaille07} model grid.  The twenty best fits are shown.  The dashed line shows the stellar photosphere, in the absence of circumstellar dust, of the central source for the best-fit model.  Histograms
of (b) the bolometric luminosity, (c) the envelope mass, and (d) the stellar
age for the 20 best-fit models (hatched histograms) and all models in the grid
(grey histograms).}
\label{g1901_robitaille_fits}
\end{figure}

\begin{figure}
\plotone{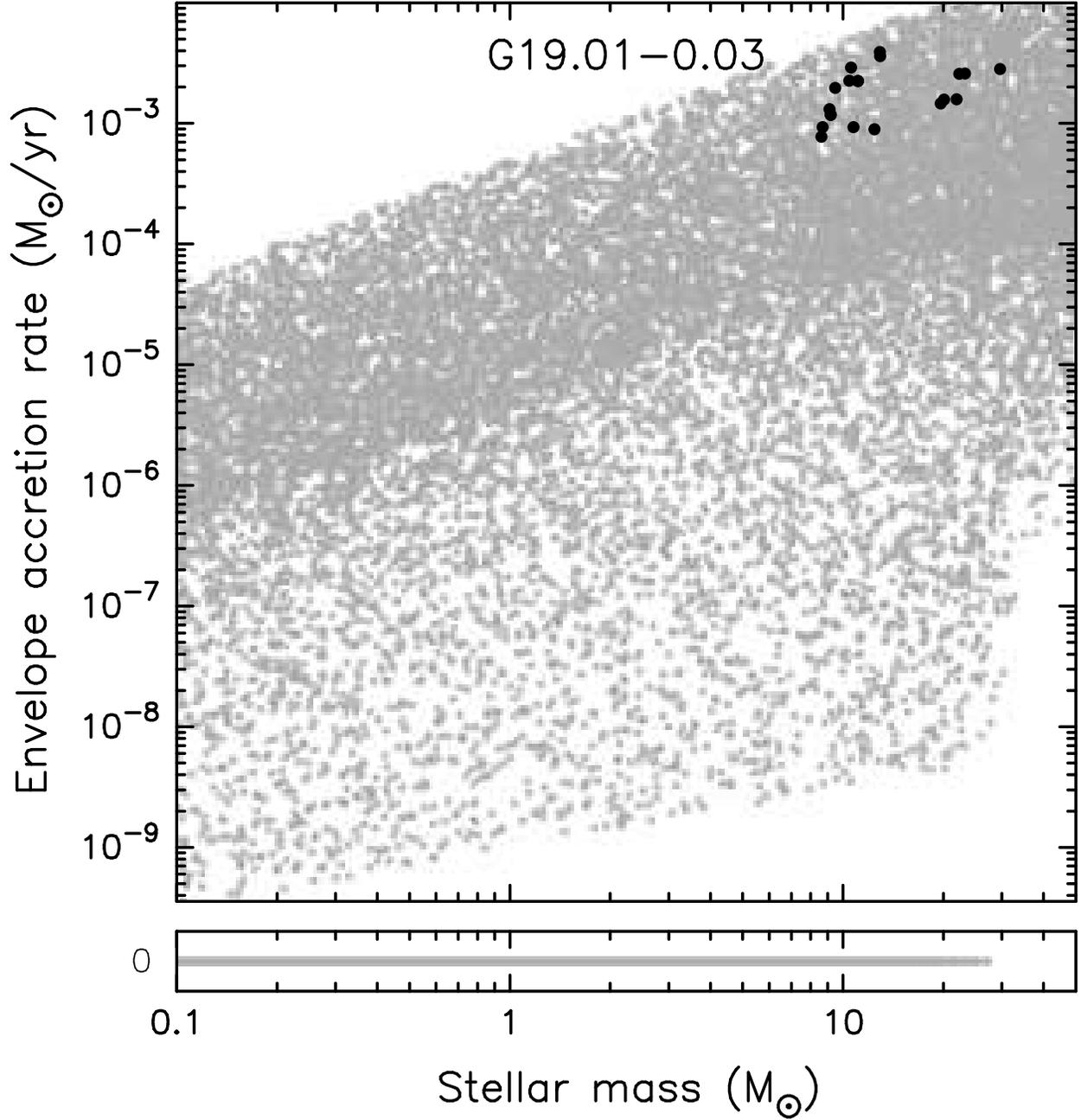}
\caption{G19.01$-$0.03: Envelope accretion rate
  v. stellar mass for the 20 best-fit models (black dots) and all models in
  the grid (grey dots).}
\label{g1901_robitaille_2}
\end{figure}

The EGO G19.01$-$0.03 is unusual in that the ``central'' source is clearly
resolved from the extended emission in IRAC images (unlike most EGOs), and is a GLIMPSE point
source \citep[see also][]{egocat}.  SED modeling can thus be used to infer the
properties of the source driving the 4.5 \um\/ outflow.  As for G11.92$-$0.61, we measured the 70
\um\/ flux density of G19.01$-$0.03 from the MIPSGAL image.  We find a flux
density of 223 Jy, with an estimated uncertainty of 50\%.  Using this
measurement, GLIMPSE catalog photometry for
the point source SSTGLMC G019.0087-00.0293, the 24 \um\/ flux density from
\citet{maserpap}, and the 1.3 mm flux density from Table~\ref{conttable}, we
fit the SED using the \citet{Robitaille07} model fitter.  As shown
in Figures~\ref{g1901_robitaille_fits}-\ref{g1901_robitaille_2}, the SED is
well-fit (Fig.~\ref{g1901_robitaille_fits}a) by models with
bolometric luminosity of \q10$^{4}$ \lsun\/
(Fig.~\ref{g1901_robitaille_fits}b), stellar mass \q10 \msun\/ (Fig.~\ref{g1901_robitaille_2}), envelope
mass \q10$^{3}$ \msun\/ (Fig.~\ref{g1901_robitaille_fits}c), and envelope accretion rate \q10$^{-3}$ \msun\/ 
yr$^{-1}$  (Fig.~\ref{g1901_robitaille_2}).  
Stellar age is also a parameter of the models, but is not
well-constrained (Fig.~\ref{g1901_robitaille_fits}d).  In
the scheme of \citet{Robitaille06}, evolutionary stage is defined by the ratio of the envelope accretion rate
to the stellar mass.  The youngest sources, Stage 0/I, are defined as having
\.{M}$_{env}$/M$_{*}>$10$^{-6}$ yr$^{-1}$.  For G19.01$-$0.03,
\.{M}$_{env}$/M$_{*}$\q10$^{-4}$ yr$^{-1}$
(Fig.~\ref{g1901_robitaille_2}), indicating that the source is likely to be very young.

\subsubsection{Temperature Estimates from Line Emission}\label{temps}

\begin{figure}
\includegraphics*[width=10.5cm, angle=-90,bb=80 50 552 705]{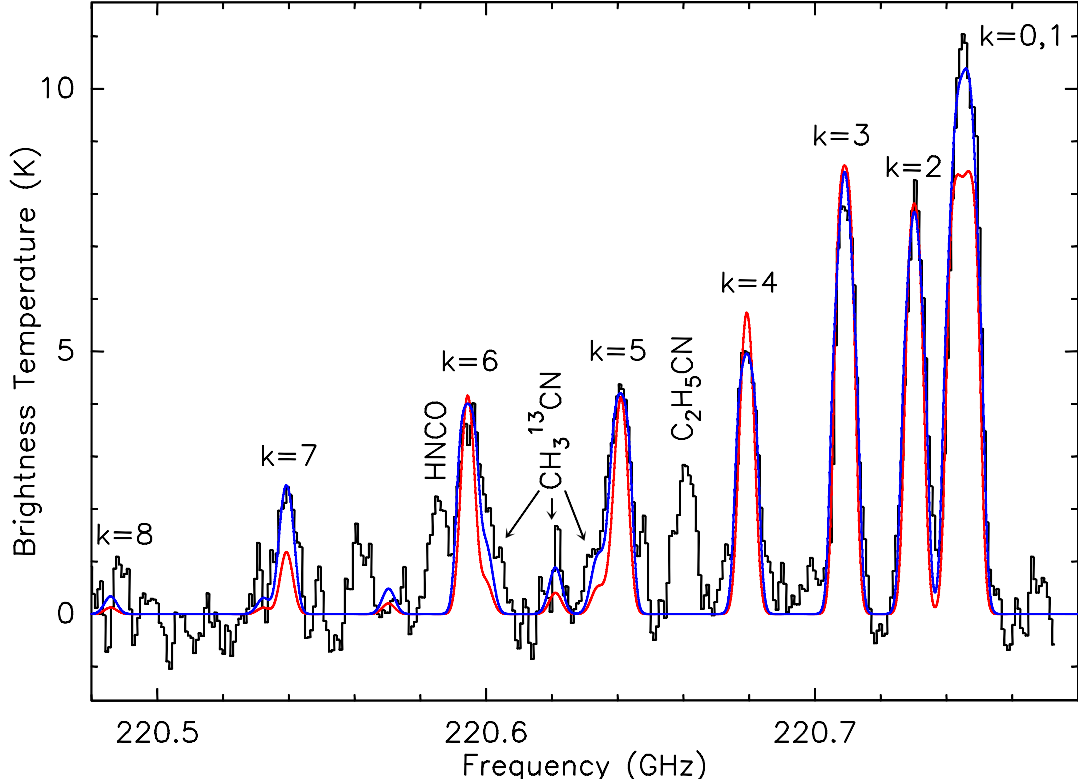}
\caption{SMA \methcyn(12-11) spectrum towards the G11.92$-$0.61-MM1 continuum peak
(black histogram) overlaid with the best-fit one component (red line) and two
component (blue line) models.  The physical parameters of the best-fit models
are summarized in Table~\ref{methcynfittable}.  For the single component
model, $\Delta$v$_{FWHM}$=6 \kms; for the two component model,
$\Delta$v$_{FWHM}$(cool)=7 \kms and $\Delta$v$_{FWHM}$(warm)=6 \kms.  For all
components, \vlsr=35.5 \kms.  The \methcyn\/ model accounts for emission from
CH$_{3}^{13}$CN, assuming CH$_{3}^{12}$C:CH$_{3}^{13}$C=60:1.  The strongest
observed CH$_{3}^{13}$CN lines, including those blended with the CH$_{3}$CN
k=5 and k=6 components, are indicated.  The C$_{2}$H$_{5}$CN line at 220.661
GHz is not fit, nor is the HNCO line at 220.585 GHz (partially blended \methcyn\/ k=6).}
\label{g11methcynfit}
\end{figure}

\begin{figure}
\includegraphics*[width=10.5cm,angle=-90,bb=80 50 552 705]{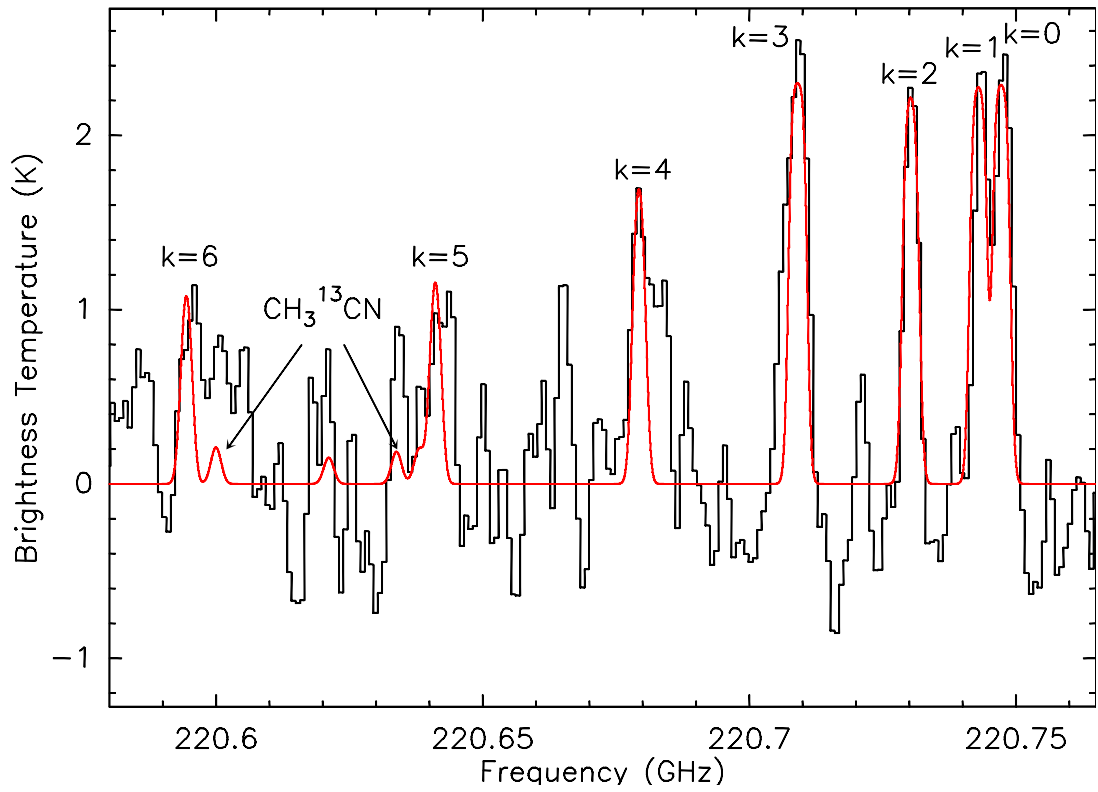}
\caption{SMA \methcyn(12-11) spectrum towards the G19.01$-$0.03-MM1 continuum peak
  (black histogram) overlaid with the best-fit single component model (
  red line).  The physical parameters of the best-fit model
are summarized in Table~\ref{methcynfittable}.}
\label{g19single}
\end{figure}

The J=12-11 \methcyn\/ ladder is well-suited for measuring the gas
temperature in hot cores
\citep[e.g.][]{Pankonin01,araya05,Zhang07,Qiu09}.
Figures~\ref{g11methcynfit} and ~\ref{g19single} show the best-fit
single-component model of the \methcyn\/ emission overlaid on the
observed spectrum at, respectively, the G11.92$-$0.61-MM1 and
G19.01$-$0.03-MM1 continuum peak.
For each \methcyn\/ emission component, the model\footnote{Developed
using the XCLASS package,
http://www.astro.uni-koeln.de/projects/schilke/XCLASS} assumes local
thermodynamic equilibrium (LTE) and the same excitation conditions for
all K components, and accounts for optical depth effects and emission
from the isotope CH$_{3}^{13}$CN.  The velocity (frequency) separations
of the K components are fixed to the laboratory values.  The
temperature, size (diameter), and \methcyn\/ column density of the
emitting region are free parameters, and the model that best fits the
observed spectrum is found by minimizing the mean squared error.  The
parameters of the best-fit models are summarized in
Table~\ref{methcynfittable}.


A single-component model provides an adequate fit to the \methcyn\/ spectrum
of 
G19.01$-$0.03-MM1 (Fig.~\ref{g19single}; T\q114 K, size\q2500 AU).  In
contrast, a single-component model is a notably poor fit to the
\methcyn\/ spectrum of G11.92$-$0.61-MM1 (Fig.~\ref{g11methcynfit}).  In
particular, the model severely underpredicts the emission from the k=7 and the
(blended) k=0/1 lines, while overpredicting the emission from most of the
intermediate k components (k=3,4,6).  To investigate this discrepancy, we
allowed for two \methcyn-emitting regions, with different temperatures, sizes,
and column densities.  As shown in Figure~\ref{g11methcynfit}, a two-component
model that includes
a compact (\q2300 AU), warm (\q166 K) component and an extended (\q11400 AU), cool
(\q77 K) component provides a much better fit to the data (see also
Table~\ref{methcynfittable}).  This combination of parameters is likely not
unique, and certainly we expect that the real emission exhibits a gradient in
temperature rather than a step function.  Even so, this result convincingly
demonstrates that both cool and warm temperatures are present.  
Interestingly, the physical scale of the warm component
(\q2300 AU) agrees remarkably well with that of \methcyn-emitting region in
G19.01$-$0.03-MM1 (\q2500 AU).

\begin{figure}
\plotone{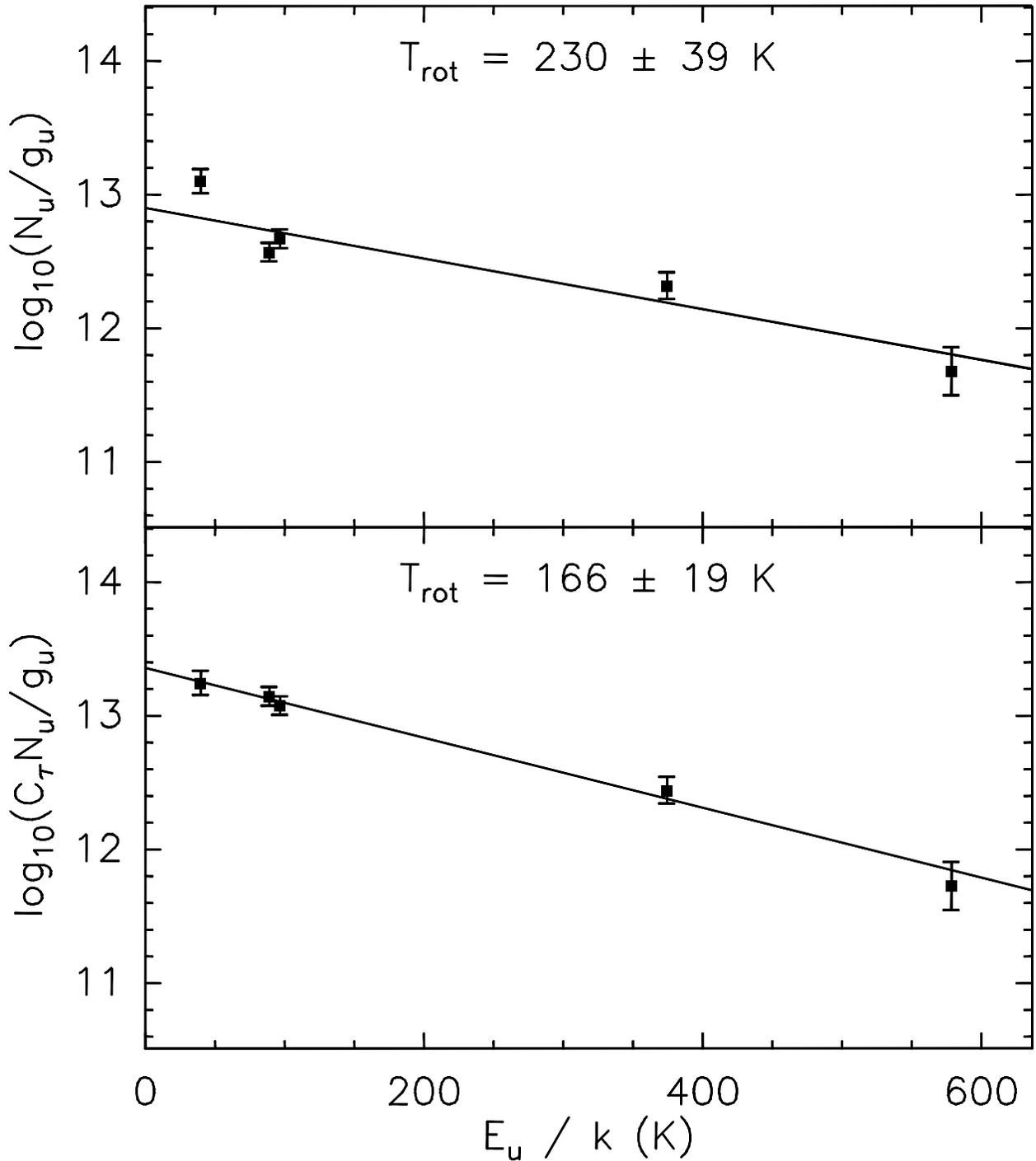}
\caption{Rotation diagrams for the \meth\/ transitions observed at the
G11.92$-$0.61-MM1 continuum peak.  The upper panel shows the fit to the raw
data.  In the lower panel, the column densities have been corrected for
optical depth effects as described in \S\ref{temps}.  The fitted temperatures
are indicated in each panel.}
\label{g11methrot}
\end{figure}

Five transitions of \meth\/ are detected towards the G11.92$-$0.61-MM1
continuum peak, with \eup=40-579 K.  This is sufficient
to obtain an independent estimate of the gas temperature by applying the rotation
diagram method \citep[e.g.][]{GL99} to the observed \meth\/ emission, using
the relations:
\begin{equation}
{\frac{N_u}{g_u}}={\frac{3k}{8\pi^{3}\nu g_I
    g_K}}{\frac{1}{\mu^{2}S}}\int{T dv},
\end{equation}
and
\begin{equation}
log(N_u)/(g_u) = log(N_{tot}/Q(T_{rot})) - 0.4343 E_{u}/kT_{rot},
\label{roteqn}
\end{equation}
where N$_{u}$ is the column density in the upper state, k is Boltzmann's
constant, $\nu$ is the line rest frequency, g$_{I}$ is the nuclear spin
degeneracy, g$_{K}$ is the K degeneracy, $\mu^{2}$S is the product of the
square of the molecular dipole moment and the line strength, $\int{T dv}$ is the
observed integrated intensity of the line, N$_{tot}$ is the total column
density, Q(T$_{rot}$) is the partition function evaluated at the rotation
temperature T$_{rot}$, and E$_{u}$ is the upper state energy of the
transition.  For each transition, the integrated line intensity was determined from a single
Gaussian fit to the line emission at the 1.3 mm continuum peak (Table~\ref{g11trans}).    
Both A and E transitions are included in the rotation diagram analysis; we do
not detect enough transitions to treat the two types separately.  The rotation
temperature derived from a weighted least-squares fit to the data is 230$\pm$39
K (Fig.~\ref{g11methrot}), somewhat higher than the temperature derived from
the \methcyn\/ fitting.  

As discussed in detail in \citet{GL99}, however, optical depth effects can
inflate the temperature derived from a rotation diagram analysis.  
We follow the procedure of \citet{Brogan07,Brogan09} in iteratively solving for
the values of $\tau$ and T$_{rot}$ that best fit the data
(\begin{math} C_{\tau}[i]=\tau[i]/(1-e^{\tau[i]}) \end{math} where i
refers to the ith spectral line).  As shown in Figure~\ref{g11methrot}, this
improves the fit considerably.  The optical depth in the line with the
highest opacity (\meth(8$_{-1,8}$-7$_{0,7}$) at 229.759 GHz) is 3.67.
With the optical depth correction, the derived temperature is
166$\pm$19 K.  This is in remarkably good agreement with the
temperature of the warm component from the \methcyn\/ analysis (166
K).  Only three \meth\/ transitions (\eup\/ 40-97 K) are detected
towards G19.01$-$0.03-MM1, too few for accurate rotation diagram
analysis, but consistent with the cooler temperature derived for this
source from \methcyn.

\subsubsection{Mass Estimates from the Dust Emission}\label{mass_dis}

At millimeter wavelengths, thermal emission from dust and free-free emission from ionized gas can both contribute to the observed continuum flux.
For our target EGOs, 
the available limits on any free-free contribution
are not very stringent.  Neither G11.92$-$0.61 nor G19.01$-$0.03 had detectable
44 GHz continuum emission in the VLA observations of \citet{maserpap}.  The
5$\sigma$ limits are 7 \mjb\/ (synthesized beam 
0\farcs99$\times$0\farcs44) and 5 \mjb\/ (synthesized beam 0\farcs69$\times$0\farcs51) respectively.  Extrapolating the
5$\sigma$ 44 GHz upper limits assuming optically thin free-free emission
(\begin{math} S_{\nu} \propto \nu^{\alpha} \end{math}, $\alpha$=-0.1) gives
upper limits at 1.3 mm of 6.0 mJy for G11.92$-$0.61 and 4.3 mJy for
G19.01$-$0.03.  For the adopted dust temperatures for G11.92$-$0.61-MM1 and
G19.01$-$0.03-MM1, a free-free contribution at this level would have a minimal
impact on the mass estimates ($\lesssim$0.4 \msun).  
If we instead extrapolate
the 5$\sigma$ 44 GHz upper limits assuming a spectral index $\alpha$=1
\citep[appropriate for a hypercompact (HC) HII region, e.g.][]{Kurtz05}, the effect
on the mass estimates is more substantial, up to \q2.5 \msun.  For the weakest mm
continuum source, G11.92$-$0.61-MM3, free-free emission from a HC HII region
($\alpha$=1) could in principle account for the entirety of the 1.4 mm
flux density observed with CARMA and the majority (\q73\%) of the 1.3 mm flux
density observed with the SMA.
Deep, high-resolution
continuum data at a range of cm wavelengths are required to constrain the
presence and properties of any ionized gas associated with our target EGOs.
In the absence of available evidence to the contrary, we assume the entirety
of the millimeter-wavelength continuum emission is attributable to thermal emission from
dust.

Table~\ref{dustmasstable} presents estimates from the thermal dust emission for the gas mass M$_{gas}$,
column density of molecular hydrogen N$_{H_2}$, and volume density of
molecular hydrogen n$_{H_2}$, for G11.92$-$0.61-MM1, MM2, and MM3 and
G19.01$-$0.03-MM1.  The gas masses are calculated from:
\begin{equation}
M_{gas}= {\frac{4.79 \times 10^{-14} R S_{\nu}(Jy) D^2(kpc)C_{\tau_{dust}}}{B(\nu,T_{dust})\kappa_{\nu}}},
\label{masseqn}
\end{equation}
where R is the gas-to-dust mass ratio (assumed to be 100), S$_{\nu}$ is the
integrated flux density from Table~\ref{conttable}, D is the distance to the
source, C$_{\tau_{dust}}$ is the correction factor for the dust
opacity \begin{math}
  C_{\tau_{dust}}=\tau_{dust}/(1-e^{\tau_{dust}}) \end{math},
B($\nu,T_{dust}$) is the Planck function, and $\kappa_{\nu}$ is the dust mass
opacity coefficient in units of cm$^{2}$ g$^{-1}$.  For gas densities of
10$^{6}$-10$^{8}$ cm$^{-3}$, $\kappa_{1.3 mm}$\q1 for dust grains with thick
or thin ice mantles \citep{OH94}.  
Scaling from $\kappa_{1.3 mm}$=1 assuming
$\beta$=1.5, we adopt $\kappa_{3.4 mm}$=0.24.  
We estimate a range of dust temperatures for each source based on its
observed spectral line properties (discussed in detail below).  The
dust opacity, ${\tau_{dust}}= -\ln (1-\frac{T_b}{T_{dust}})$, is
derived using the beam-averaged brightness temperature ($T_b$) and
assumed dust temperature ($T_{dust}$) for each source and listed in
Table~\ref{dustmasstable}.  The calculated values of $\tau_{dust}$ are
generally small ($<$0.1), indicating that the dust emission is
optically thin.  The column densities and volume densities presented
in Table~\ref{dustmasstable} are also beam-averaged values.

As noted above, estimating gas masses using equation~\ref{masseqn} requires an estimate of the
dust temperature.  For G11.92$-$0.61-MM1 and G19.01$-$0.03-MM1 we use the
values of T$_{gas}$ derived from the \methcyn\/ and \meth\/ emission
(\S\ref{temps}).  At the high gas densities implied by our observations
($\gtrsim$10$^{6}$ cm$^{-3}$), the gas and dust temperatures are expected to
be well-coupled \citep[e.g.][]{Ceccarelli96,Kaufman98}.  For
G11.92$-$0.61-MM1, the situation is complicated by the presence of two
temperature components, implied by the \methcyn\/ fits (\S\ref{temps}).  Both
the compact warm (size \q0\farcs6) and more extended cool (size \q3\farcs0)
emission regions are similar in scale to the 3\farcs2 $\times$ 1\farcs8 SMA beam.  A step-function temperature structure is physically
unrealistic, but the sensitivity and spatial resolution of the present
observations are insufficient to resolve the temperature gradient in MM1.  
In the future, the sensitivity and high spatial resolution attainable with 
ALMA will allow detailed investigation of the temperature structure.  
Since the observed millimeter continuum is likely a mix of emission from the warm and cool components, 
we adopt a broad temperature range (70-190 K) for the estimates in Table~\ref{dustmasstable}.

Constraining the temperatures of G11.92$-$0.61-MM2 and G11.92$-$0.61-MM3 is
more difficult because of the paucity of associated line emission.  MM2 lacks
clear MIR counterparts at 3.6-24 \um, is completely devoid of mm-wavelength
line emission, and has no known maser emission.  In contrast, MM3 emits at 24
\um\/ and is associated with 6.7 GHz Class II \meth\/ masers and possibly with
a C$^{18}$O clump.  MM3 is also associated with the brightest 8 \um\/ emission
in the region (Fig.~\ref{contfig}a,b).  Taken together, the evidence strongly suggests that
MM3 is warmer than MM2.  For MM2, we adopt a temperature range
T$_{dust}$=20-40 K based on the absence of associated molecular line emission.
The 6.7 GHz \meth\/ masers associated with MM3 are quite weak (peak
T$_{b}$\q16500 K, 1\farcs94 $\times$ 0\farcs96 synthesized beam), as are the \meth\/ masers associated with MM1 \citep[peak
T$_{b}$\q7400 K,][]{maserpap}.  Class II \meth\/ masers are radiatively pumped
by infrared photons emitted by warm dust \citep[e.g.][]{Cragg05}.
\citet{Cragg92} found that a blackbody with T$<$50 K was sufficient to excite
moderate 6.7 GHz \meth\/ maser emission (T$_{b}<$6$\times$10$^{4}$ K).  More
detailed investigations of Class II \meth\/ maser excitation have focused
primarily on the parameter space that gives rise to bright (T$_{b}>$10$^{4}$
K) maser emission \citep[e.g.][who invoke dust temperatures $>$100
K]{Cragg05}.  No high-excitation molecular lines (\eup$>$100 K) are observed towards
MM3.  
In sum, the multiwavelength data suggest two possibilities: MM3 may be
of intermediate temperature, or may be hotter (and more evolved) and
simply have very little molecular material left around it.  Additional
data are required to constrain the nature and evolutionary state of
MM3 (\S\ref{evolution}); we adopt a range of T$_{dust}$=30-80 K for
the estimates in Table~\ref{dustmasstable}.

The physical parameters listed in Table~\ref{dustmasstable} can be calculated
from two independent datasets for each core (SMA 1.3 mm and CARMA 1.4 mm for
G11.92$-$0.61, SMA 1.3 mm and CARMA 3.4 mm for G19.01$-$0.03).  For each
compact millimeter continuum source in Table~\ref{dustmasstable}, the mass estimate derived from the lower
resolution dataset is greater than that derived from the higher resolution dataset.
Conversely, a larger beam-averaged column density and volume density are
calculated from the higher resolution data.  These trends are consistent
with the lower-resolution data being more sensitive to emission on larger
spatial scales.  We note that the mass estimates derived from the dust continuum emission
include only circum(proto)stellar material, and \emph{not} the mass of any
protostar or ZAMS star that has already formed within a compact core.

For comparison, Table~\ref{methcynfittable} presents estimates of N(\h),
n(\h), and M$_{gas}$ derived from the best-fit source size and \methcyn\/
column density for the hot cores G11.92$-$0.61-MM1 and G19.01$-$0.03-MM1.
Estimates are presented for \methcyn/\h\/ abundances of 10$^{-7}$, 10$^{-8}$, and
10$^{-9}$.  Values for the abundance of \methcyn/\h\/ in hot cores reported in
the literature span at least an order of magnitude, from \q10$^{-8}$-10$^{-7}$
\citep{tony04, Zhang07, bisschop07}.  
Lower abundances (\q 10$^{-9}$) may also be possible
even at relatively high temperatures ($>$100 K) in massive hot cores, depending on the
warm-up timescale driving the gas-grain chemistry \citep{Garrod08}.
Given the uncertainty in the \methcyn\/ abundance, the gas mass estimates
derived from the \methcyn\/ emission (Table~\ref{methcynfittable}) and from
the millimeter dust continuum emission (Table~\ref{dustmasstable}) are broadly consistent.

\subsubsection{Nature of the Continuum Sources}

In summary, the millimeter continuum sources
G11.92$-$0.61-MM1, G11.92$-$0.61-MM2, and G19.01$-$0.03-MM1 are
dominated by thermal dust emission.  The circum(proto)stellar gas
masses of these cores range from \q8-62 \msun\/ (based on the SMA
data, resolution \q3$\times$2\pp).  G11.92$-$0.61-MM1 and
G19.01$-$0.03-MM1 are hot cores, with derived gas temperatures of
166$\pm$ 20 K and 144$\pm$ 15 K, respectively.  SED modeling indicates
that a central (proto)star of \q10 \msun\/ is present within the
G19.01$-$0.03-MM1 core.  The properties of individual members of the
G11.92$-$0.61 (proto)cluster cannot be constrained by this method, as
the sources MM1, MM2, and MM3 are unresolved in available data at
wavelengths $<$ 1.3 mm.  However, the bolometric luminosities of
G19.01$-$0.03 and of the G11.92$-$0.61 (proto)cluster as a whole are
comparable (\q10$^{4}$ \lsun).  The nature of G11.92$-$0.61-MM3 is
less clear.  In principle, an HCHII region undetected in previous
observations could account for the majority of the G11.92$-$0.61-MM3
mm flux density (\S\ref{mass_dis}), but additional observations at cm
wavelengths are needed to investigate this possibility.  If the mm flux
density of G11.92$-$0.61-MM3 is dominated by dust emission, the compact gas
mass is \q3-9 \msun, the lowest of the observed
cores.  The relative evolutionary states of the members of the
G11.92$-$0.61 (proto)cluster, and of G11.92$-$0.61 and G19.01$-$0.03,
are discussed further in \S\ref{evolution}.

Based on the SMA 1.3 mm data, the total mass in \emph{compact} cores
is \q37-94 \msun\/ in G11.92$-$0.61 and \q12-16 \msun\/ in
G19.01$-$0.03.  Additional low-mass sources may also be present, but
undetected in our observations; the 5$\sigma$ sensitivity limit of the
SMA data corresponds to a mass limit of a few \msun\/ for moderate
dust temperatures (Table~\ref{dustmasstable}).  \citet{atlasgal}
calculate a mass for the larger-scale (40$\times$34\pp) G19.01$-$0.03
gas/dust clump of 1070 \msun, based on ATLASGAL 870 \um\/ data and an
\ammonia\/ T$_{kin}$ of 19.5 K.  This suggests that only \q1\% of the
total mass is in the compact core we observe with the SMA, and a
considerable reservoir of material is in an extended envelope that is mostly
resolved out in the continuum as in the \co\/ line emission
(\S\ref{g19_extended}).  
The compact cores in the G11.92$-$0.61
protocluster constitute a
larger
percentage of the total mass reservoir.  From the 850 \um\/ SCUBA flux
\citep[12 Jy,][]{Walsh03}, we estimate a total mass for the clump of
\q 780 \msun\/ for T$_{dust}$=20 K \citep[typical of the \ammonia\/
temperatures reported for ATLASGAL sources by][]{atlasgal} and
$\kappa_{850 \mu m}$=2.2 \citep[interpolated from the values tabulated
by][]{OH94}.  Based on this estimate, the compact SMA cores in
G11.92$-$0.61 comprise \q5-12\% of the total mass, with a remaining
large-scale gas reservoir of several hundred \msun\/ for the
G11.92$-$0.61 (proto)cluster.

Single dish surveys of massive star forming regions have revealed
spectroscopic signatures of parsec-scale infall in cluster forming
environments \citep[e.g.][]{WuEvans03}.  In addition, new high
resolution observations of the G20.08$-$0.14 N cluster detect infall
at the scale of both cluster forming clumps and massive star forming
cores, all part of a continuous, hierarchical accretion flow
\citep{Roberto09}.  Recent simulations also indicate the importance
of accretion from large-scale gas reservoirs in massive star and
cluster formation, particularly for determining the final stellar
masses \citep{Smith09,Peters10,Wang10}.  Since the presence of an
active outflow indicates ongoing accretion, the masses of the members
of the G11.92$-$0.61 (proto)cluster may grow significantly with time.
For G19.01$-$0.03, the SED modeling is consistent with a central YSO
of mass \q 10 \msun\/ that is actively accreting at a rate of \q
10$^{-3}$ \msun\/ year$^{-1}$.  This central source is associated with
a compact gas and dust core of mass \q 12-16 \msun.  However, with a
substantial (\q 1000 \msun) extended reservoir of material from which
to draw, the final mass of G19.01$-$0.03 may be substantially higher.


\subsection{Outflows}\label{outflow_dis}

A single dominant bipolar molecular outflow is associated with each of our
targeted EGOs.  These outflows are traced by high-velocity, well-collimated \co(2-1) and
\hco(1-0) emission.  
In both EGOs, the red and blue
outflow lobes clearly trace back to a driving source identified with a compact
mm continuum core (Figs.~\ref{g11redbluefig}, \ref{g19redbluefig}).  
This relative clarity is somewhat unusual.  In
many massive star-forming regions, multiple outflows are observed,
with complex kinematics that can make it
difficult to identify driving source(s) \citep{Zhang07,Shepherd07,Brogan09}.  Indeed, since YSOs of all masses
drive bipolar molecular outflows during the formation process
\citep[e.g.][]{Richer00}, one would expect multiple outflows in a protocluster
such as G11.92$-$0.61.

A second outflow may indeed be present in G11.92$-$0.61.  Blueshifted
\co(2-1), \hco(1-0),and SiO(2-1) emission are present NE of the mm
continuum cores, and redshifted emission to the SW
(\S\ref{g11_extended}).  This is opposite the velocity gradient of the
dominant outflow,
and this emission may trace a second outflow.  If so, the driving
source is likely the continuum source MM3, which is approximately
equidistant between the two lobes (Fig.~\ref{g11redbluefig}c; the
possible second outflow is most prominent at moderate velocities,
see also \S\ref{g11_extended}).  Alternatively, the observed
morphology may be attributable to orientation effects.  An outflow
nearly in the plane of the sky may appear to have overlapping red and
blue-shifted lobes \citep[e.g.][]{CB90}.  Another possible explanation
is outflow precession.  For an outflow axis close to the plane of the
sky, precession can produce the appearance of inversions between
blue/red-shifted emission along the outflow axis
\citep[e.g.][]{Beuther08}, such as the pattern seen in G11.92$-$0.61.
In addition, the \co\/ and \hco\/ data hint at the possible presence
of a third, low-velocity outflow along a SE-NW axis.  As shown in
Figures~\ref{g11_12cochannels}-\ref{g11_hcochannels}, moderately
redshifted gas is present SE of the continuum sources, and moderately
blueshifted gas to the NW (26.7, 39.9, and 46.5 \kms\/ panels).  The
interpretation of this emission as an outflow is, however, very
uncertain.  The moderate-velocity \co\/ emission appears to correlate
with extended 4.5 \um\/ emission and 44 GHz \meth\/ masers, but the
\hco\/ emission (which is subject to less spatial filtering) is much
more extended, suggesting confusion with the ambient cloud, and the
SiO(2-1) emission (Fig.~\ref{g11_siochannels}) does not show the same
velocity pattern.  There is no clear evidence in our data for an
outflow driven by the continuum source MM2.





\subsubsection{Outflow Properties}

We estimate the physical properties of the molecular outflows in
G11.92$-$0.61 and G19.01$-$0.03 independently from the SMA \co(2-1)
and the CARMA \hco(1-0) data.  As discussed in \S\ref{g11_extended}
and \S\ref{g19_extended}, outflow gas is confused with diffuse
emission from the surrounding cloud at velocities near the source
\vlsr.  This problem is particularly acute for \co, because of its
high abundance.  To avoid including contributions from the ambient
cloud, we consider only high velocity gas in our estimates of the
outflow physical properties (Table~\ref{outflowtable}).  To further
isolate the outflow gas, a polygonal mask is defined for each red or
blueshifted outflow lobe in Figures~\ref{g11redbluefig} and
~\ref{g19redbluefig}.  The polygonal masks are drawn to encompass all
outflow emission in the integrated intensity images of the
high-velocity gas, and checked against the datacubes.  The appropriate
mask is applied to each channel in which the outflow dominates over
emission from the ambient cloud.  Assuming optically
thin emission, the gas mass of the
outflow is then calculated from
\begin{equation}
M_{out}={\frac{1.186 \times 10^{-4} Q(T_{ex}) e^{\frac{E_{upper}}{T_{ex}}} D^{2} \int
  S_{\nu} dv}{\nu^3 \mu^2 S \chi}}
\end{equation}
where M$_{out}$ is the outflow gas mass in \msun, T$_{ex}$ is the excitation
temperature of the transition in K, Q(T$_{ex}$) is the partition function,
\eup\/ is the upper energy level of the transition in K, $\nu$ is the
frequency of the transition in GHz, $\chi$ is the abundance of the observed
molecule relative to \h, D is the distance to the source in
kpc, and $S_{\nu}$ is the line flux in Jy.  Following \citet{qiu09g230}, for
\co\/ we adopt an abundance ($\chi$) relative to \h\/ of 10$^{-4}$, an excitation
temperature of 30 K, and a mean gas atomic weight of 1.36 (included in the
constant in equation (1)).  For \hco, we adopt the same excitation
temperature (T$_{ex}$=30 K), and an abundance of 10$^{-8}$
relative to \h\/ \citep{Vogel84,Rawlings04,KW07}.  We use Q(30 K)=11.19 for
\co\/ and Q(30 K)=14.36 for \hco, interpolating from the values provided in the
Cologne Database for Molecular Spectroscopy \citep[CDMS,][]{Muller01, Muller05}
and $\mu^2$S=0.02423 debye$^{2}$ for \co(2-1) and $\mu^2$S=15.21022 debye$^{2}$ for \hco(1-0)
from the Splatalogue\footnote{http://www.splatalogue.net/} spectral line database.
Following \citet{qiu09g230}, we estimate the outflow momentum and energy using
\begin{equation}
P_{out}=\Sigma M_{out}(\Delta v) \Delta v
\end{equation}
and
\begin{equation}
E_{out}={\frac{1}{2}} \Sigma M_{out}(\Delta v) (\Delta v)^{2}
\end{equation}
where M$_{out}$($\Delta$v) is the outflow mass in a given channel and $\Delta$v
=$|$v$_{center,channel}-$\vlsr$|$.  For these calculations, we adopt \vlsr=35
\kms\/ for G11.92$-$0.61 and \vlsr=60 \kms\/ for G19.01$-$0.03.  
We estimate the dynamical timescale
from \begin{math} t_{dyn}=L_{outflow}/v_{max} \end{math} where the length L$_{outflow}$
and the maximum velocity v$_{max}$ are determined separately for the red and blue lobes of each
outflow (Table~\ref{outflowtable}).  In estimating L$_{outflow}$, we measured
the extent of the red/blueshifted emission from the driving mm continuum
source.  For G11.92$-$0.61, we assumed
that the main outflow was driven by MM1, and the possible
second (``northern'') outflow by MM3.
Using the dynamical timescales, we also
estimate the mass and momentum outflow rates,  \.{M}$_{out} =
  M_{out}/t_{dyn}$  and  \.{P}$_{out} =
  P_{out}/t_{dyn}$. 
For each outflow, the outflow parameters are listed in Table~\ref{outflowtable}, along
with the velocity ranges used.  
For G19.01$-$0.03, the ``NE blue clump'' 
(Table~\ref{outflowtable}) is the easternmost knot of blueshifted \co\/ emission 
(Fig.~\ref{g19redbluefig}a).  This knot is offset from the main \co\/ jet, and
a separate mask was defined for it.  However, the \hco\/ and
SiO morphology indicate that this \co\/ emission is likely part of the
outflow, so we include it in our estimates of the outflow properties.

Several salient points are reflected in Table~\ref{outflowtable}: (1) channels
nearest the systemic velocity disproportionately affect the outflow mass
estimates; (2) the estimates derived from \hco\/ and \co\/ observations differ
by approximately an order of magnitude; and (3) there is considerable
uncertainty in the estimate of the dynamical timescale, and hence of the mass
and momentum outflow rates.  Each of these points is discussed in more detail
below.

As noted above, dominant, unconfused outflow emission is the primary criterion
for choosing the velocity (channel) ranges over which to integrate outflow
mass, momentum, and energy.  In general, column densities are highest near the
systemic velocity of a cloud.  As a result, estimates of outflow mass and
other properties are \emph{extremely} sensitive to how closely the velocities
considered approach the \vlsr, e.g. to the minimum value of
$\Delta$v=$|$v$_{center,channel}-$\vlsr$|$.  For this reason, where practicable
we choose velocity ranges such that min.($\Delta$v$_{blue}$)\q
min.($\Delta$v$_{red}$) and used the same velocity range for \co\/ and \hco\/
mass estimates.  
In G19.01$-$0.03, it is possible to follow the
outflow much closer to the systemic velocity in \hco(1-0) than in \co(2-1),
with minimal confusion from ambient diffuse gas.  As an illustrative example,
in Table~\ref{outflowtable} we present estimates of the G19.01$-$0.03 outflow
properties derived from \hco\/ using velocity ranges beginning \q$\pm$6\kms,
$\pm$8\kms, and $\pm$10\kms\/ from the systemic velocity.  The difference in the estimated total outflow mass (and
consequently in \.{M}$_{out}$) is about a factor of 2.  The estimates of the
outflow momentum, energy, and momentum outflow rate are less severely affected
because the channels in question are near the systemic velocity (low
$\Delta$v).  In G11.92$-$0.61, the situation is complicated by the possible
second outflow, so we restrict the \hco\/ mass estimates to the same velocity
range considered for \co.

For a given outflow lobe, the mass estimate derived from 
\hco(1-0) is roughly an order of magnitude larger than that derived from
\co(2-1).  There are two possible explanations for this discrepancy: spatial filtering and uncertain \hco\/ abundance.  
For the massive outflow in G240.31+0.07 (D=6.4 kpc), \citet{qiu09g230} found that
their SMA compact configuration \co\/ observations recovered $<$10\%
of the single-dish flux at line center, \q80\% in the ``line wing''
($\Delta$v\q 13 \kms), and nearly 100\% at more extreme velocities
($\Delta$v$\gtrsim$ 15 \kms).  
Our CARMA \hco\/ data are more sensitive to larger-scale emission than our SMA \co\/ data, and
the linear resolution of the
\citet{qiu09g230} SMA observations is comparable to that of our CARMA EGO data. 
It is plausible that our CARMA observations are picking up outflow emission
on larger spatial scales, to which the SMA is insensitive.  In this case, the
outflow parameters estimated from the \hco\/ emission would be more
representative of the true outflow properties.  
If, however, the
\hco\/ abundance in our target EGOs is enhanced above our adopted value of
10$^{-8}$, this could lower the mass estimates from the \hco\/ emission into
better agreement with those from \co.
Modeling by \citet{Rawlings04} found best-fit \hco\/ abundances of 10$^{-10}$ in
the envelope, 10$^{-9}$ in the jet/cavity, and 10$^{-7}$ in the boundary
layer, though the models were optimized for the low-mass source L1527.  
Moderate optical depth corrections would also increase the \co\/ mass
estimates (see below), bringing them into better agreement with those from \hco.


There is considerable ambiguity in the determination of the dynamical
timescale, particularly for asymmetric and/or clumpy outflows such as those
observed towards our target EGOs.  We have followed other recent high-resolution interferometric
outflow studies \citep[e.g.][]{Qiu09,qiu09g230} in defining \begin{math}
  t_{dyn}=L_{outflow}/v_{max} \end{math}, and calculate t$_{dyn}$
independently for the red and blue lobes of each outflow.  As
Table~\ref{outflowtable} illustrates, these estimates can differ
significantly.  
Some single-dish studies instead
use \begin{math} t_{dyn}=R/V \end{math}, where R is the distance between the
\emph{peaks} of the red and blue outflow lobes, and V is the mean outflow
velocity, calculated as the outflow momentum divided by the outflow mass \citep[P/M,
e.g.][]{Zhang05}.  
Both approaches assume, however, that an outflow (or an outflow lobe) can be
well-characterized by a single velocity.  The G19.01$-$0.03 outflow is clumpy,
and characterized by discrete knots of high velocity gas.  If  \begin{math}
  t_{dyn}=L_{knot}/v_{max,knot} \end{math} is
calculated individually for each of the three blueshifted knots in Figure~\ref{g19redbluefig}a,
the values are (from north to south) \q3600 years, 1100 years, and 600 years.
This range suggests the limitations of attempting to evaluate the age of
a flow using a single velocity.  Estimating t$_{dyn}$ is further complicated
by the unknown effects of ambient density and inclination angle.  The
expressions for t$_{dyn}$ above assume free expansion.    
Finally, the inclination of the outflows is unknown.  
The extended morphologies of the 4.5 \um\/ emission and the
high-velocity molecular gas in our target EGOs suggest that the
outflows may lie near the plane of the sky
(Figs.~\ref{g11redbluefig},~\ref{g19redbluefig}).  However,
intermediate inclination angles are also plausible, since the red and
blueshifted outflow lobes are spatially distinct, particularly in
G19.01$-$0.03 \citep[e.g.][]{CB90}, and very high velocity gas (\q 60
\kms\/ from \vlsr\/ in G11.92$-$0.61 and $>$100 \kms\/ from \vlsr\/ in
G19.01$-$0.03) is observed.  Table~\ref{outflowtable} presents
estimates of outflow parameters without correction for inclination,
and for $\theta$=10\degree, 30\degree, and 60\degree, where $\theta$
is the inclination angle of the outflow \emph{to the plane of the
sky}.  In the extreme case of $\theta$=10\degree, correcting for
inclination increases the estimated \.{M}$_{out}$ and P$_{out}$ by a
factor of \q6, and \.{P}$_{out}$ and E$_{out}$ by a factor of \q30.
For an intermediate inclination
$\theta$=30\degree, the increases are more moderate: a factor of \q2
for \.{M}$_{out}$ and P$_{out}$, and \q4 for \.{P}$_{out}$ and E$_{out}$.
For outflows in which the
red and blue lobes give very different estimates of t$_{dyn}$,
Table~\ref{outflowtable} presents estimates of \.{M}$_{out}$ and
\.{P}$_{out}$ for the outflow as a whole using an
intermediate timescale.  In calculating \.{M}$_{out}$ and \.{P}$_{out}$
from the \hco\/ data we adopt the dynamical timescales
calculated from \co, since the highest velocity outflow gas extends
beyond the limited velocity coverage of the CARMA observations.

In general, our estimates of outflow mass are lower limits, and likely extreme lower
limits.  As a result, the other physical parameters (which depend on the
outflow mass) will also be underestimated.  There are three main contributing factors: 
(1) extended emission
missed by the interferometers; (2) outflow emission near the systemic velocity
excluded by our conservative velocity cuts; and (3) the assumption of optically
thin emission.  
Our estimates of the outflow mass 
assume optically thin emission in both
\co(2-1) and \hco(1-0).  While this assumption is likely valid for \hco, it is
more problematic for \co, and some recent interferometric outflow
studies have made detailed corrections for \co\/ optical depth
\citep[e.g.][]{qiu09g230}.  
Because we were conservative in selecting the velocity ranges over
which to calculate outflow parameters, 
significant ($\ge$5$\sigma$) $^{13}$CO(2-1) emission is detected in only one
channel that contributes to the estimates presented in
Table~\ref{outflowtable}, for one outflow: the main outflow in G11.92$-$0.61
(v=48.1 \kms).  If we calculate the $^{12}$CO/$^{13}$CO line ratio at the
$^{13}$CO peak in this channel, the implied optical depth correction factor is \q6.7 for a
$^{12}$CO/$^{13}$CO abundance ratio of 40 \citep[for a Galactocentric distance
of 4.7 kpc,][]{WilsonRood94}.  Applying this factor would increase the contribution
of this single channel to the outflow mass from \q0.07 \msun\/ to \q0.5 \msun,
the mass of the red outflow lobe from \q0.2 to \q 0.6 \msun, and the total
mass of the outflow from \q 0.8 \msun to \q 1.2 \msun.  Applying this single
correction factor, however, would likely result in an overestimate.  The correction
factors derived at two other positions in the outflow with detected $^{13}$CO
emission are more modest  \citep[\q3 and 4.5, see also][]{CB90}.    
The signal-to-noise of the $^{13}$CO data are not sufficient to accomodate
attempting an opacity correction as a function of position, particularly given
the overwhelming lack of detected $^{13}$CO emission in the other channels
considered.  By assuming optically thin emission, our mass estimates based on
\co\/ will definitively be lower limits, without the ambiguity of possibly
overcorrecting. 

We do not estimate outflow parameters from the SiO(2-1) emission
observed with CARMA because of the uncertainty in the SiO abundance in
the emitting region.  Values in the literature for the SiO abundance
in molecular outflows and massive star-forming regions vary over at
least two orders of magnitude \citep[from \q10$^{-6} $ to
10$^{-8}$,][]{pdf97,Caselli97,Schilke97}.  Models indicate that the
SiO abundance depends sensitively on the shock conditions
\citep[including velocity, ambient density, and time since the passage
of the shock,][]{pdf97,Schilke97}, which are not constrained by our
single-transition SiO observations.  
Since our CARMA data show that the SiO emission is
extended well beyond the beamsize of the JCMT SiO(5-4) spectra (\S\ref{g11_previous},\S\ref{g19_previous}),
we cannot constrain the physical conditions in the SiO emitting gas.

\subsubsection{Comparison with Other Objects}

Properties of outflows from MYSOs reported in the literature, based on
high angular resolution observations, range over several orders of
magnitude.  As indicated by the discussion above, some of this range
may be attributable to differences in spatial filtering and to the
(large) uncertainties inherent in assuming tracer abundances and
correcting (or not) for optical depth and inclination effects.  At the
low end, \citet{Qiu09} calculate M$_{out}$ of 0.22 \msun, P$_{out}$ of
4.9 \msun\/ \kms, \.{M}$_{out}$ of 10$^{-4}$ \msun\/ yr$^{-1}$, and
\.{P}$_{out}$ of 2.2$\times$10$^{-3}$ \msun\/ \kms\/ yr$^{-1}$ based
on SMA \co\/ data for the extremely high velocity outflow in HH80-81
(D=1.7 kpc). 
At the high end, outflow masses of several tens \citep[27 \msun,
\emph{IRAS} 18566+0408, D=6.7 kpc:][]{zhang07b} to $\gtrsim$ 100 \msun\/
\citep[98 \msun, G240.31+0.07, D=6.4 kpc; 124 \msun,
Orion-KL:][]{qiu09g230,beuthernissen08} have been reported.  These studies,
however, use tracers with uncertain abundance in outflows
\citep[SiO;][]{zhang07b}, or combine single dish and interferometric
data \citep{qiu09g230,beuthernissen08}.  Also, except for Orion-KL, the
estimated dynamical timescales for these more massive outflows are
longer ($>$10$^{4}$ years), so the estimated mass outflow rates,
\.{M}$_{out}$, are still only a few $\times$ 10$^{-3}$ \msun\/ yr$^{-1}$. 
The estimated parameters of the molecular outflows in our target EGOs
(Table~\ref{outflowtable}) are roughly in the middle of the range reported in the literature.
The main outflow in G11.92$-$0.61 and the outflow in G19.01$-$0.03
have broadly similar characteristics: each has t$_{dyn}$ of a few
$\times 10^3$ years, and (based on the \hco\/ data) M$_{out}$ of a few \msun, P$_{out}$ of a few
hundred \msun\/ \kms, E$_{out}$ of tens to a hundred $\times 10^{45}$
ergs, \.{M}$_{out}$ of a few $\times 10^{-3}$ \msun\/ yr$^{-1}$, and
\.{P}$_{out}$ of a few hundredths to one \msun\/ \kms\/ yr$^{-1}$
(the estimates of E$_{out}$ and \.{P}$_{out}$
are most severely affected by the uncertainty in the inclination
angle). 
These parameters are generally comparable to those for the total
high-velocity gas (attributed to three separate outflows) in the
massive star-forming region \emph{IRAS} 17233-3606 derived from
high-resolution SMA \co\/ observations by \citet{leurini09} (opacity
correction applied), though for \emph{IRAS} 17233-3606 the estimated
dynamical timescale is somewhat shorter (\q300-1600 years) 
than for our target EGOs.  The EGO outflow properties are also quite
similar to those of the outflows in the AFGL5142 protocluster, as
estimated from OVRO \hco(1-0) and SMA \co(2-1) observations
\citep[particularly accounting for the different assumed \hco\/
abundance, 10$^{-9}$;][]{Hunter99,Zhang07}.  As noted in
\S\ref{g11compact}, the frequency coverage of the \citet{Zhang07} SMA
data is comparable to that of our observations, and the SMA spectrum
of AFGL5142 MM2--the probable driving source of the north-south
outflow studied by \citet{Hunter99}--is very similar to that of
G11.92$-$0.61-MM1.
Even the least massive outflow observed towards our target EGOs
(the ``northern outflow'' in G11.92$-$0.61) has values of M$_{out}$,
\.{M}$_{out}$, etc. at least an order of magnitude greater than those
typical of low-mass outflows observed at high angular
resolution\citep[e.g.][]{Arce2006}.

Several large scale single-dish surveys of molecular outflows have
shown correlations between the properties of the outflow and those of
the driving source \citep[in particular its bolometric luminosity and
core mass, e.g.][]{CB92,ShC96b,Hunter97,Beuther02}, though other recent studies
have found considerable scatter and weak or no evidence of any
correlations \citep[e.g.][]{RidgeMoore01,Zhang05}.  The applicability
of these relations to parameters derived from interferometric
observations is unclear, since, as discussed above, interferometers
resolve out a significant fraction of the extended emission and so
underestimate the outflow mass and other parameters.  The mass and
momentum outflow rates for our target EGOs (Table~\ref{outflowtable})
do generally agree, within the considerable scatter, with the
\.{M}$_{out}$ and \.{P}$_{out}$ expected for a driving source of
luminosity \q10$^{4}$ \lsun\/ \citep[\.{M}$_{out}$ \q a few $\times$
10$^{-5}$ to a few $\times$ 10$^{-3}$ \msun\/ year$^{-1}$,
\.{P}$_{out}$\q a few $\times$ 10$^{-3}$ to 10$^{-1}$ \msun\/ \kms\/
year$^{-1}$:][]{CB92,ShC96b,Beuther02,Zhang05}.  For G11.92$-$0.61 and
G19.01$-$0.03, the mass in the outflow (as derived from the \hco\/
data, Table~\ref{outflowtable}) is roughly comparable to the
circum(proto)stellar gas mass of the driving compact millimeter core
(Table~\ref{dustmasstable}), and, for G19.01$-$0.03, to the mass of
the central (proto)star inferred from the SED modeling (\S\ref{seds}).
It has been suggested for some time that the mass in MYSO outflows is
accumulated from the larger-scale environment
\citep[e.g.][]{Churchwell97}.  Notably, the properties of the outflows in our
target EGOs are consistent with the single-dish relations with respect
to ``core mass'' only if the total masses of the single-dish clump (as
opposed to the masses of the compact cores resolved with the SMA and
CARMA) are considered.
  

\subsection{Diversity within the EGO Sample}


Our high resolution mm observations suggest considerable
diversity within the EGO sample in the clustering properties and evolutionary
states of the outflow driving sources.  Since extended 4.5 \um\/ emission specifically
targets a population with ongoing outflow activity and active, rapid
accretion, the range in source properties is of interest in the context of
recent theoretical work on feedback effects in cluster-scale models of massive
star formation \citep[e.g.][]{KrumholzMatzner09,Bate09,Wang10,Peters10}.

\subsubsection{Multiplicity}\label{clustering}

The EGO G11.92$-$0.61 was chosen as a target for high resolution mm
observations in part because its 24 \um\/ morphology and  two associated 6.7 GHz Class II \meth\/
maser spots indicated the possible presence of multiple MYSOs.  
Our high-resolution SMA and CARMA data indeed reveal three compact millimeter
continuum sources associated with the EGO.  The clustering scale of these
cores is \q 0.1 pc, comparable to that in S255N, a massive protocluster that
is also associated with a 4.5 \um\/ outflow \citep{s255n}.  No additional
structure is seen in the CARMA 1.4 mm image (resolution 1\farcs44 $\times$
0\farcs87 \q 0.03$\times$ 0.02 pc \q 5500 $\times$ 3300 AU)  as compared to
the lower resolution SMA 1.3 mm image (resolution 3\farcs2 $\times$1\farcs8 \q
0.06 $\times$ 0.03 pc \q 12000 $\times$ 6800 AU).  
Further structure may, however, be suggested by the 
complicated kinematics of the \co, \hco,  and particularly the SiO emission.
In NGC6334I(N), a tight (\q 800 AU separation) binary is thought to cause the
precession of the molecular outflow: both components of the binary are
contained within a hot core of diameter \q 1400 AU \citep{Brogan09}.  
If outflow precession is responsible for the complex velocity structure in
G11.92$-$0.61, it would suggest that the outflow driving
source, G11.92$-$0.61-MM1, could be an unresolved (proto)binary. 

In contrast to G11.92$-$0.61, our interferometric observations of
G19.01$-$0.03 reveal only a single continuum source.  For
G19.01$-$0.03, our SMA 1.3 mm continuum image provides the highest
resolution: 3\farcs2$\times$1\farcs7 \q0.07$\times$0.03 pc \q 13400
$\times$7100 AU.  This is sufficient to resolve clustering at the
scale seen in G11.92$-$0.61 and S255N, though not proto-Trapezia \citep[such as NGC6334I and I(N),][]{Hunter06}.  The
\methcyn(12-11) spectrum of G19.01$-$0.03-MM1 is unusual, in that the
k=0, k=1, and k=2 components are nearly equal in strength, and the k=3
component is even stronger (Fig.~\ref{g19single}).  This is likely
indicative of either unresolved multiplicity or unresolved
density/temperature gradients, but the current data are insufficient to
distinguish between these scenarios.  

If G19.01$-$0.03 is truly a
single (proto)star, it would be an unusual example of a massive star
forming in isolation.  Some recent models suggest that an apparently
lone MYSO could also be indicative of the very earliest stages of
cluster formation.  \citet{Peters10} find that the most massive star
begins to grow early in the cluster formation process, and finishes
accreting while its surrounding cluster is still forming.  Accretion
is clearly ongoing in G19.01$-$0.03, and our mm observations provide
several other indicators of its youth (\S\ref{evolution}).
Higher resolution and more sensitive (sub)mm observations
of G19.01$-$0.03 are needed to determine whether sub-10000 AU
clustering or low-mass (proto)stars are present.


\subsubsection{Relative Evolutionary State}\label{evolution}

The three millimeter continuum sources in the G11.92$-$0.61 (proto)cluster
exhibit a range of MIR-mm properties and maser associations, suggestive of a
range in mass and/or evolutionary state.
The strongest mm continuum source,    
G11.92$-$0.61-MM1, is a hot core, and the driving source of the dominant
bipolar outflow in the region.
In contrast, G11.92$-$0.61-MM2 is devoid of
millimeter-wavelength line emission, and shows no evidence for associated
outflow or maser activity.  The gas mass of G11.92$-$0.61-MM2, calculated from
its arcsecond-scale thermal dust emission, is \q17-42 \msun, sufficient to form an
intermediate to high-mass star.  The compact gas mass associated with
G11.92$-$0.61-MM1 is smaller, \q5-13 \msun.  This does not, however, include the
mass of central (proto)star(s), which the associated MIR and maser emission
indicate are likely present.  MM2, which exhibits no signs of maser or
outflow activity, may be a massive protostellar core.  High-resolution
observations in a thermometer of cold, dense gas (e.g. \ammonia) would help to
determine whether this source is internally heated by a central YSO or
externally heated by feedback processes in the cluster environment.

The evolutionary state of G11.92$-$0.61-MM3, the weakest millimeter
continuum source in the region, is somewhat ambiguous.  It is
associated with 6.7 GHz Class II \meth\/ maser and 24 \um\/ emission,
and with the strongest 8.0 \um\/ emission in the region
(Fig.~\ref{contfig}a,b).  
The nature of the 8 \um\/ counterpart is unclear, as it appears somewhat
extended and it is possible that \h\/ line emission contributes significantly
to the broadband flux.
Very
little compact mm molecular line emission is detected coincident with
MM3, and no emission in high excitation
(\eup$>$100 K) molecular lines.  Like MM1, the compact gas mass
associated with MM3 is relatively small (\q2-6 \msun), but the MIR
and maser emission suggest that a central (proto)star has likely
already formed.  \citet{debuizer06} suggested that Class II \meth\/
maser emission may be excited in the MIR-bright walls of an outflow
cavity.  The morphology of the 24 \um\/ emission, however, appears
inconsistent with MM3 being a hotspot on an outflow cavity wall.
Another possibility is that MM3 is more evolved, and so has dispersed
most of the molecular gas in its immediate vicinity.  This would be
consistent with its MIR properties.  We also note that MM3 is the only
compact millimeter continuum source for which the 
arcsecond-scale 1.4 mm emission observed with CARMA could conceivably
be due to a HC HII region that fell below the \citet{maserpap} 44 GHz
continuum detection limit (\S\ref{mass_dis}).  
High
resolution observations over a range of cm wavelengths are required to
constrain the evolutionary state of MM3.

Several  lines of evidence suggest that G19.01$-$0.03 is younger than the
G11.92$-$0.61 massive star-forming region.  Compared to G11.92$-$0.61-MM1, the
millimeter molecular line emission from G19.01$-$0.03-MM1 is sparse and weak.
The relative lack of chemical complexity observed towards G19.01$-$0.03-MM1,
and its cooler derived temperature, are both consistent with a less-advanced
hot core chemisty \citep[e.g.][]{vd98}.  There have been a number of
suggestions in the literature that the relative abundances of various
sulfur-bearing species may provide chemical clocks
for hot cores \citep[e.g.][]{Charnley97,Hatchell98,Herpin09}.  Transitions of
several species invoked in these models fall within the frequency coverage of
our SMA observations, namely SO, OCS, and SO$_{2}$.
Unfortunately, the most diagnostic abundance ratios are those relative to
 H$_{2}$S \citep[e.g.][]{Charnley97}, which does not have transitions within
 our SMA bandwidth.  
 
The calculation and comparison of abundances for the sulfur-bearing
species in the EGO hot cores must be regarded with considerable caution.
Such comparisons rely on an assumption that all relevant emission arises from
the same excitation conditions in the same physical volume of gas.  
From our data, it is evident that in G11.92$-$0.61 the outflow contributes to
the excitation of SO emission, and even the compact OCS emission may have an
outflow contribution (\S\ref{g11compact}).  Considering the two-component
temperature model required to fit the \methcyn\/ emission, it is also
plausible that the low-excitation SO emission arises from a larger volume than
the higher-excitation OCS emission (both unresolved by the SMA beam).
Similarly, the thermal dust emission (from which we infer N(H$_{2}$)) may arise
from a different volume than the molecular line emission.  If we nonetheless
compute the OCS and SO abundances for G11.92$-$0.61-MM1 assuming T$_{ex}$=166 K and
N(H$_{2}$)=2.7$\times$10$^{23}$ cm$^{-2}$ (beam-averaged column density for T$_{dust}$=166 K and the SMA beam), we find abundances of \q 1.9
$\times$ 10$^{-8}$ and 1.3
$\times$ 10$^{-8}$ relative to \h, respectively.  
For G19.01$-$0.03-MM1, (T$_{ex}$=115 K, 
N(H$_{2}$)=3.5$\times$10$^{23}$ cm$^{-2}$, Table~\ref{dustmasstable}), we calculate abundances 
of \q 3$\times$10$^{-9}$ for OCS and \q 6$\times$10$^{-10}$ for SO.
These estimates correspond to
a [SO/OCS] ratio of \q 0.2 for G19.01$-$0.03 and \q 0.7 for
G11.92$-$0.61-MM1, consistent with G19.01$-$0.03 being younger and less evolved \citep{Herpin09}.
Other, more global indicators also point to the relative youth of
G19.01$-$0.03 compared to G11.92$-$0.61.  Since MYSOs dissipate their
natal envelopes as they grow, the presence of an extended envelope is
suggestive of youth.  Compact mm core(s) account for a smaller
fraction of the total clump mass in G19.01$-$0.03 than in
G11.92$-$0.61 (\S\ref{mass_dis}), consistent with indications from the
\co\/ emission and SED modeling that G19.01$-$0.03-MM1 is still
embedded in a massive (\q 1000 \msun) large-scale envelope.



\subsubsection{Future Work}\label{future}

These initial results demonstrate the potential of the EGO sample for probing
the importance of protostellar feedback in the formation of massive stars and
star clusters.  
Recent theoretical work has just begun to include realistic feedback effects
in cluster-scale models, including protostellar outflows \citep{Wang10},
photoionization/HII regions \citep{Peters10,KrumholzMatzner09}, and radiative
feedback \citep{Bate09}.  However, current models do not include
all feedback mechanisms, and hence do not address the question of which
mechanism(s) are most important at which stages of (proto)cluster formation,
or of how these mechanisms interact.  EGOs are an outflow-selected sample.
Hence, characterizing their outflows as well as other possible forms of
feedback (ionized jets/winds, gas heating) will help to establish whether
there is an outflow-dominated stage of (proto)cluster formation.  Addressing
this question in a statistically meaningful way will require high resolution
(sub)mm wavelength data similar to those presented here for a wide range of
EGO sources.  High-resolution cm continuum data are also necessary to
constrain the presence and physical properties of ionized gas, while
cm-wavelength line observations (e.g. of \ammonia) are needed to constrain gas
temperatures and densities in cool cores that lack mm-wavelength line
emission.  Accumulating uniform datasets for large samples is
essential for comparisons of different objects.  The capabilities of
the EVLA and ALMA put such surveys within reach, and EGOs will be
promising targets for these instruments.  One fortunate characteristic
of EGOs is the probable 229.759 GHz \meth\/ Class I maser line which we
detect toward both objects.  These are, to our knowledge, the 
fifth and sixth reported examples of this maser in the literature, although
higher resolution observations are still required to unambiquously
confirm the maser nature of the emission based on the brightness
temperature.  As with the 44 GHz Class I masers, these features may be
common in massive star-forming regions.  If so, they will be of
interest as self-calibration targets to help enable future very
high-resolution 1.3 mm observations (e.g. with ALMA).

\section{Conclusions}\label{conclusions}

Our high-resolution millimeter observations of two EGOs unambiguously
show that they are young MYSOs driving massive bipolar outflows.  The
spatial coincidence of high velocity \co(2-1) and \hco(1-0) emission
with the extended 4.5 \um\/ lobes supports the outflow hypothesis for the
4.5 \um\/ emission.  A single dominant outflow is identified in each EGO,
with tentative evidence for multiple outflows in one source
(G11.92$-$0.61).  The morphology and kinematics of the SiO(2-1) emission
differ from the other outflow tracers in that some of the strongest
red and blueshifted features are offset from the extended 4.5 \um\/
emission, and may trace the impact of outflow shocks on dense gas in
the surrounding cloud.  The morphology of the high-velocity gas with
respect to 44 GHz Class I \meth\/ maser emission further solidifies the
association of this type of maser with outflows.  Anomalously intense
and narrow components of 229.759 GHz \meth\/ emission are also detected
in the outflow lobes from both objects, suggesting additional Class I
maser activity.  The outflow driving sources appear as compact cores
of millimeter continuum emission and dense gas, including the hot core
molecules \meth, \methcyn\/ and OCS.  Coincident with 22 GHz water maser
emisson, G11.92$-$0.61-MM1 shows considerably richer and stronger hot
core line emission than G19.01$-$0.03-MM1, consistent with its warmer
temperature derived from the multi-transition analysis of the \methcyn\/
and \meth\/ emission (166$\pm$20 v. 114$\pm$15 K).  Both hot cores exhibit
24 \um\/ and 70 \um\/ emission in MIPSGAL images and contain 6.7 GHz Class
II \meth\/ masers, all consistent with their identification as MYSOs.

Our observations also reveal considerable diversity within the EGO
sample.  Although observed at the same spatial resolution, G19.01$-$0.03
appears as a single MYSO while G11.92$-$0.61 resolves into a cluster of
three compact dust cores.  In addition to the difference in
multiplicity, several other factors point to G19.01$-$0.03 being in a
earlier evolutionary stage: SED modeling, the relative weakness of its
hot core emission, and the dominance of the extended envelope of
molecular gas.  In contrast, G11.92$-$0.61 appears to have already
formed a protocluster whose members span a range of ages -- one is a
hot core and two are almost entirely devoid of line emission.  These
initial results demonstrate the potential of the EGO sample for
probing the importance of protostellar feedback in the formation of
massive stars and star clusters.  The future capabilities of the EVLA and
ALMA will enable uniform surveys of a statistically meaningful number
of regions which will enable the relative importance of outflows,
photoionization, and radiative feedback to be assessed.

\acknowledgments

This research has made use of NASA's Astrophysics Data System
Bibliographic Services and the SIMBAD database operated at CDS,
Strasbourg, France.  This research made use of the myXCLASS program
(http://www.astro.uni-koeln.de/projects/schilke/XCLASS), which
accesses the CDMS (http://www.cdms.de) and JPL
(http://spec.jpl.nasa.gov) molecular data bases.  C.J.C. would like to
thank B. Whitney for helpful discussions about YSO models and SEDs,
and R. Indebetouw for helpful discussions about MIR photometry.
Support for this work was provided by NSF grant AST-0808119.
C.J.C. was partially supported during this work by a National Science
Foundation Graduate Research Fellowship, and is currently supported by
an NSF Astronomy and Astrophysics Postdoctoral Fellowship under award
AST-1003134.

{\it Facilities:}  \facility{SMA ()}, \facility{CARMA ()}, \facility{Spitzer ()}


\begin{deluxetable}{ccccccccc}
\tablewidth{0pt}
\tablecaption{Observational Parameters\label{obstable}}
\tablehead{ 
\colhead{Obs. Date} & 
\colhead{Syn. Beam} & 
\colhead{Prim. Beam} & 
\colhead{Cont.} &
\multicolumn{2}{c}{Line Observations} & 
\multicolumn{3}{c}{Calibrators} \\
\colhead{} &
\colhead{\pp $\times$ \pp} &
\colhead{FWHP} &
\colhead{rms} & 
\colhead{$\Delta$v$_{chan.}$} & 
\colhead{rms\tablenotemark{a}} & 
\colhead{Gain} & 
\colhead{Bandpass} & 
\colhead{Flux}\\
\colhead{} & 
\colhead{} & 
\colhead{\pp} &
\colhead{\mjb} & 
\colhead{\kms} & 
\colhead{\mjb} & 
\colhead{} & 
\colhead{} &
\colhead{}  
}
\tablecolumns{9}
\tabletypesize{\scriptsize}
\setlength{\tabcolsep}{0.05in}
\startdata
\cutinhead{G11.92$-$0.61}
\sidehead{1.3 mm SMA}
2008 June 23 & 3.2$\times$1.8 & 55 & 3.5
& 1.1 & 100 & J1733-130,J1911-201 & 3C454.3 & Uranus \\
\sidehead{1.4 mm CARMA}
2008 April 25 & 1.44$\times$0.87 & 31 & 4.3 & \nodata & \nodata
& J1733-130,J1911-201 & 3C454.3 & J1733-130\tablenotemark{b} \\ 
\sidehead{3.4 mm CARMA}
2008 July 29 & 6.8$\times$5.1 & 80 & \nodata & 1.7 & 10 &
J1733-130,J1911-201 & 3C454.3 & Uranus\\ 
\cutinhead{G19.01$-$0.03}
\sidehead{1.3 mm SMA}
2008 June 23 & 3.2$\times$1.7 & 55 & 3.5
& 1.1 & 100 & J1733-130,J1911-201 & 3C454.3 & Uranus \\
\sidehead{3.4 mm CARMA}
2008 July 30 & 5.7$\times$5.1 & 80 & 0.5 & 1.7
 & 10 &
J1733-130,J1911-201 & 3C454.3 & Uranus\\ 
\enddata
\tablenotetext{a}{Hanning smoothed}
\tablenotetext{b}{Assuming S(1.4 mm)= 2.69 Jy}
\end{deluxetable}

\begin{deluxetable}{lccccc}
\tablewidth{0pt}
\tablecaption{Observed Properties of Millimeter Continuum Sources \label{conttable}}
\tablehead{ 
\colhead{} & 
\multicolumn{2}{c}{J2000.0 Coordinates\tablenotemark{a}} &
\colhead{Peak Intensity\tablenotemark{b}} & 
\colhead{Integ. Flux Density\tablenotemark{b}} & 
\colhead{Size\tablenotemark{c}} \\
\colhead{Source} & 
\colhead{$\alpha$ ($^{\rm h}~~^{\rm m}~~^{\rm s}$)} &
\colhead{$\delta$ ($^{\circ}~~{\arcmin}$~~\pp)} &
\colhead{\mjb} &
\colhead{mJy} & 
\colhead{\pp $\times$ \pp\/ [P.A. $^{\circ}$]} 
}
\tablecolumns{6}
\tabletypesize{\scriptsize}
\startdata
\cutinhead{G11.92$-$0.61}
\sidehead{1.3 mm SMA\tablenotemark{d}}
MM1 & 18 13 58.103 & $-$18 54 20.19 & 219 (6) & 338 (14) & 1.7$\times$1.7 [126] \\
MM2 & 18 13 57.860 & $-$18 54 14.04 & 136 (6) & 203 (14) & 1.9$\times$0.8 [161] \\ 
MM3 & 18 13 58.21    & $-$18 54 16.2   & 49 (6) & \nodata & \nodata \\
Total &                    & & &590 (34) &\\ 
\sidehead{1.4 mm CARMA\tablenotemark{e}}
MM1 & 18 13 58.116 & $-$18 54 20.14 & 96 (5) & 169 (13) & 1.26$\times$0.76 [26]\\
MM2 & 18 13 57.860 & $-$18 54 14.17 & 78 (6) & 121 (13) & 1.20$\times$0.55 [36]\\
MM3 & 18 13 58.159  & $-$18 54 16.0   & 25 (6) & 36 (13)   & 1.59$\times<$0.57 [40]\\ 
Total &                    & & &326 (39) &\\
\cutinhead{G19.01$-$0.03}
\sidehead{1.3 mm SMA\tablenotemark{f}}
MM1 & 18 25 44.790 & $-$12 22 45.86 & 207 (3) & 275 (7) & 1.9$\times$0.9 [78]\\
\sidehead{3.4 mm CARMA\tablenotemark{g}}
MM1 & 18 25 44.80 & $-$12 22 46.8 & 9.4 (0.9) & 27 (3) & 8.5$\times$5.9 [134]\\
\enddata
\tablenotetext{a}{Centroid position determined by fitting a single
  two-dimensional Gaussian
  component to each source.  The statistical uncertainty is indicated by the
  number of significant figures; absolute positional uncertainties are at
  least an order of magnitude larger (see \S\ref{obs}).}
\tablenotetext{b}{Estimated statistical uncertainties are given in
  parentheses.  Unresolved sources are indicated by \nodata\/ in the
  integrated flux density column.}
\tablenotetext{c}{Deconvolved source size determined by fitting a single
  Gaussian component to each source.  Unresolved sources are indicated by \nodata}
\tablenotetext{d}{Synthesized beam 3\farcs2$\times$1\farcs8
  (P.A.$=$59$^{\circ}$)}
\tablenotetext{e}{Synthesized beam 1\farcs44$\times$0\farcs87 (P.A.$=$25$^{\circ}$)}
\tablenotetext{f}{Synthesized beam 3\farcs2$\times$1\farcs7 (P.A.$=$63$^{\circ}$)}
\tablenotetext{g}{Synthesized beam 5\farcs7$\times$5\farcs1 (P.A.$=$-27$^{\circ}$)}
\end{deluxetable}


\begin{deluxetable}{llcccccc}
\tablewidth{0pc}
\tablecolumns{8}
\tabletypesize{\scriptsize}
\tablecaption{1.3 mm Spectral Lines: G11.92$-$0.61 \label{g11trans}}
\tablehead{
\colhead{} & 
\colhead{} & 
\colhead{} & 
\colhead{} & 
\multicolumn{4}{c}{Fitted Line Parameters} \\
\colhead{Species} & 
\colhead{Transition} & 
\colhead{Frequency}  & 
\colhead{E$_{\rm upper}$/k} &
\colhead{Intensity\tablenotemark{a}}  & 
\colhead{V$_{center}$\tablenotemark{a}} &
\colhead{Width\tablenotemark{a}} & 
\colhead{$\int$ S dv\tablenotemark{a}}\\
\colhead{}  & 
\colhead{} &   
\colhead{(GHz)}    &  
\colhead{(K)} & 
\colhead{\jb}    &  
\colhead{\kms} &     
\colhead{\kms} &
\colhead{\jb \kms}
}    
\startdata
\cutinhead{MM1}
HC$_{3}$N(V$_{7}$=1f) & 24-23 & 219.174 & 452 & 0.53 (0.05) & 33.6 (0.6) &
13.1 (1.3) & 7.4 (1.0)\\ 
C$_{2}$H$_{5}$CN & 24$_{2,22}$-23$_{2,21}$ & 219.506 & 136 & 0.42 (0.05) & 36.6 (0.6) & 10.9
(1.4) & 4.8 (0.8)\\ 
C$^{18}$O & 2-1 & 219.560 & 16 & 1.6 (0.1) & 34.9 (0.1) & 4.3 (0.3) & 7.2 (0.7) \\ 
HNCO & 10$_{2,9}$-9$_{2,8}$\tablenotemark{b} & 219.734 & 228 & \nodata &
\nodata & \nodata & \nodata \\
HNCO & 10$_{2,8}$-9$_{2,7}$\tablenotemark{c} & 219.737 & 228 & \nodata &
\nodata & \nodata & \nodata \\
HNCO & 10$_{0,10}$-9$_{0,9}$\tablenotemark{d} & 219.798 & 58 & 1.10 (0.06) & 35.2 (0.3) & 9.9 (0.6) & 11.5 (0.9)\\
H$_{2}^{13}$CO & 3$_{1,2}$-2$_{1,1}$\tablenotemark{d} & 219.909 & 33 & 0.59 (0.03) & 35.1 (0.2)  & 9.1 (0.5)  &
5.7 (0.5) \\
SO & 6$_{5}$-5$_{4}$\tablenotemark{d} & 219.949 & 35 & 2.62 (0.05) & 35.3 (0.1) & 10.5 (0.2) & 29.2 (0.9)\\
\methanol & 8$_{0,8}$-7$_{1,6}$ & 220.078 & 97 & 1.25 (0.06) & 35.3 (0.2) & 8.0 (0.4) & 10.6 (0.7)\\
$^{13}$CO & 2-1 &  220.399 & 16 & \nodata & \nodata & \nodata & \nodata \\
CH$_{3}$CN & 12$_{8}$-11$_{8}$\tablenotemark{e} & 220.476 & 526 & \nodata & \nodata & \nodata & \nodata\\
CH$_{3}$CN & 12$_{7}$-11$_{7}$ & 220.539 & 419 & 0.55 (0.05) & 35.5 (0.5) &9.7 (1.1) & 5.7 (0.9) \\
HNCO & 10$_{1,9}$-9$_{1,8}$\tablenotemark{f} & 220.585 & 102 & 0.50 (0.04) & 33.4 (0.5)  & 10.9
(1.2) &	5.8 (0.8)\\
CH$_{3}$CN & 12$_{6}$-11$_{6}$\tablenotemark{f} & 220.594 & 326 & 0.84 (0.03) & 33.8 (0.3) & 15.2 (0.8) & 13.6
(0.8) \\ 
CH$_{3}$CN & 12$_{5}$-11$_{5}$ & 220.641 & 247 & 0.95 (0.05) & 36.3 (0.4) & 14.1 (0.9) &14.2 (1.3)\\
C$_{2}$H$_{5}$CN & 25$_{2,24}$-24$_{2,23}$ & 220.661 & 143 & 0.67 (0.03) & 35.8 (0.2) & 9.6 (0.6) &  6.8
(0.5)\\
CH$_{3}$CN & 12$_{4}$-11$_{4}$ & 220.679 & 183 & 1.19 (0.06) & 35.2 (0.2) & 10.4 (0.6) & 13.2 (1.0) \\
CH$_{3}$CN & 12$_{3}$-11$_{3}$ & 220.709 & 133 & 1.95 (0.08) & 35.0 (0.2) & 9.4 (0.4)  & 19.5 (1.2) \\
CH$_{3}$CN & 12$_{2}$-11$_{2}$ & 220.730 & 97 & 1.86 (0.03) & 34.9 (0.1) & 9.9 (0.2) & 19.6
(0.5)\\
CH$_{3}$CN & 12$_{1}$-11$_{1}$\tablenotemark{g} & 220.743 & 76 & \nodata & \nodata & \nodata & \nodata\\
CH$_{3}$CN & 12$_{0}$-11$_{0}$\tablenotemark{h} & 220.747 & 69 & \nodata & \nodata & \nodata & \nodata\\
C$_{2}$H$_{5}$CN & 26$_{2,25}$-25$_{2,24}$ & 229.265 & 154 & 0.56(0.05) & 35.7 (0.4) & 8.7
(0.9) & 5.2 (0.7)\\ 
\methanol & 15$_{4,11}$-16$_{3,13}$E & 229.589 & 374& 0.58 (0.05) & 35.1 (0.4) & 10.4 (1.0) &
6.5 (0.8) \\
\methanol & 8$_{-1,8}$-7$_{0,7}$E & 229.759 & 89 & 1.67 (0.07) & 35.5 (0.1) & 7.2 (0.3) & 12.8 (0.8)\\
\methanol & 19$_{5,14}$-20$_{4,17}$A$^{-}$ & 229.939 & 579 & 0.43 (0.09) &
32.1 (0.4) & 4.2 (1.0) & 1.9 (0.6)\\ 
\methanol & 3$_{-2,2}$-4$_{-1,4}$E & 230.027 & 40 & 0.83(0.06) & 35.8 (0.2) & 7.2 (0.6) & 6.3
(0.7)\\
$^{12}$CO & 2-1 & 230.538 & 17& \nodata & \nodata & \nodata & \nodata \\
OCS & 19-18 & 231.061 & 111 & 1.80 (0.08) & 35.0 (0.2) & 7.9 (0.4) & 15.2
(1.0)\\
\cutinhead{MM3}
C$^{18}$O & 2-1 & 219.560 & 16 & 1.54 (0.09) & 34.4 (0.1) & 3.5 (0.2) & 5.7 (0.5)\\
\enddata
\tablenotetext{a}{Formal errors from the single Gaussian fits are given in ().}
\tablenotetext{b}{Blended with HNCO(10$_{2,8}$-9$_{2,7}$)}
\tablenotetext{c}{Blended with HNCO(10$_{2,9}$-9$_{2,8}$)}
\tablenotetext{d}{Profile not Gaussian--skewed to red.}
\tablenotetext{e}{For reference, the frequency of the k=8
\methcyn(12-11) transition is listed here and marked in Figure~\ref{g1192_smaspec}, though it is detected at $<$3$\sigma$.}
\tablenotetext{f}{Partially blended with nearby line.}
\tablenotetext{g}{Blended with CH$_{3}$CN(12$_{0}$-11$_{0}$)}
\tablenotetext{h}{Blended with CH$_{3}$CN(12$_{1}$-11$_{1}$)}
\end{deluxetable}

\begin{deluxetable}{llcccccc}
\tablewidth{0pc}
\tablecolumns{8}
\tabletypesize{\scriptsize}
\tablecaption{1.3 mm Spectral Lines: G19.01$-$0.03MM1 \label{g19trans}}
\tablehead{
\colhead{} & 
\colhead{} & 
\colhead{} & 
\colhead{} & 
\multicolumn{4}{c}{Fitted Line Parameters} \\
\colhead{Species} & 
\colhead{Transition} & 
\colhead{Frequency}  & 
\colhead{E$_{\rm upper}$/k} &
\colhead{Intensity\tablenotemark{a}}  & 
\colhead{V$_{center}$\tablenotemark{a}} &
\colhead{Width\tablenotemark{a}} & 
\colhead{$\int$ S dv\tablenotemark{a}}\\
\colhead{}  & 
\colhead{} &   
\colhead{(GHz)}    &  
\colhead{(K)} & 
\colhead{\jb}    &  
\colhead{\kms} &     
\colhead{\kms} &
\colhead{\jb \kms}
}    
\startdata
C$^{18}$O & 2-1 & 219.560 & 16 & 1.30 (0.08) & 60.3 (0.1) &  3.8 (0.3) & 	5.3 (0.5)\\ 
HNCO & 10$_{0,10}$-9$_{0,9}$ & 219.798 & 58 & 0.34 (0.04) & 59.2 (0.6) &	10.1 (1.5) &
3.6 (0.7) \\ 
H$_{2}^{13}$CO & 3$_{1,2}$-2$_{1,1}$ & 219.909 & 33 & 0.36 (0.04) &59.3 (0.3) &	5.3
(0.7) & 	2.0 (0.3) \\ 
SO & 6$_{5}$-5$_{4}$\tablenotemark{b} & 219.949 & 35 & 0.51 (0.06) & 	59.0 (0.2) & 	3.8 (0.5) &	2.1 (0.4)\\
\methanol & 8$_{0,8}$-7$_{1,6}$ \tablenotemark{b}& 220.078 & 97& 0.59 (0.06) & 59.9 (0.3) & 5.0 (0.6) & 3.1
(0.5) \\
$^{13}$CO & 2-1 &  220.399 & 16 & \nodata & \nodata & \nodata & \nodata \\
CH$_{3}$CN & 12$_{4}$-11$_{4}$ & 220.679 & 183& 0.34 (0.03) & 58.5 (0.4) & 10.0 (1.1) & 3.6
(0.5)  \\ 
CH$_{3}$CN & 12$_{3}$-11$_{3}$\tablenotemark{b} & 220.709 & 133 & 0.61 (0.05) & 60.6 (0.3) & 6.3 (0.6)	& 4.1
(0.5) \\
CH$_{3}$CN & 12$_{2}$-11$_{2}$ & 220.730 & 97 & 0.58 (0.05) & 60.1 (0.2) &  3.7 (0.4)	& 2.3
(0.3)\\
CH$_{3}$CN & 12$_{1}$-11$_{1}$\tablenotemark{c} & 220.743 & 76 & 0.58 (0.06) & 59.5 (0.2) &
3.9 (0.6) & 2.4 (0.4)\\
CH$_{3}$CN & 12$_{0}$-11$_{0}$\tablenotemark{c} & 220.747 & 69 & 0.61 (0.05) & 59.6 (0.2) & 4.3 (0.6) & 2.8
(0.5)\\
\methanol & 8$_{-1,8}$-7$_{0,7}$E & 229.759 & 89 & 0.8 (0.1) & 	60.4 (0.2) & 3.6 (0.5)
& 3.2 (0.6)\\
\methanol & 3$_{-2,2}$-4$_{-1,4}$E\tablenotemark{b} & 230.027 & 40 & 0.43 (0.06) & 60.0 (0.4) & 5.6 (1.0) &
2.56 (0.6)\\
$^{12}$CO & 2-1 & 230.538 & 17 & \nodata & \nodata & \nodata & \nodata \\
OCS & 19-18 & 231.061 & 111 & 0.60 (0.06) & 60.0 (0.3) & 5.0 (0.6) & 3.2 (0.5)\\
\enddata
\tablenotetext{a}{Formal errors from the single Gaussian fits are given in ().}
\tablenotetext{b}{Profile not Gaussian.  \meth\/ lines have blue wings.
  \methcyn\/ k=3 line is skewed to blue of fit.}
\tablenotetext{c}{CH$_{3}$CN (12$_{0}$-11$_{0}$) and CH$_{3}$CN
  (12$_{1}$-11$_{1}$) are partially blended.  Parameters for
  \methcyn\/ k=0 and k=1 lines derived from a two-component Gaussian fit.}
\end{deluxetable}

\begin{deluxetable}{lcccccccccc}
\tablewidth{0pc}
\setlength{\tabcolsep}{0.05in}
\tablecaption{\methcyn\/ Fit Parameters \label{methcynfittable}}
\tablehead{
\colhead{Source} &
\colhead{Distance} &
\colhead{T} & 
\multicolumn{3}{c}{Size} & 
\colhead{N(\methcyn)} & 
\colhead{Abundance} & 
\colhead{N(\h)} & 
\colhead{n(\h)} & 
\colhead{M$_{gas}$} \\
\colhead{} & 
\colhead{kpc} &
\colhead{K} & 
\colhead{\pp} & 
\colhead{pc} &
\colhead{AU} & 
\colhead{cm$^{-2}$} & 
\colhead{} & 
\colhead{cm$^{-2}$} & 
\colhead{cm$^{-3}$} &
\colhead{\msun} 
}
\tablecolumns{11}
\tabletypesize{\scriptsize}
\startdata
\sidehead{Single Component Fit}
G11.92$-$0.61-MM1 & 3.8 & 140 & 1.0 & 0.02 & 3800 & 3.3$\times$10$^{16}$ &
10$^{-7}$ & 3.3$\times$10$^{23}$ & 1.2$\times$10$^{7}$ & 1.9\\
 &  &&  &  &  &  & 10$^{-8}$ & 3.3$\times$10$^{24}$ & 1.2$\times$10$^{8}$ & 19\\
 &  &&  &  &  &  & 10$^{-9}$ & 3.3$\times$10$^{25}$ & 1.2$\times$10$^{9}$ & 192\\
G19.01$-$0.03-MM1 & 4.2 & 114 & 0.6 & 0.01 & 2500 & 1.7$\times$10$^{16}$ &
10$^{-7}$ & 1.7$\times$10$^{23}$ & 8.8$\times$10$^{6}$ & 0.4\\
&  &&  &  &  &  & 10$^{-8}$ & 1.7$\times$10$^{24}$ & 8.8$\times$10$^{7}$ & 4.2\\
&  &&  &  &  &  & 10$^{-9}$ & 1.7$\times$10$^{25}$ & 8.8$\times$10$^{8}$ & 42\\
\sidehead{Two Component Fit}
\sidehead{G11.92$-$0.61-MM1}
Cool & 3.8 & 77 & 3.0 & 0.06 & 11400 & 1.1$\times$10$^{15}$ &
10$^{-7}$ & 1.1$\times$10$^{22}$ & 1.3$\times$10$^{5}$ & 0.6\\
&  &&  &  &  &  & 10$^{-8}$ & 1.1$\times$10$^{23}$ & 1.3$\times$10$^{6}$ &
5.6\\
&  &&  &  &  &  & 10$^{-9}$ & 1.1$\times$10$^{24}$ & 1.3$\times$10$^{7}$ &
56\\
Warm & 3.8 & 166 & 0.6 & 0.01 & 2300 & 2.3$\times$10$^{17}$ & 10$^{-7}$ & 2.3$\times$10$^{24}$ & 1.4$\times$10$^{8}$ & 4.9\\
&  &&  &  &  &  & 10$^{-8}$ & 2.3$\times$10$^{25}$ & 1.4$\times$10$^{9}$ &
49\\
&  &&  &  &  &  & 10$^{-9}$ & 2.3$\times$10$^{26}$ & 1.4$\times$10$^{10}$ &
490\\
\enddata
\end{deluxetable}

\begin{deluxetable}{lccccc}
\tablewidth{0pc}
\setlength{\tabcolsep}{0.05in}
\tablecaption{Derived Properties of Millimeter Continuum Sources \label{dustmasstable}}
\tablehead{
\colhead{Source} & 
\colhead{T$_{dust}$} & 
\colhead{$\tau_{dust}$} & 
\colhead{M$_{gas}$} & 
\colhead{N$_{H_{2}}$\tablenotemark{a}} & 
\colhead{n$_{H_{2}}$\tablenotemark{a}} \\
\colhead{} & 
\colhead{K} & 
\colhead{} & 
\colhead{\msun} & 
\colhead{10$^{23}$ cm$^{-2}$} & 
\colhead{10$^{7}$ cm$^{-3}$} 
}
\tablecolumns{6}
\tabletypesize{\scriptsize}
\startdata
\cutinhead{SMA 1.3 mm}
G11.92$-$0.61-MM1 & 70-190 & 0.026-0.010 & 23.5-8.2 & 6.6-2.3 & 1.0-0.3 \\
G11.92$-$0.61-MM2 & 20-40 & 0.056-0.028 & 61.6-26.3 & 17.4-7.4 & 2.6-1.1 \\
G11.92$-$0.61-MM3 & 30-80 & 0.009-0.003 &8.8-2.9 & 2.5-0.8 & 0.4-0.1 \\
Total & & &93.9-37.4 & & \\
5$\sigma$ & 20/50& & 5.2/1.7 & & \\
\hline
G19.01$-$0.03-MM1 &100-130 & 0.016-0.012 & 15.9-12.0 & 3.9-3.0 & 0.5-0.4 \\
5$\sigma$ & 20/50 && 6.3/2.1 & & \\
\cutinhead{CARMA 1.4 mm}
G11.92$-$0.61-MM1 & 70-190 &0.066-0.024 & 12.9-4.5 & 16.8-5.8 & 5.3-1.8 \\
G11.92$-$0.61-MM2 & 20-40 & 0.175-0.084 &41.6-17.3 & 53.9-22.4 & 16.9-7.1 \\
G11.92$-$0.61-MM3 & 30-80 & 0.033-0.012 &7.0-2.3 & 9.1-3.0 & 2.9-1.0 \\ 
Total & & &61.5-24.1 & & \\
5$\sigma$ & 20/50 && 6.9/2.3 & & \\
\cutinhead{CARMA 3.4 mm}
G19.01$-$0.03-MM1 &100-130 & 0.002-0.001 &40.2-30.7 & 1.9-1.5 & 0.1-0.09 \\
5$\sigma$ & 20/50& & 20.6/7.7 & & \\
\enddata
\tablenotetext{a}{Beam-averaged quantities}
\end{deluxetable}

\begin{deluxetable}{llcccccccccc}
\tablewidth{0pc}
\setlength{\tabcolsep}{0.04in}
\tablecaption{Outflow Parameters \label{outflowtable}}
\tablehead{
\colhead{} &
\colhead{v$_{min}$\tablenotemark{a}} &
\colhead{v$_{max}$\tablenotemark{a}} &
\colhead{Min($|v-$\vlsr$|$)} & 
\colhead{M$_{out}$ } & 
\colhead{$\theta$\tablenotemark{b}} &
\colhead{P$_{out}$} & 
\colhead{E$_{out}$} & 
\colhead{Length} &
\colhead{T$_{dyn}$} & 
\colhead{\.{M}$_{out}$} &
\colhead{\.{P}$_{out}$} \\ 
\colhead{} & 
\colhead{\kms} &
\colhead{\kms} &
\colhead{\kms} &
\colhead{\msun} & 
\colhead{\degree} &
\colhead{\msun\/ \kms} & 
\colhead{ergs} &
\colhead{pc} &  
\colhead{yrs} & 
\colhead{\msun\/ yr$^{-1}$} & 
\colhead{ \msun\/ \kms\/ yr$^{-1}$} \\
\colhead{} & 
\colhead{} & 
\colhead{} & 
\colhead{} & 
\colhead{} & 
\colhead{} &
\colhead{}  & 
\colhead{$\times$10$^{45}$} & 
\colhead{} & 
\colhead{} & 
\colhead{$\times$10$^{-4}$} & 
\colhead{} 
\\}
\tablecolumns{12}
\tabletypesize{\scriptsize}
\startdata
\cutinhead{G11.92$-$0.61: Main Outflow}
\sidehead{\textbf{\co(2-1)}}
Red & 48.1 & 71.2 & 13 & 0.2 & - & 3.1 & 0.6 & 0.2 & 5000 & 0.4 & 0.001 \\
    &      &      &    &     & 10 & 17.7 & 19.0 & 0.2 & 900 & 2.0 & 0.020 \\
    &      &      &    &     & 30 & 6.1 & 2.3 & 0.2 & 2900 & 0.6 &  0.002 \\
    &      &      &    &     & 60 & 3.5 & 0.8 & 0.3 & 8700 & 0.2 & 0.0004 \\
Blue & $-$24.4 & 21.8 & 13 & 0.6 & - & 12.6 & 3.2 & 0.4 & 6000 & 1.0 &  0.002 \\
   &      &      &    &          & 10 & 72.6 & 107 & 0.4 & 1100 &  5.6 & 0.069  \\
    &      &      &    &        & 30 & 25.2 & 12.9 & 0.4 & 3500 & 1.7 & 0.007  \\
    &      &      &    &         & 60 & 14.6 & 0.4 & 0.8 & 10400 & 0.6 & 0.001  \\
Total\tablenotemark{c} & & & & 0.8 & - & 15.7 & 3.8 &0.5 & & 1.3 &  0.003  \\
                     & &&    &     & 10 & 90.3 & 126 & 0.6 & &7.6 &  0.089  \\
                     & &&    &     & 30 & 31.4 & 15.2 & 0.6 & &2.3 & 0.009  \\
                     & &&    &     & 60 & 18.1 & 5.1 & 1.1 & &0.8 & 0.002  \\
\hline
\sidehead{\textbf{\hco(1-0)}}
Red & 48.1 & 71.1 & 13 & 2.6 & - &55.2 & 12.9 & 0.2 & 5000\tablenotemark{d}& 5.2 & 0.011  \\
    &      &      &    &     & 10& 317.7 &  427 & 0.2 & 900 & 29.5 & 0.360  \\
    &      &      &    &     & 30& 110.3 & 51.5 & 0.2 & 2900 & 9.0 & 0.038  \\
    &      &      &    &     & 60& 63.7 & 17.2 &  0.4 & 8700 & 3.0 & 0.007  \\
Blue & -14.3 & 21.8 & 13 & 5.2 & - & 134.2 & 40.2 & 0.4 & 6000\tablenotemark{d} & 8.7 & 0.022  \\
    &      &      &    &     & 10& 773.0 & 1330 & 0.4 & 1100 & 49.4 &  0.731  \\
    &      &      &    &     & 30& 268.5 & 161 & 0.4 & 3500 & 15.1 & 0.078  \\
    &      &      &    &     & 60& 155.0 & 53.6 & 0.7 & 10400 & 5.0 & 0.015  \\
Total\tablenotemark{c} & & & & 7.8 & - & 189.4 & 53.1 & 0.6 & & 13.9 & 0.033  \\
    & & & &       & 10& 1090.7 & 1760 & 0.6 & & 78.9 & 1.091  \\
    && &    &     & 30& 378.8 & 212 & 0.6 &  & 24.1 & 0.116  \\
    && &    &     & 60& 218.7 & 70.8 & 1.1 & & 8.0 & 0.022  \\

\cutinhead{G11.92$-$0.61: Northern Outflow}
\sidehead{\textbf{\co(2-1)}}
Red & 48.1 & 64.6 & 13 & 0.14 & - & 2.9 & 0.6 & 0.4 & 12000 & 0.1 & 0.0002  \\
    &      &      &    &      & 10& 16.6 & 20.7 & 0.4 & 2100 & 0.7 & 0.008  \\
    &      &      &    &      & 30& 5.8 & 2.5 & 0.4 &  6900 & 0.2 & 0.001  \\
    &      &      &    &      & 60& 3.3 & 0.8 & 0.7 & 20800 & 0.07 & 0.0002  \\
Blue & 8.6 & 21.8 & 13 & 0.05 & - & 0.8 & 0.1 & 0.1 & 5000 & 0.09 & 0.0002  \\
    &      &      &    &      & 10& 4.6 & 4.8 & 0.1 & 900 & 0.5 & 0.005  \\
    &      &      &    &      & 30& 1.6 & 0.6 & 0.2 & 2900 & 0.2 & 0.001  \\
    &      &      &    &      & 60& 0.9 & 0.2 &  0.3 & 8700 & 0.05 & 0.0001  \\
Total & & & &  0.2 & - & 3.7 & 0.7 &  0.5 & 8500\tablenotemark{e} & 0.2 & 0.0004  \\
      & & & &       &10& 21.3 & 25.5 & 0.5 & 1500 & 1.3 & 0.014  \\
      & & & &       &30& 7.4 & 3.1 & 0.6 &  4900 & 0.4 & 0.002  \\
      & & & &       &60& 4.3 & 1.0 & 1.0 & 14700 & 0.1 & 0.0003  \\
\hline
\sidehead{\textbf{\hco(1-0)}}
Red & 48.1 & 59.6 & 13 & 0.8 & - & 13.4 & 2.4 & & 12000\tablenotemark{d}& 0.7 & 0.001  \\
    &      &      &    &     & 10 & 77.1 & 79.3 & & 2100 & 3.7 & 0.036  \\
    &      &      &    &     & 30 & 26.8 & 9.6 & & 6900 & 1.1 & 0.004  \\
    &      &      &    &     & 60 & 15.5 & 3.2 & & 20800 & 0.4 & 0.001  \\
Blue & 15.3 & 21.8 & 13 & 0.3 & - &4.0 & 0.6 & & 5000\tablenotemark{d} & 0.5 & 0.001  \\
    &      &      &    &      & 10 & 23.1 & 21.2 & & 900 & 2.9 & 0.026  \\
    &      &      &    &      & 30 & 8.0 & 2.6 & & 2900 & 0.9 & 0.003  \\
    &      &      &    &      & 60 & 4.6 & 0.9 & & 8700 & 0.3 & 0.001  \\
Total & & & & 1.0 & - & 17.4 & 3.0 & & 8500\tablenotemark{d} & 1.2 &  0.002  \\
& &      &    &   & 10 & 110.2 & 100 & & 1500 & 6.9 & 0.067  \\
& &    &    &     & 30 & 34.8 & 12.1 & & 4900 & 2.1 & 0.007  \\
& &    &    &     & 60 & 20.1 & 4.0 & & 14700 & 0.7 & 0.001  \\
\cutinhead{G19.01$-$0.03}
\sidehead{\textbf{\co(2-1)}}
Red & 75.6 & 88.8 & 16 & 0.1 & - & 2.3 & 0.5 & 0.4 & 16400 & 0.07 & 0.0001  \\
    &      &      &    &     & 10 & 13.2 & 15.3 & 0.4 & 2900 & 0.41 & 0.005  \\
    &      &      &    &     & 30 & 4.6 & 1.9 & 0.5 & 9500 & 0.13 & 0.0005  \\
    &      &      &    &     & 60 &  2.6 & 0.6 & 0.9 & 28400 & 0.004 & 0.00009  \\
Blue\tablenotemark{f} & -46.4 & 39.3 & 21 & 0.3 & - &8.4 & 3.3 & 0.3 & 3000 & 1.0& 0.003 \\
    &      &      &    &                        & 10 & 48.6 & 110 & 0.3 & 500 & 6.0 & 0.09 \\
    &      &      &    &                        & 30 & 16.9 & 13.2 & 0.4 &  1700 & 1.8 & 0.01 \\
    &      &      &    &                        & 60 & 9.8 & 4.4 &  0.6 & 5200 & 0.6 &  0.002 \\
NE Blue clump & 9.6 & 39.3 & 21 & 0.1 & - & 2.2 & 0.7 & & & & \\
Total\tablenotemark{g} & & & & 0.5 & - & 12.9 & 4.5 & 0.7& 5000 &1.1 & 0.003  \\
    &      &      &    &           & 10 & 74.3 & 149 & 0.7 & 900 & 6.4 & 0.08  \\
    &      &      &    &           & 30 & 25.8 & 17.9 & 0.8 & 2900 & 2.0 & 0.009 \\
    &      &      &    &           & 60 &  14.9 & 6.0 & 1.4 & 8700 & 0.7 & 0.002 \\

\hline
\sidehead{\textbf{\hco(1-0)}}
Red & 66.9 & 83.4 & 6.9 & 3.6 & - & 40.4 & 5.4 & 0.4 &16400\tablenotemark{d} & 2.2 & 0.002\\
    &      &      &     &     & 10 &232.4& 177 & 0.4 & 2900 & 12.5 & 0.080 \\
    &      &      &     &     & 30 &80.7& 21.4 & 0.4 & 9500 & 3.8 & 0.009\\
    &      &      &     &     & 60 &46.6& 7.1 & 0.7 & 28400 & 1.3 & 0.002\\
Blue & 7.9 & 53.8 & 6.2 & 9.5 & - & 155.7 & 42.3 &0.4 &3000\tablenotemark{d} & 31.5 & 0.052\\
     &      &      &     &     & 10 &896.4& 1400 & 0.4 & 500 & 179 & 1.695\\
     &      &      &     &     & 30 &311.3& 169 & 0.5 & 1700 & 54.5 & 0.180\\
     &      &      &     &     & 60 &179.7& 56.4 & 0.8 & 5200 & 18.2 & 0.035\\
Total & & & & 13.1 & - & 196.0 & 47.7 & 0.8 & 5000& 26.1& 0.039\\
      & & & &      & 10 &1128.7& 1580 & 0.8 & 900 & 148 & 1.280\\
      & & & &      & 30 &392.0& 191 & 0.9 & 2900 & 45.2 & 0.136\\
      & & & &      & 60 &226.3& 63.6 & 1.5 & 8700 & 15.1 & 0.026\\
\hline
Red & 68.6 & 83.4 & 8.6 & 2.4 & - &  31.9 & 4.8 & & &1.5 & 0.002\\
    &      &      &     &     & 10 & 183.7 & 158 & & & 8.3 & 0.064\\
    &      &      &     &     & 30 & 63.8 & 19.1 & & & 2.5 & 0.007\\
    &      &      &     &     & 60 & 36.8 & 6.4 & & & 0.8 & 0.001\\
Blue & 7.9 & 52.2 & 7.8 & 6.5 & - &  137.2 & 41.2 & & &21.5 & 0.046\\
      &      &      &     &     & 10 & 790.1 & 1370 & & & 122 & 1.494\\
      &      &      &     &     & 30 & 274.4 & 165 & & & 37.3 & 0.158\\
      &      &      &     &     & 60 & 158.4 & 54.9 & & & 12.4 & 0.030\\
Total & & & & 8.9 & - & 169.1 & 45.9 & & & 17.7 & 0.034\\
      & & & &     & 10 &973.8& 1520 & & & 100 & 1.104\\
      & & & &     & 30 &338.2& 184 & & & 30.7 & 0.117\\
      & & & &     & 60 &195.2& 61.3 & & & 10.2 & 0.023\\
\hline
Red & 70.2 & 83.4 & 10.2 & 1.8 & - & 26.4 & 4.3 & & &1.1& 0.002\\
    &      &      &     &     & 10 & 151.8 & 142 & & & 6.0 & 0.052\\
    &      &      &     &     & 30 & 52.7 & 17.2 & & & 1.8 & 0.006\\
    &      &      &     &     & 60 & 30.4 & 5.7 & & & 0.6 & 0.001\\
Blue & 7.9 & 50.5 & 9.5 & 5.0 & - & 125.6 & 40.3 & & &16.6 & 0.042\\
      &      &      &     &     & 10 & 723.4 & 1340 & & & 94.2 & 1.367\\
      &      &      &     &     & 30 & 251.2 & 161 & & & 28.8 & 0.145\\
      &      &      &     &     & 60 & 145.0 & 53.7 & & & 9.6 & 0.028\\
Total & & & & 6.7 & - & 152.0 & 44.6 & & &13.5 & 0.030\\
      & & & &     & 10 &875.2& 1480 & & & 76.3 & 0.993\\
      & & & &     & 30 &303.9& 178 & & & 23.3 & 0.105\\
      & & & &     & 60 &175.5& 59.4 & & & 7.8 & 0.020\\
\enddata
\tablenotetext{a}{Central velocity of first/last channel used for
  calculation of outflow properties.}
\tablenotetext{b}{Inclination angle \emph{from the plane of the sky}.}]
\tablenotetext{c}{Sum of red and blue lobes.}
\tablenotetext{d}{From \co}
\tablenotetext{e}{Calculated from the end-to-end outflow length divided by 2
  and v$_{max,outflow}$=28 \kms.}
\tablenotetext{f}{Excluding channels v=26.1 and 22.8 \kms.  These channels are
  dominated by image artifacts from resolved-out large scale emission.}
\tablenotetext{g}{Including NE blue clump}
\end{deluxetable}

\clearpage

\clearpage

\clearpage

\clearpage


\end{document}